\documentclass[12pt]{article}

\usepackage{bm, commath}
\usepackage{natbib}
\usepackage{caption}
\usepackage{cancel}
\usepackage{subcaption}
\usepackage{amsmath,graphicx}
\usepackage{amsmath, amsfonts, amsthm}
\usepackage{float}
\usepackage{booktabs,siunitx}
\usepackage{url}
\usepackage{multirow}
\usepackage{amssymb}
\usepackage{blkarray}
\usepackage{bbm}
\usepackage{multicol}
\usepackage{authblk}
\usepackage{natbib}
\usepackage{hyperref}
\usepackage[shortlabels]{enumitem}
\usepackage{longtable}
\usepackage{adjustbox}
\usepackage{threeparttable}
\usepackage{hhline}
\usepackage{colortbl}
\usepackage{lscape}
\usepackage{xr}
\usepackage{cleveref}
\usepackage{listings}
\usepackage{bbm}
\usepackage{textcomp}
\usepackage{authblk}
\usepackage[ruled]{algorithm2e}
\SetKwInput{KwParam}{Parameter}
\SetAlgoCaptionLayout{centerline}

\usepackage{sectsty}
\usepackage[doublespacing]{setspace}
\usepackage[utf8]{inputenc}
\usepackage{float}
\usepackage{tikz}
\usetikzlibrary{shapes,decorations,arrows,calc,arrows.meta,fit,positioning}
\usepackage{comment}
\usepackage{mwe}
\usepackage{graphicx}

\newtheorem{assumption}{Assumption}
\newtheorem*{assumption*}{Assumption}

\newtheorem{proposition}{Proposition}

\newtheorem{lemma}{Lemma}

\newtheorem{innercustomgeneric}{\customgenericname}
\providecommand{\customgenericname}{}
\newcommand{\newcustomtheorem}[2]{%
  \newenvironment{#1}[1]
  {%
   \renewcommand\customgenericname{#2}%
   \renewcommand\theinnercustomgeneric{##1}%
   \innercustomgeneric
  }
  {\endinnercustomgeneric}
}

\newcustomtheorem{customthm}{Theorem}
\newcustomtheorem{customassumption}{Assumption}

\usepackage{mathtools}

\usepackage{xcolor}

\usepackage{accents}

\makeatletter
\renewcommand{\algocf@captiontext}[2]{#1\algocf@typo. \AlCapFnt{}#2} % text of caption
\def\@algocf@capt@plain{top}
\renewcommand{\algocf@makecaption}[2]{%
  \addtolength{\hsize}{\algomargin}%
  \sbox\@tempboxa{\algocf@captiontext{#1}{#2}}%
  \ifdim\wd\@tempboxa >\hsize%     % if caption is longer than a line
  \hskip .5\algomargin%
  \parbox[t]{\hsize}{\algocf@captiontext{#1}{#2}}% then caption is not centered
  \else%
  \global\@minipagefalse%
  \hbox to\hsize{\box\@tempboxa}% else caption is centered
  \fi%
  \addtolength{\hsize}{-\algomargin}%
}
\makeatother

\begin{document}

\def\spacingset#1{\renewcommand{\baselinestretch}%
{#1}\small\normalsize} \spacingset{1}

\sectionfont{\bfseries\large\sffamily}%

\subsectionfont{\bfseries\sffamily\normalsize}%

%\begin{center}
%\noindent
%{\sffamily\bfseries\LARGE 
%}%

%\noindent

%\end{center}

\title{Debiasing hazard-based, time-varying vaccine effects using vaccine-irrelevant infections:\\  
\Large An observational extension of a pivotal Phase 3 COVID-19 vaccine efficacy trial
}

\author[1,2]{Ethan Ashby}
\author[3]{Dean Follmann}
\author[2]{Holly Janes}
\author[2]{Peter B. Gilbert}
\author[1]{Ting Ye}
\author[4]{Lindsey R Baden}
\author[5]{Hana M El Sahly}
\author[2]{Bo Zhang}

\affil[1]{\small{Department of Biostatistics, University of Washington, Seattle, WA, USA}}
\affil[2]{\small{Vaccine and Infectious Disease Division, Fred Hutchinson Cancer Center, Seattle, WA, USA}}
\affil[3]{\small{Biostatistics Research Branch, National Institute of Allergy and Infectious Diseases, Bethesda, MD, USA}}
\affil[4]{\small{Brigham and Womens' Hospital, Harvard Medical School, Boston, MA, USA}}
\affil[5]{\small{Departments of Molecular Virology and Microbiology and Medicine, Baylor College of Medicine, Houston, TX, USA}}

\date{\today}
\maketitle

%\author[]{}
%\date{}

\iffalse
\bigskip
  \bigskip
  \bigskip
  \begin{center}
    {\Large\bf Debiasing hazard-based, time-varying vaccine effects using vaccine-irrelevant infections} \\ \vspace{0.1cm} {\large An observational extension of a pivotal Phase 3 COVID-19 vaccine efficacy trial}
\end{center}
  \medskip
\bigskip
\fi

\begin{abstract}
Understanding how vaccine effectiveness (VE) changes over time can provide evidence-based guidance for public health decision making. While commonly reported by practitioners, time-varying VE estimates obtained using Cox regression are vulnerable to hidden biases. To address these limitations, we describe how to leverage vaccine-irrelevant infections to identify hazard-based, time-varying VE in the presence of unmeasured confounding and selection bias. We articulate assumptions under which our approach identifies a causal effect of an intervention deferring vaccination and interaction with the community in which infections circulate. We develop sieve and efficient influence curve-based estimators and discuss imposing monotone shape constraints and estimating VE against multiple variants. As a case study, we examine the observational booster phase of the Coronavirus Vaccine Efficacy (COVE) trial of the Moderna mRNA-1273 COVID-19 vaccine which used symptom-triggered multiplex PCR testing to identify acute respiratory illnesses (ARIs) caused by SARS-CoV-2 and 20 off-target pathogens previously identified as compelling negative controls for COVID-19. Accounting for vaccine-irrelevant ARIs supported that the mRNA-1273 booster was more effective and durable against Omicron COVID-19 than suggested by Cox regression. Our work offers an approach to mitigate bias in hazard-based, time-varying treatment effects in randomized and non-randomized studies using negative controls.
\end{abstract}
\vspace{0.3 cm}

\newpage
\spacingset{1.6}

\setlength\abovedisplayskip{2pt}
\setlength\belowdisplayskip{1pt}%

\section{Introduction}\label{s:intro}\vspace{-4mm}

\subsection{The promise and perils of hazard-based time-varying vaccine effects}\label{ss:intro promise and perils}

A key goal of vaccine research is to developing novel vaccines or adapt existing vaccines to be long-lasting and broadly protective against many pathogen strains. Additionally, understanding the breadth and durability of vaccine effectiveness (VE) can inform decisions regarding whether and when to offer vaccines to the public. For example, evidence of waning influenza vaccine effectiveness supports that vaccinating high-risk groups closer to the flu season may offer greater protection than vaccination well in advance of the flu season \citep{BELONGIA2015246, Ferdinands2017}. Hence, understanding \textit{time-varying VE}, or vaccine effectiveness in preventing disease $X$ days after vaccination, is of interest to both vaccine science and public health. A common approach to estimate time-varying VE is to fit a Cox proportional hazards model with a coefficient that varies in time-since-vaccination \citep{lin_effectiveness_2022, lin_durability_2023}. However, hazard-based VE estimates are vulnerable to subtle biases that compromise their reliability in both randomized and observational studies. We describe these sources of bias below.

\begin{figure}
    \centering
    \includegraphics[width=0.85\linewidth]{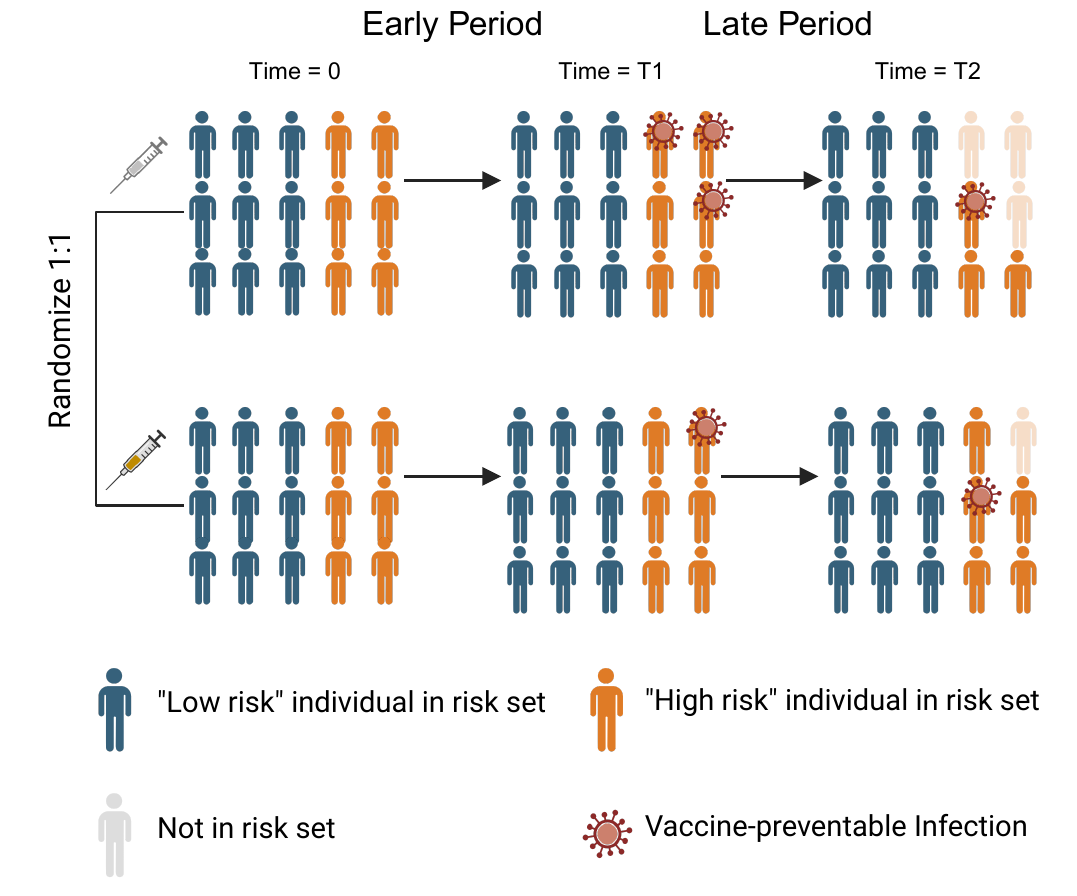}
    \caption{Illustration of depletion of susceptibles in a randomized trial comparing placebo (top) to an effective vaccine (bottom) in the presence of heterogeneous infection risk.}
    \label{fig:depletion_susceptibles}
\end{figure}

Even in randomized trials -- the gold-standard study design for evaluating vaccines -- hazard-based, time-varying vaccine effects can suffer from a built-in selection bias \citep{Hernan2010} commonly referred to as ``depletion of susceptibles bias" \citep{Lipsitch_2019}. In Figure \ref{fig:depletion_susceptibles}, we illustrate how depletion of susceptibles bias arises in a randomized trial of a moderately effective vaccine conducted in a population with heterogeneous infection risk. For illustrative purposes, suppose the study is conducted in a mixed population of ``low-risk" (blue) and ``high-risk" (orange) individuals. By virtue of randomization, the proportions of low and high risk individuals are balanced at baseline. Suppose we are interested in estimating the vaccine's efficacy during the late period (T1 to T2) and comparing it to the early period (0 to T1) to determine if VE is waning. One approach is to compare the hazard ratios of infection between the early and late periods. By the definition of the hazard function, the late period hazard ratio compares infection rates between vaccinated and unvaccinated persons \textit{who were uninfected during the early period}. Conditioning on no early period infection selects for a vaccinated risk set with higher average infection risk (i.e., higher proportion of orange participants) than the unvaccinated risk set because high-risk participants from the unvaccinated group were infected more rapidly during the early period. This example exhibits how hazard-based, time-varying vaccine effects can become entangled with time-varying imbalances in infection risk between vaccinated and unvaccinated risk sets. The result of depletion of susceptibles bias is exaggerated estimates of VE waning. Previous efforts to mitigate depletion of susceptibles bias included serological testing to detect prior natural immunity \citep{kahn_identifying_2022}, exploiting seasonal variation in infections to approximate a trial where exposure is deferred until all participants are vaccinated \citep{Lipsitch_2019, ray_depletion_susceptibles_2020}, and proposing a causal VE waning parameter where exposure is deferred until an analyst-chosen time point \citep{Janvin2024-ly}.

Conducting long, placebo-controlled vaccine trials may not be feasible in some circumstances due to resource limitations, lack of clinical equipoise, or pathogen evolution in the population. Observational studies contain rich data on how vaccine effectiveness varies across different settings, variants, and over time. However, VE estimates from observational studies are particularly vulnerable to bias. World Health Organization guidance documents \citep{world_health_organization_evaluation_2021, world_health_organization_evaluation_2022} highlight the myriad sources of confounding that could impact observational vaccine studies including differences in behaviors (such as mask-wearing and social distancing) influencing contact rates with infected community members \citep{Halloran_1996_concep, Tsiatis_2021}, immune function influencing per-contact susceptibility to infection \citep{Halloran_1996_concep, kahn_identifying_2022}, and healthcare seeking behaviors impacting case ascertainment \citep{Jacksonetal2005, jackson2013test}. These concerns are not theoretical: several observational studies in winter 2021 reported low and negative VE estimates, which are hypothesized to be the result of hidden bias \citep{bodner_observed_2023}.

\subsection{Negative control outcomes in vaccine studies}\label{ss:intro negative outcome}

Common strategies for bias mitigation like observed confounder adjustment are ill-suited to vaccine studies because the sources of bias are hard to measure, overlapping, and potentially time-varying. For example, a strong predictor of an individual's infection risk --- and therefore a potential source of bias --- is their rate of contact with infected individuals in the broader community \citep{Halloran_1996_concep}. Accurate measurement of contact rates is nearly impossible for several reasons \citep{Unkel_2014}. Contact rates change over time and measuring contacts requires knowledge about which members of the broader community are infected. Moreover, it is nearly impossible to establish a standard definition of ``contact" for community-transmitted respiratory pathogens. In practice, analysts typically measure and adjust for demographic factors (e.g., age, employment as a healthcare worker, etc.) with the hope that they are correlated with the true, unmeasured predictors of infection.

Another strategy for bias mitigation is to measure and adjust for a negative control outcome (NCO), or an auxiliary endpoint unaffected by the intervention but that shares unmeasured causes with the primary outcome \citep{rosenbaum_1984,lipsitch_negative_2010, shi_selective_2020}. A common choice of NCO in vaccine effectiveness studies is infection/illness caused by a pathogen/serotype antigenically distinct enough from the pathogen of interest such that the vaccine is assumed to have no effect -- we refer to such infections as \textit{``off-target" or ``vaccine-irrelevant"}. NCOs have been used to detect \citep{Lewnard_2020, Hitchings_2022} and eliminate \citep{broome_pneumococcal_1980, Etievantetal2023} confounding bias in time-invariant VE estimates from observational studies. However, prior negative control approaches do not address selection bias caused by differential depletion of susceptibles and do not extend naturally to estimands based on time-varying hazards, where both the treatment effect and confounding structures may vary over time. Herein, we describe an approach to identify and estimate hazard-based, time-varying vaccine effects under hidden confounding and selection bias by leveraging vaccine-irrelevant infections as NCOs.

\subsection{Case study: an observational study fused to a pivotal COVID-19 vaccine trial}\label{ss:intro case study}

\begin{figure}
    \centering
    \begin{subfigure}[b]{0.95\textwidth}
    \centering
    \includegraphics[width=0.7\linewidth]{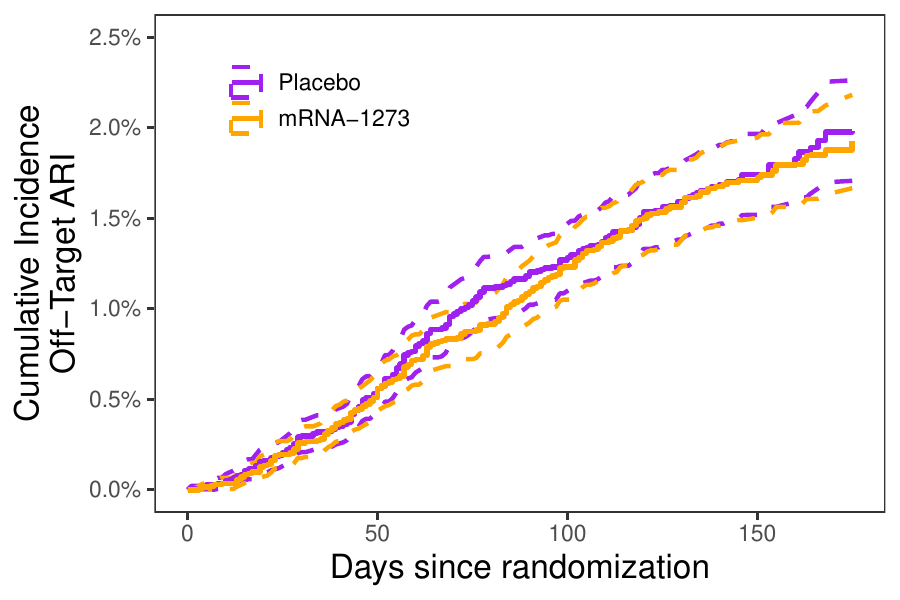}
    \caption{Acute respiratory illnesses (ARIs) caused by off-target pathogens were unaffected by mRNA-1273 vaccination in COVE's blinded phase, highlighting their validity as negative control outcomes. Cumulative incidence estimators adjusted for sex, racial/ethnic underrepresented minority status, randomization stratum, and a continuous baseline risk score \citep{gilbert_immune_2022, westling_inference_2023}.}
    \label{fig:top}
    \end{subfigure}
    \hfill
    \begin{subfigure}[b]{0.95\textwidth}
    \centering
    \includegraphics[width=0.85\linewidth]{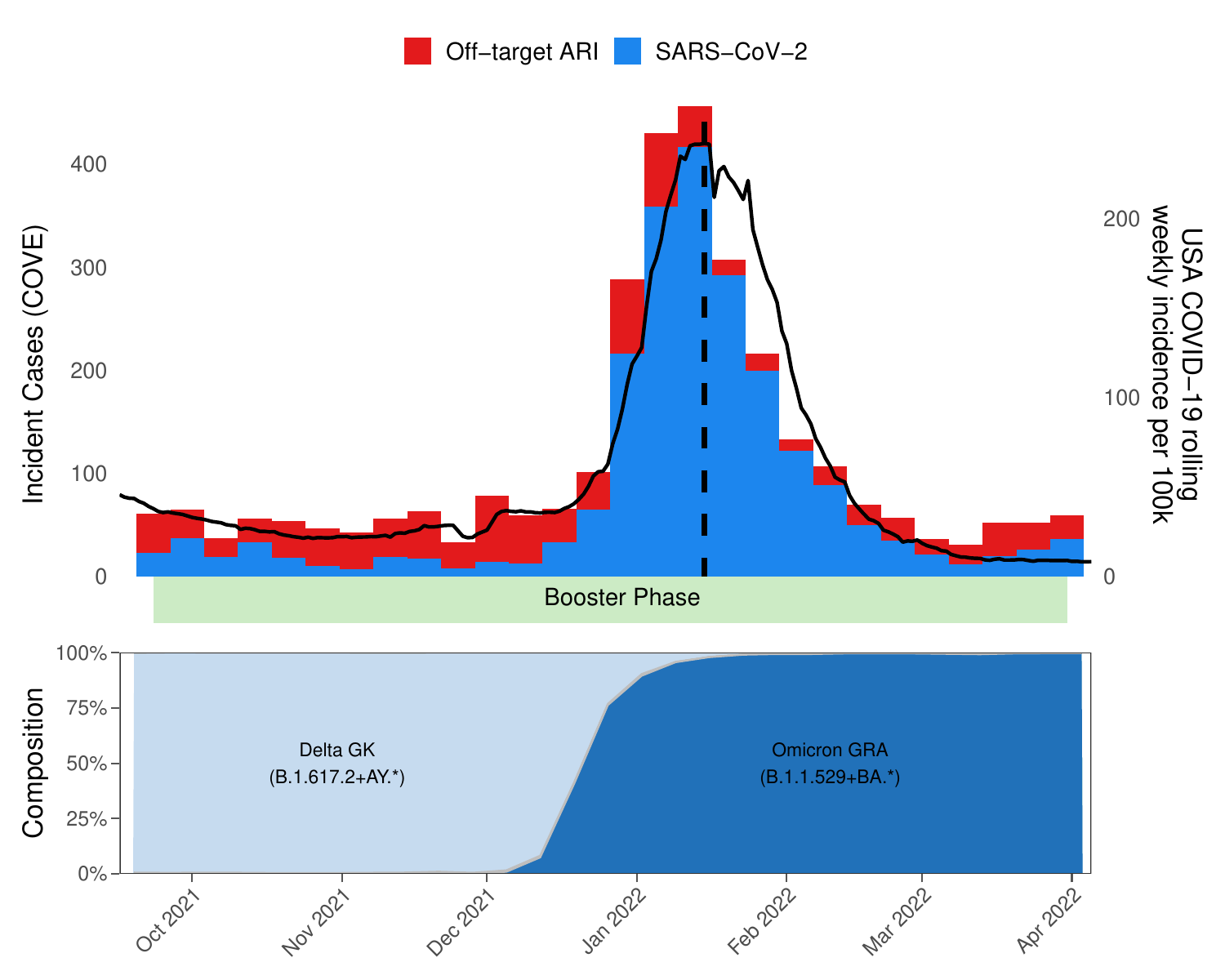}
    \caption{Stacked histogram showing weekly incident COVID-19 and off-target ARI endpoints during the booster phase of COVE. Also shown are rolling weekly COVID-19 incidence per 100,000 US residents from NYT/JHU \citep{tsueng_outbreakinfo_2023} SARS-CoV-2 variant compositions in US obtained from Global Initiative on Sharing All Influenza Data (GISAID) \citep{khare_gisaids_2021}}
    \label{fig:bottom}
    \end{subfigure}
    \caption{}\label{fig:COVE}
\end{figure}

Herein, we examine the observational, booster phase of the the Coronavirus Vaccine Efficacy (COVE) study, a phase 3, randomized, double-blind, placebo-controlled study of the Moderna mRNA-1273 COVID-19 vaccine in the US adults aged $\geq$18 years. Details of COVE's blinded and open-label phases are presented elsewhere \citep{baden2021efficacy, Elsahlyetal2021, baden_long-term_2024}. On September 23, 2021, the study protocol was amended to offer a single 50-$\mu$g mRNA-1273 booster to participants. COVE's ``booster phase" spanned from September 23, 2021 to the data cutoff date (April 5, 2022) and followed participants for a median of 5.3 months post-boost. The dominant SARS-CoV-2 lineage was Delta followed by Omicron, and COVID-19 trends in COVE mirrored those observed in the United States (US) (Figure \ref{fig:top}).

COVE was an ideal setting to study the potential of vaccine-irrelevant infections for bias elimination. COVE's booster phase emulated a prospective observational vaccine study because participants voluntarily elected if and when to receive the booster. This posed challenges to estimating VE, since vaccination status, vaccination date, and the post-vaccination period may have been associated with differences in exposure, susceptibility, and/or testing behavior that could confound VE estimates. COVE also used a multiplex PCR assay (Biofire$^{\circledR}$ RP2.1) to capture symptomatic acute respiratory illnesses (ARIs) caused by SARS-CoV-2 and a panel of 20 other ``off-target"  viral/bacterial respiratory pathogens. Since adaptive immunity is antigen-specific and the mRNA-1273 vaccine construct was targeted specifically to SARS-CoV-2, it was highly plausible that the vaccine would not affect the off-target ARIs. An analysis of COVE's blinded phase by \citet{Ashby_2025} adapted in Figure \ref{fig:top} tested this hypothesis and confirmed that mRNA-1273 vaccination had a negligible effect on the incidence of off-target ARIs. However, participants who experienced a vaccine-irrelevant ARI had 3-fold higher odds of experiencing COVID-19 during the same study period \citep{Ashby_2025}, suggesting that vaccine-irrelevant ARIs and COVID-19 were positively associated through a set of unmeasured common causes. Together, these results supported off-target ARIs as a compelling choice of NCO for COVID-19. Subsequently, \citet{Ashby_2025} proposed a strategy to \textit{detect} unmeasured confounding and selection bias in hazard-based, time-varying VE estimates by estimating the time-varying association between vaccination and off-target ARI conditional on COVID-19-free risk sets. However, their analysis did not articulate a strategy to use vaccine-irrelevant ARIs to \textit{remove} bias from time-varying VE estimates, a key goal to reliably evaluate VE durability and inform public health decision-making.

In Section 2 of this article, we propose an approach to identify hazard-based, time-varying vaccine effects in the presence of unmeasured confounding and selection bias by leveraging vaccine-irrelevant infections as NCOs. In Section 3, we discuss a set of assumptions under which our approach identifies a causal effect under a hypothetical joint intervention deferring dates of vaccination and interaction with the community in which infections circulate. In Section 4, we develop two strategies for inference and briefly discuss imposing monotone shape constraints and estimation of time-varying VE against multiple variants. In Section 5, we briefly summarize results of numerical experiments comparing the performance of our proposed estimators to standard Cox regression analyses. In Section 6, we study the COVE booster phase and estimate time-varying VE estimates against Delta and Omicron COVID-19.
\section{Methods}
\label{s:model}

\subsection{Notation}
\label{ss:notation}
Consider a study measuring two infection outcomes: a vaccine-preventable infection (denoted $J=1$) and a vaccine-irrelevant infection (denoted $J=0$). Because infection rates vary over calendar time, all subsequent developments use a calendar time scale instead of a study time scale. Throughout we will assume non-informative censoring of infection times. Define $T$ as the earliest infection date of either type and $V$ as the vaccination date. Let $U(t)$ denote an unmeasured frailty summarizing a potentially large, time-varying collection of participant characteristics that influence infection risk, although we do not need to specify what these characteristics are (see next subsection for additional discussion). We define the cause-specific hazard for infection type $J=j$ as
\begingroup
  \setlength{\abovedisplayskip}{6pt}   % space above
  \setlength{\belowdisplayskip}{6pt}
\begin{align*}
    h_{j}(t \mid V=v, U(t)=u) := \lim_{\epsilon \downarrow 0} \frac{P(J=j, T \in [t, t+\epsilon) \mid V=v, U(t) = u, T \geq t)}{\epsilon}
\end{align*}
\endgroup
where we omit stratification or adjustment for measured covariates for brevity, although they can be readily incorporated. In the developments below, let $E_{U \sim p_{U}}[\cdot]$ denote taking the expectation with respect to $U$ drawn from a distribution with density function $p_{U}$. 

\subsection{Motivating example: a multiplicative, time-varying frailty model}\label{subsec:motivating_example}

We motivate our approach using particular specifications of the cause-specific hazards of infection that extend those of \citet{fintzi2021assessing} to include a common, time-varying unmeasured frailty $U(t)$ \citep{Unkel_2014}. Let $h_{01}$ and $h_{00}$ denote the baseline hazards of vaccine-preventable ($J=1$) and vaccine-irrelevant ($J=0$) infections respectively. Let $Z(t) := I(V \leq t)$ denote the current vaccination status and $\tau=\max(0, t-V)$ denote time-since-vaccination. Suppose the cause-specific hazards take the following forms.
\begingroup
  \setlength{\abovedisplayskip}{6pt}   % space above
  \setlength{\belowdisplayskip}{6pt}
    \begin{equation}\label{eq:frailty}
    \begin{split}
    &h_{1}(t \mid V, U(t)) = h_{01}(t) U(t) \exp\{Z(t) f(\tau)\}, \\
    &h_{0}(t \mid V, U(t)) = h_{00}(t) U(t).
    \end{split}
    \end{equation}
\endgroup
Model \ref{eq:frailty} postulates the hazard of off-target infections are unaffected by vaccination and that vaccine effectiveness against the vaccine-preventable infection varies in time-since-vaccination, $\tau$, according to an unknown function $f(\cdot): \mathbb{R}^{+} \rightarrow \mathbb{R}$. Further discussion of VE's dependence on time-since-vaccination is deferred until the next section. Importantly, model \ref{eq:frailty} assumes that both vaccine-preventable and vaccine-irrelevant infections depend on an unmeasured, time-varying frailty $U(t)$. Under the conceptualization of infection risk of \citet{Halloran1991-pa}, $U(t)$ represents the product of the participant-level contact rate with infected individuals in the community, the per-contact probability of infection, and per-infection case ascertainment probability. More generally, $U(t)$ can be conceptualized as a univariate, principal frailty summarizing a potentially high-dimensional set of confounders and prognostic factors influencing infection risk. The shared dependence of the primary and negative control outcomes on a common set of unmeasured factors is a fairly standard assumption in the negative control literature referred to as \textit{$U$-comparability} \citep{lipsitch_negative_2010, shi_selective_2020}. 

By including a frailty, model \ref{eq:frailty} acknowledges the existence of unmeasured heterogeneity in infection risk which can contribute to differential depletion of susceptibles \citep{fay_risk_2022}. Compared to static frailty models, a time-varying frailty model is more realistic by capturing the \textit{dynamic} behavioral and biological factors influencing infectious disease risk. The time-varying frailty also flexibly accommodates different mechanisms of time-varying confounding. For example, high-risk persons may receive vaccination earlier due to personal motivation or prioritization by mass-vaccination campaigns -- this is captured by a negative association between $V$ and $U(t)$. Another example is behavioral disinhibition: a positive association between $Z(t)$ and $U(t)$ accommodates situations where vaccinated participants cease non-pharmaceutical interventions (e.g., masking, social distancing) that reduce exposure to respiratory pathogens.

How do we avoid bias caused by $U(t)$ and identify $f(\cdot)$? A key insight is that while the frailties are not directly observed, vaccine-irrelevant infections can be useful proxy measurements for the frailties. We can re-express the hazards in \ref{eq:frailty} on a common scale defined by $h_{00}(t)$ and $U(t)$ using a change-of-scale parameter $\alpha_0(t) = \log\{h_{01}(t)/h_{00}(t)\}$.
\begingroup
  \setlength{\abovedisplayskip}{6pt}   % space above
  \setlength{\belowdisplayskip}{6pt}
\begin{align*}
    h_1(t \mid V, U(t)) &= h_{00}(t) U(t)\exp\{Z(t)f(\tau) + \alpha_0(t)\} \\
    h_0(t \mid V, U(t)) &= h_{00}(t) U(t)
\end{align*}
\endgroup
To proceed, we highlight a connection between our parametrization and those used in matched-pair/twin experiments. In twin studies, a unique baseline hazard is assigned to each pair which absorbs unmeasured variables shared within the pair, while each twin’s hazard differs in the parametric component pertaining to treatment \citep{Holt1974}. In our case, the two hazards describe \textit{``twin" infection processes measured within the same individual}, where the vaccine-preventable and vaccine-irrelevant infections are akin to the ``treated" and ``untreated" twin respectively. In twin study analyses, it is common to control for within-pair unmeasured factors by pair stratification. In our setting, we stratify the analysis at the individual participant level to eliminate the common unmeasured causes of vaccine-preventable and irrelevant infections encoded by $U(t)$. For example, conditional on a participant experiencing an infection at time $T$, the probability that the infection was vaccine-preventable ($J=1$) can be expressed using only observed data quantities.
\begingroup
  \setlength{\abovedisplayskip}{6pt}   % space above
  \setlength{\belowdisplayskip}{6pt}
\begin{align*}
    \frac{\cancel{h_{00}(T)U(T)} \exp\{Z(T)f(T-V) + \alpha_0(T)\}}{\cancel{h_{00}(T)U(T)}[1 + \exp\{Z(T)f(T-V) + \alpha_0(T)\}]} = \frac{\exp\{Z(T)f(T-V) + \alpha_0(T)\}}{1 + \exp\{Z(T)f(T-V) + \alpha_0(T)\}}
\end{align*}
\endgroup
As shown in Section S1.1 of the Supplement, $f(\cdot)$ is identified as the maximizer of the following \textit{individually-stratified partial likelihood} which does not involve $U(t)$.
\begin{equation}\label{eq:ISPL}
\resizebox{0.9\textwidth}{!}{%
    $L(f, \alpha) = \underset{n\uparrow \infty}{\lim} \overset{n}{\underset{i=1}{\prod}} \left(\frac{\exp\{Z_i(T_i) f(T_i-V_i) + \alpha_0(T_i)\}}{1+\exp\{Z_i(T_i) f(T_i-V_i) + \alpha_0(T_i)\}}\right)^{J_i} \left(\frac{1}{1+\exp\{Z_i(T_i) f(T_i-V_i) + \alpha_0(T_i)\}}\right)^{1-J_i}$
}
\end{equation}
While the typical Cox partial likelihood compares the event time ranks \textit{between} individuals, the individually-stratified partial likelihood makes pairwise comparisons of event time ranks \textit{within} individuals. Within-individual comparisons avoid confounding and selection since frailties are implicitly controlled for and no comparisons are possible after the first event. Our approach is similar to self-controlled case series methods \citep{Whitaker_SCCS_2006, Lin_Henley_2016}, but accommodates terminal events and time-varying confounding by measuring negative control and primary outcomes in parallel over the same time period.

\subsection{General development}\label{s:general}

While we used particular hazard parametrizations to motivate our approach, we describe a more general set of assumptions for nonparametric identification of time-varying VE. We define VE against outcome $J=j$ comparing vaccination on date $V=v$ (for $v < t$) to vaccination at some future date $V > t$ (or equivalently, unvaccinated at time $t$) as follows.
\begingroup
  \setlength{\abovedisplayskip}{6pt}   % space above
  \setlength{\belowdisplayskip}{6pt}
\begin{align*}
    \text{VE}_j(t \mid V=v, U(t)=u) &:= 1 - \exp\left\{\log\left[ \frac{h_j(t \mid V=v, U(t)=u)}{h_j(t \mid V > t, U(t)=u)}\right]\right\}
\end{align*}
\endgroup

Our interest lies in identifying $\text{VE}_1(\cdot)$. In all subsequent developments, we assume the hazards for each infection type are nonzero such that the associated hazard ratios are well-defined. This requires that both on and off-target infections co-circulate during a common calendar time period. Suppose we make the following assumptions.

\begin{assumption}[Weak overlap] $P(0 < P(t < V|U(t)) < 1)=1$ for all $t>0$. 
\end{assumption}

\begin{assumption}[Negative control outcome]
    The hazard of off-target infection conditional on unmeasured characteristics at time $t$ does not depend on the vaccination date, or equivalently, the vaccination status and time-since-vaccination at time $t$: $h_0(t|V, U(t)) = h_0(t|U(t))$. This implies $\text{VE}_0(t \mid V=v, U(t)=u)=0$ for all $u, v$.
\end{assumption}

\begin{assumption}[VE depends only on time-since-vaccination]
    For all $t>v>0$, there exists a function $f(\cdot): \mathbb{R}^{\geq 0} \rightarrow \mathbb{R}$ satisfying,
    \begin{align*}
    \text{VE}_1(t \mid V=v, U(t) ) &= 1-\exp\{I(v \leq t)f(t-v)\}
    \end{align*}

\end{assumption}

\begin{assumption}[Bias equivalence] Define $p_{U1}$ and $p_{U0}$ as the density of $U(t)$ among those disease-free at time $t$ and vaccinated $\tau$ days ago and unvaccinated respectively. For all $t \geq 0$ 
\[
\resizebox{0.85\textwidth}{!}{$
\begin{aligned}
    &\lim_{\epsilon \downarrow 0} \frac{E_{U \sim p_{U1}}[P(J=1, T \in [t, t + \epsilon) | Z(t)=0, \tau = 0, U(t)=U, T \geq t)]}{E_{U \sim p_{U0}}[P(J=1, T \in [t, t + \epsilon) | Z(t)=0, \tau = 0, U(t) = U, T \geq t)]} = \\
    &\lim_{\epsilon \downarrow 0} \frac{E_{U \sim p_{U1}}[P(J=0, T \in [t, t + \epsilon) | Z(t)=0, \tau = 0, U(t)=U, T \geq t)]}{E_{U \sim p_{U0}}[P(J=0, T \in [t, t + \epsilon) | Z(t)=0, \tau = 0, U(t)=U, T \geq t)]}
\end{aligned}
$} \]
\end{assumption}

Assumption 1 requires sufficient variability in vaccination statuses of participants throughout the study. This assumption is required to ensure that all conditional expectations and hazards are well-defined and ensure identifiability of model parameters. Assumption 1 could be violated in a study where all participants are rapidly vaccinated. In Section S6.2 in the Supplementary Materials, we discuss assumptions to recover identifiability of model parameters using external, population-level disease surveillance data.

Assumption 2 asserts that conditional on $U(t)$, vaccination has no influence on the off-target infection hazard. This will hold when $U(t)$ summarizes the relevant confounders and unmeasured sources of heterogeneity in infection risk. The high antigenic specificity of the adaptive immune response and the absence of a vaccine effect on off-target ARIs in COVE's blinded phase \citep{Ashby_2025}, provide support for this assumption. 

Assumption 3 states that VE against the vaccine-preventable infection varies only in time-since-vaccination. Two consequences of Assumption 3 is that VE does not vary in calendar time nor the unmeasured frailty $U(t)$.  VE could depend on calendar time in cases where a new viral variant emerges over time for which the vaccine is less effective \citep{Follmann2022}. For example, suppose that at time $t_1$, viral variant A is predominant, but $\delta$ days later, a new variant B emerges for which the vaccine is less effective. Vaccine efficacy $\tau$ days after vaccination on date $t_1$ may differ from the vaccine efficacy $\tau$ days after vaccination on date $t_1 + \delta$ due to the shift in circulating variants: $\text{VE}_1(t_1 \mid V=v, U(t)=u) \neq \text{VE}_1(t_1 + \delta | V = v + \delta, U(t)=u)$. In our case study, to avoid violations of this assumption, we conduct analyses stratified by time periods where Delta or Omicron were predominant and compute variant-specific, time-varying VE estimates against Delta and Omicron COVID-19 (see Section \ref{subsec:strain-specific} and Supplementary Section S6.3 for details). Requiring that VE does not depend on $U(t)$ is similar to a ``no effect modification by unmeasured confounders" assumption proposed previously \citep{wang_bounded_2018}, as this would pose significant identifiability challenges. Assumption 3 suggests that if interest lies in $\text{VE}_1(t|v,u)$, it suffices to focus on $f(t-v)$, a parsimonious summary of time-varying VE that is transportable to settings with different distributions of $U(t)$.

Assumption 4 generalizes standard equiconfounding assumptions to include selection effects \citep{TchetgenTchetgen_2015, TchetgenTchetgen_2024, Boyer2025-cm}. Assumption 4 will be most plausible when the vaccine-preventable and vaccine-irrelevant infections depend on a common set of factors with similar effects on the hazards of each endpoint. The multiplicative frailty model \ref{eq:frailty} satisfies Assumption 4, because substituting $h_{01}(t) U(t)$ and $h_{00}(t) U(t)$ into the interior of the expectations on the left-hand and right-hand sides of Assumption 4 respectively both yield $E[U(t) | Z(t)=1, \tau = t-v, T \geq t]/E[U(t)|Z(t)=0, \tau = 0, T \geq t]$. In fact, we do not strictly require that the frailties for vaccine-preventable and vaccine-irrelevant infections be the same. Hazard models of the form, $h_1(t) = h_{01}(t) U(t) \exp\{Z(t) f(t-v)\}$ and $h_{0}(t) = h_{00}(t) W(t)$ could also satisfy Assumption 4 when $E[U(t) | Z(t)=z, \tau, T \geq t]$ is proportional to $E[W(t) | Z(t)=z, \tau, T \geq t]$ for all $z = \{0,1\}$ and $\tau > 0$. A sufficient condition for this to hold is if $U(t) \overset{d}{=} c_0 W(t)$ for a scalar $c_0$, implying that the frailties belong to the same scale family.

As a starting point for identification, we examine the vaccine-preventable hazard ratio, which by arguments in the Section S1.2 in the Supplement, can be expressed as follows.
\begingroup
  \setlength{\abovedisplayskip}{6pt}   % space above
  \setlength{\belowdisplayskip}{6pt}
\[
\resizebox{\textwidth}{!}{$
\begin{aligned}
\frac{h_1(t | Z(t)=1, \tau = t-v)}{h_1(t | Z(t)=0, \tau = 0)} = \exp\{f(t-v)\} \left(\lim_{\epsilon \downarrow 0} \frac{E_{U \sim p_{U1}}[P(J=1, T \in [t, t + \epsilon) | Z(t)=0, \tau = 0, U(t)=U, T \geq t)]}{E_{U \sim p_{U0}}[P(J=1, T \in [t, t + \epsilon) | Z(t)=0, \tau = 0, U(t)=U, T \geq t)]}\right)
\end{aligned}
$}
\]
\endgroup
where $p_{U1}$ and $p_{U0}$ are defined in Assumption 4. The vaccine-preventable hazard ratio equals the target parameter times a bias term resulting from marginalizing over different, imbalanced distributions of $U(t)$. While we cannot directly measure the frailty imbalance directly, as shown in Section S1.2 of the Supplement, the cause-specific hazard ratio for vaccine-irrelevant infections at time $t$ indirectly measures the frailty imbalance.
\begingroup
  \setlength{\abovedisplayskip}{6pt}   % space above
  \setlength{\belowdisplayskip}{6pt}
{\footnotesize
\begin{align*}
    &\frac{h_0(t | Z(t)=1, \tau = t-V)}{h_0(t | Z(t)=0, \tau = 0)} = \lim_{\epsilon \downarrow 0} \frac{E_{U \sim p_{U1}}[P(J=0, T \in [t, t + \epsilon) | Z(t)=0, \tau = 0, U(t)=U, T \geq t)]}{E_{U \sim p_{U0}}[P(J=0, T \in [t, t + \epsilon) | Z(t)=0, \tau = 0, U(t)=U, T \geq t)]}
\end{align*}
}%
\endgroup
If the vaccine-preventable and irrelevant pathogens are comparable outcomes, it is plausible that the bias inflicted on both endpoints by marginalizing over imbalanced frailty distributions will be equal. This intuition is formalized by Assumption 4, and establishes that the time-varying VE function is identified as a ratio of cause-specific hazard ratios.
\begingroup
  \setlength{\abovedisplayskip}{6pt}   % space above
  \setlength{\belowdisplayskip}{6pt}
\[
\resizebox{\textwidth}{!}{$
\begin{aligned}
    \exp\{f(t-v)\} &= \lim_{\epsilon \downarrow 0} \frac{P(J=1, T \in [t, t + \epsilon) | Z(t)=1, \tau = t-v, T \geq t)}{P(J=1, T \in [t, t + \epsilon) | Z(t)=0, \tau = 0, T \geq t)} \times \left(\frac{P(J=0, T \in [t, t + \epsilon) | Z(t)=1, \tau = t-v, T \geq t)}{P(J=0, T \in [t, t + \epsilon) | Z(t)=0, \tau = 0, T \geq t)}\right)^{-1} 
\end{aligned}
$}
\]
\endgroup
In principle, one could estimate the time-varying VE function by estimating each cause-specific hazard or hazard ratio nonparametrically (e.g., using kernel smoothing) and computing the contrast \citep{GilbertWeiKosorokClemens2002}. However, this approach will suffer from slow convergence rates and instability in finite samples. Based on conditional probability arguments shown in Supplementary Section S1.2, we can express the target parameter equivalently as an odds ratio of infection causes among those infected at time $t$.
\begingroup
  \setlength{\abovedisplayskip}{6pt}   % space above
  \setlength{\belowdisplayskip}{6pt}
\[
\resizebox{\textwidth}{!}{$
\begin{aligned}
    \exp\{f(t-v)\} = \frac{P(J=1 | Z(t)=1, \tau = t-v, T = t)/P(J=0 | Z(t)=1, \tau = t-v, T = t)}{P(J=1 | Z(t)=0, \tau = 0, T = t)/P(J=0 | Z(t)=0, \tau = 0, T = t)}
\end{aligned}
$}\]
\endgroup
Expressing the target parameter as an odds ratio motivates a logistic regression model as an estimating equation for the time-varying VE function.
\begingroup
  \setlength{\abovedisplayskip}{6pt}   % space above
  \setlength{\belowdisplayskip}{6pt}
\begin{equation}\label{eq:nlr}
    \log\left\{\frac{P(J=1 | Z(t)=1, \tau = t-v, T = t)}{P(J=0 | Z(t)=1, \tau = t-v, T = t)}\right\} = f(t-v) + \alpha_0(t)
\end{equation}
\endgroup
Where $\alpha_0(t) :=  \log\{P(J=1 | Z(t)=0, \tau=0, T=t)/P(J=0 | Z(t)=0, \tau=0, T=t)\}$ is the odds of failure from the vaccine-preventable infection among those unvaccinated and infected at time $t$. Importantly, the likelihood associated with the logistic model in \ref{eq:nlr} is precisely the individually-stratified partial likelihood in \ref{eq:ISPL}, confirming the applicability of our method beyond particular hazard parametrizations like those in Section \ref{subsec:motivating_example}.

\section{Connection to a causal VE waning parameter}
\label{sec: connection to causal parameter}
We motivated our target parameter using hazard functions depending on time-varying, latent frailty. However, hazard-based treatment effects have been critiqued for unclear causal interpretations \citep{Hernan2010}, and others have sought to clarify useful hazard-based causal contrasts \citep{Martinussen2020-qm, fay_causal_2024}. In this section, we study the causal interpretation of our statistical parameter under appropriate assumptions.

We consider potential infection outcomes that depend on a joint intervention that defers vaccination and exposure until specific dates. While administration of a particular vaccine regimen (e.g., single-dose mRNA-1273) on a specific date is a clear and well-defined intervention, we must carefully define ``exposure" and consider what it means to intervene upon exposure. One possible definition of ``exposure" is \textit{virological}: a controlled administration (e.g., via intranasal pipette) of a known quantity of virus particles capable of causing infection in a naive patient \citep{Janvin2024-ly}. One can conceptualize intervening on virological exposure in a challenge trial where the timing of challenge exposure is under the investigator's control. However, applying the virological definition of exposure to real-world studies is complicated by differences between idealized ``challenge" exposures and real-world exposures due to changing viral strains and variable viral loads in the population. Instead, we adopt an \textit{epidemiological} definition of exposure \citep{Follmann_nonpar_2025} where a participant forgoes a period of strict isolation and may interact with the community in which pathogens circulate. While it is difficult and perhaps unethical to isolate participants from the community, one can conceptualize a trial that randomizes participants to receive vaccination and forgo isolation on different dates. Trials that randomize participants to receive a vaccine on different dates before the flu season approximate deferred exposure designs under an epidemiological definition, as participants can effectively be considered ``isolated" when influenza circulation is low \citep{Lipsitch_2019, ray_depletion_susceptibles_2020}.

Let the time origin $t_0 = 0$ refer to the study start date. Let $T_1(v, d)$ denote the hypothetical time-to-vaccine-preventable infection under an intervention which defers vaccination until date $V=v \geq 0$ and epidemiological exposure (i.e., interaction with the community) until date $d \geq 0$. Let $T_0(v,d)$ denote the analogous potential vaccine-irrelevant infection time. Let $T(v,d)$ denote $\min\{T_1(v,d), T_0(v,d)\}$. We define the following causal time-varying VE parameter.
\begingroup
  \setlength{\abovedisplayskip}{6pt}   % space above
  \setlength{\belowdisplayskip}{6pt}
\begin{equation}\label{eq:causalestimand}
    \text{VE}_C(t-v) = 1-\theta_C(t-v) = \underset{\epsilon \downarrow 0}{\lim} \; 1 - \frac{P(T_1(v,0) \in [t, t+\epsilon)| T(v,0) \geq t, V=v)}{P(T_1(\infty,t) \in [t, t+\epsilon)| T(v,0) \geq t, V=v)}
\end{equation}
\endgroup
Unlike the classic hazard ratio which has an unclear causal interpretation because it contrasts two different groups of participants \citep{Hernan2010}, \ref{eq:causalestimand} describes a single, well-defined group of participants: persons uninfected at time $t$ when vaccinated on date $V=v$ and permitted to interact with the community at study entry. Hence, \ref{eq:causalestimand} has a familiar interpretation akin to a ``treatment effect on the treated" estimand. The estimand \ref{eq:causalestimand} avoids depletion of susceptibles bias because all participants are susceptible to failure at time $t$ under both ``vaccinated and immediately exposed" (numerator) and ``unvaccinated but delayed exposure" (denominator) interventions. Intuitively, the parameter addresses the question: \textit{``Among patients vaccinated and exposed to infection immediately who are still uninfected at time $t$, what is their risk of infection compared to their hypothetical risk if they were unvaccinated but avoided infection by isolating from the community until $t$?"}

To identify \ref{eq:causalestimand}, we assume the hazard ratio of vaccine-preventable and irrelevant infections is stable across subgroups when the unvaccinated, deferred exposure regime is enforced.

\begin{assumption}[Cross-cause control stability]
    Define $\alpha_0(t)$ as the ratio of community infection pressures from the vaccine-preventable infection and vaccine-irrelevant infection over $[t, t+\epsilon)$: $\alpha_0(t) := \int_t^{t+\epsilon} h_{01}(s) ds / \int_t^{t+\epsilon} h_{00}(s) ds$. Assume that under deferred exposure and no vaccination, the ratio of cause-specific hazards is determined by the community force of infection and is invariant to the subgroup. That is, for all $t > 0$, and $v > 0$,
    \begingroup
    \setlength{\abovedisplayskip}{6pt}   % space above
    \setlength{\belowdisplayskip}{6pt}
    \[
    \resizebox{0.99\textwidth}{!}{$
    \begin{aligned}
        \underset{\epsilon \downarrow 0}{\lim} \frac{P(T_1(\infty, t) \in [t, t+\epsilon) | T(v,0) \geq t, V=v)}{P(T_0(\infty, t) \in [t, t+\epsilon) | T(v,0) \geq t, V=v)} = \underset{\epsilon \downarrow 0}{\lim} \frac{P(T_1(\infty, t) \in [t, t+\epsilon) | T(\infty,0) \geq t, V=\infty)}{P(T_0(\infty, t) \in [t, t+\epsilon) | T(\infty,0) \geq t, V=\infty)} = \alpha_0(t)
    \end{aligned}$
    }
    \]
    \endgroup
\end{assumption}

We use an analogy to justify this assumption. Suppose that two groups play dodgeball for $\epsilon$ minutes under identical conditions. Players are ``out" when they are hit by a red or blue dodgeball, and all players are eligible to be ``out" at the start of the playing period. Suppose red balls are twice as abundant as blue balls in the gym. While one group may have better dodging skills than the other group, it is reasonable the number of players hit by a red ball should be twice the number of players hit by a blue ball in both groups. We adapt this analogy to our setting by letting red and blue balls represent vaccine-preventable and vaccine-irrelevant infections, changing the two groups of players to the two subgroups--$\{T(v,0) \geq t, V=v\}$ and $\{T(\infty, 0) \geq t, V=\infty\}$--, and recasting the game duration as the exposure increment $\epsilon$. Importantly, Assumption 5 does not require that the infection rates be identical for different subgroups. Instead, Assumption 5 requires the \textit{ratio} of vaccine-preventable to vaccine-irrelevant infections is stable across the different groups.

Given the specificity of the adaptive immune response and empirical evidence from COVE's blinded phase \citep{Ashby_2025}, it is reasonable to assume that the vaccine-irrelevant infections do not depend on the vaccination date.

\begin{assumption}[NCO]
    Assume that the vaccine-irrelevant infection is unaffected by vaccination date, such that for all $t>0$, $v>0$, and any $a, a'>0$ such that $a \neq a'$,
    \begingroup
    \setlength{\abovedisplayskip}{6pt}   % space above
    \setlength{\belowdisplayskip}{6pt}
    \begin{align*}
        P(T_0(a, t) \in [t, t+\epsilon) | T(v,0) \geq t, V=v) = P(T_0(a', t) \in [t, t+\epsilon) | T(v,0) \geq t, V=v)
    \end{align*}
    \endgroup
\end{assumption}

A key roadblock to identification of our target parameter is that in real-world studies, all participants interact with the community at study entry. In other words, we never observe data under the ``deferred exposure" world where participants are isolated. However, it may be reasonable that among participants who were \textit{naturally} uninfected by time $t$, their failure rates just after time $t$ would be unchanged had they been isolated until $t$. Assumption 7 provides the crucial link between the counterfactual ``deferred exposure" world and the factual ``immediate exposure" world. 

\begin{assumption}[Exposure deferral irrelevance] for both $j \in \{0,1\}$ and any $t >0$,
    \begin{align*}
        P(T_j(v,t) \in [t, t+\epsilon) | T(v,0) \geq t, V=v) = P(T_j(v,0) \in [t, t+\epsilon) | T(v,0) \geq t, V=v)
    \end{align*}
\end{assumption}

Assumption 7 is a strong assumption analogous to Assumption 19 in \citet{Janvin2024-ly} that contends that being naturally uninfected by time $t$ simulates isolation from the community until $t$. Assumption 7 could be violated if there are a significant number of sub-infectious pathogen exposures or asymptotic infections during the period from $0$ to $t$ that prime participants' immune systems and reduce infection risk at $t$. Assumption 7 may be plausible if sub-infectious pathogen exposures or asymptomatic infections are rare or are unlikely to influence subsequent infection risk. Assumption 7 is plausible if $[0,t]$ was a period of very low pathogen circulation, as participants can be effectively considered isolated \citep{Lipsitch_2019, ray_depletion_susceptibles_2020}. 

As shown in Section S1.3 of the Supplement, under Assumptions 5-7 and consistency of potential outcomes, the causal quantity $\theta_C(t-v)$ is identified by the ratio of cause-specific hazard ratios. As before, applying conditional probability rules shows that the causal parameter is also identified by the odds ratio of infection causes at time $t$.
\begingroup
\setlength{\abovedisplayskip}{6pt}   % space above
\setlength{\belowdisplayskip}{6pt}
\begin{align*}
    \theta_C(t-v) &= \frac{P(J=1 | T = t, V=v)/P(J=1 | T = t, V=\infty)}{P(J = 0 | T = t, V=v)/P(J=0 | T = t, V=\infty)}
\end{align*}
\endgroup
In some cases, it may be implausible to assume that the ratio of failure rate from vaccine-preventable to vaccine-irrelevant infections are identical across groups (Assumption 5). In Section S1.3.1 of the Supplement, we introduce a sensitivity analysis that relaxes Assumption 5 to assume proportional failure rate ratios.

\section{Model specification and inference}
\label{sec:inference}

One strategy for inference on $f(\cdot)$ is to propose parametric models for $f(\cdot)$ and $\alpha_0(t)$ and maximize the associated logistic likelihood. However, parametric models are likely too simplistic in epidemic settings with unpredictable, fluctuating forces of infection and highly nonlinear $\alpha_0(t)$. An alternative approach is to pursue nonparametric modeling. However, nonparametric modeling of $f(\cdot)$ leads to undesirable statistical properties; $f(\cdot)$ is not a pathwise differentiable parameter in \ref{eq:nlr} meaning that root-$n$ consistent and asymptotically normal (CAN) estimators are not attainable without additional assumptions.

Results from semiparametric theory offer a compromise, and show that CAN estimators of $f(\cdot)$ can be obtained with parametric models while leaving $\alpha_0(t)$ unspecified. This is an acceptable concession in many cases, as $f(\cdot)$ likely follows relatively simple shapes. Therefore, we base inference on the \textit{semiparametric} logistic regression (SLR) model \citep{Tchetgen_Tchetgen2010-yi, tan_doubly_2019, liu_doubledebiased_2020, vdl_gilbert_2025}.
\begingroup
\setlength{\abovedisplayskip}{6pt}   % space above
\setlength{\belowdisplayskip}{6pt}
\begin{equation}\label{eq:slr}
    \log\left\{\frac{P(J=1 | Z(t)=z, \tau = t-v, T = t)}{P(J=0 | Z(t)=z, \tau = t-v, T = t)}\right\} = z \beta_0^T \boldsymbol{\psi}(t-v) + \alpha_0(t)
\end{equation}
\endgroup
Where $\beta_0 \in \mathbb{R}^d$ is the parameter of interest, $\boldsymbol{\psi} = \{\psi_1, \ldots, \psi_d\}$ is a finite collection of $d$ basis functions with support on $\mathbb{R}^+$, and $\alpha_0(t)$ is an unspecified function of calendar time.

\subsection{Inference via method of sieves}
\label{subsec:inference sieve}
One strategy for inference in model \ref{eq:slr} is to approximate the unknown function $\alpha_0(t)$ with a basis expansion that grows with the sample size, an approach known as sieve estimation \citep{Grenander1981}. Suppose we approximate $\alpha_0(t)$ via
\begin{align*}
    \alpha_0(t) \approx \sum_{j=1}^{M} \alpha_j \phi_{j, v}(t)
\end{align*}
where $\{\phi_{j, K, v}\}_{j=1}^{M}$ is a collection of B-splines of degree $v$ with $K$ equally placed knots and coefficients $\alpha_{M}^T \in \mathbb{R}^{M}$. Let $\hat{\beta}_{K,n}$ refer to the maximum likelihood estimator for $\beta_0$ when $\alpha_0(t)$ is approximated using the sieve of dimension $M$. As the sample size grows, we permit the number of knots $K_n \rightarrow \infty$, enabling better approximation of the unknown function. If $\alpha_0(t)$ is sufficiently smooth, the sieve's approximation error will vanish at a fast enough rate such that $\hat{\beta}_{K,n}$ will be CAN for $\beta_0$. Our formal smoothness condition is $\alpha_0(t)$ is uniformly bounded and belongs to a Hölder class $\mathcal{G}(v_0, \gamma, c_0)$, consisting of functions $g$ satisfying
\begin{align*}
    |g^{(v_0)}(x_1) - g^{(v_0)}(x_0)| \leq c_0 |x_1 - x_0|^{\gamma}
\end{align*}
For arbitrary finite $\gamma$ and $c_0$ and for all $x_1, x_0 \in [0,1]$. Some common examples of Hölder classes are Lipschitz functions ($v_0=0, \gamma=1$), functions with bounded derivatives ($v_0=1, \gamma=0$), and functions with bounded higher order derivatives ($v_0=r, \gamma=0$). Let $p = v_0 + \gamma$ refer to the \textit{smoothness index} of the function class and let $M_n \approx n^{\lambda}$ where $\lambda$ is the \textit{sieve dimension growth rate}. Before introducing asymptotic results, we first introduce the following quantities. Define $Q_0(v,t) := \mathbb{P}(J=1|V=v, T=t) \equiv \text{expit}\{I(v \leq t) \beta_0^T \boldsymbol{\psi}(t-v) + \alpha_0(t)\}$ as the probability of vaccine-preventable infection under \ref{eq:slr}. Define the following $d$-dimensional nuisance parameter vector.
\begingroup
\setlength{\abovedisplayskip}{6pt}   % space above
\setlength{\belowdisplayskip}{6pt}
\begin{align*}
    \boldsymbol{r}_0(T) &:= \frac{\mathbb{E}_V[I(V \leq T) \boldsymbol{\psi}(T-V) Q_0(1-Q_0)(V,T)|T]}{\mathbb{E}_V[Q_0(1-Q_0)(V,T)|T]}
\end{align*}
\endgroup
The following proposition, proven in Section S2 of the Supplement, establishes that $\hat{\beta}_{K,n}$ is CAN for $\beta_0$.

\begin{proposition}[$\hat{\beta}_{K,n}$ is CAN]
    Assume that Conditions S1-S5 in the Supplement hold. Specifically, let $\alpha_0(t)$ lie in a Hölder class with smoothness index $p=v_0 + \gamma$. Let the number of knots $K_n \approx n^{\lambda}$. If $1/3 > \lambda$ and $p > 1/2$, then
    \begin{align*}
        \sqrt{n} (\hat{\beta}_{K,n} - \beta_0) \rightsquigarrow N(0, \Sigma^{-1})
    \end{align*}
    Where $\Sigma \in \mathbb{R}^{d \times d}$ with entries
\begingroup
  \setlength{\abovedisplayskip}{6pt}   % space above
  \setlength{\belowdisplayskip}{6pt}
\[
\resizebox{\textwidth}{!}{$
    \begin{aligned}
        \sigma_{jk} &= \mathbb{E}\left[Q_0(V,T)(1-Q_0(V,T)) \left(I(V \leq T) \psi_j(T-V) - r_{0j}(T)\right) \left(I(V \leq T) \psi_k(T-V) - r_{0k}(T)\right)\right]
    \end{aligned}
$}
\]
\endgroup
Where $Q_0$ and $r_{0}$ are defined as above.
\end{proposition}

\subsection{Inference via Efficient Influence Curve (EIC)}
\label{subsec:inference EIF}

An alternative inferential strategy exploits influence curves, which characterize the sensitivity of the target parameter to perturbations of the statistical model and can be used to construct asymptotically linear estimators. Popular approaches for semiparametric estimation include one-step estimation \citep{Bickel_semi_1993} and influence-curve-based estimating equations \citep{RobinsRotnitzkyZhao1994, VanderLaanRobins2003}. Here, we will discuss targeted maximum likelihood estimation (TMLE) \citep{vanderLaanRose2011}, which provides a general framework for using the efficient influence curve (EIC) to construct efficient, robust plug-in estimators with favorable finite-sample properties.

A key step to obtaining the TMLE estimator $\hat{\beta}_{TMLE}$ of $\beta_0$ is the derivation of the EIC. Recall the definitions of $Q_0(V,T)$ and $\boldsymbol{r}_0(T)$ from the previous section. The following Lemma, proven in Section S3.1 of Supplement, establishes the EIC of $\beta_0$ in model \ref{eq:slr}.

\begin{lemma}[EIC of $\beta_0$]
    The vector-valued EIC of $\beta_0$ with respect to the semiparametric statistical model in \ref{eq:slr} is equal to the following up to a scaling matrix.
    \begin{align*}
    D_P(Q_0, r)(j,t,v) &\propto \left(I(v \leq t) \boldsymbol{\psi}(t-v) - \boldsymbol{r}_0(t)\right) \left(j - Q_0(v,t)\right) 
    \end{align*}
\end{lemma}

At a high-level, the TMLE algorithm works in three steps (see Section S3.2 of Supplement for details). In step one, we obtain initial estimates of the relevant components of the distribution $P_0$ ($Q_0$ and $r_0$ in this case) which we denote $P^{0}_n \equiv (\hat{Q}^{0}_n, \hat{r}^{0}_n)$. In step two, we use the EIC to identify a ``clever covariate": $H := (I(v \leq t) \boldsymbol{\psi}(t-v) - \boldsymbol{r}_0(t))$. In step three, starting at $k=0$, we iteratively update the nuisance parameter fits $P^{k+1}_n$ using a logistic model for $J$ with main effect $\hat{H}^k$ and offset $\text{logit}(\hat{Q}^{k}_n)$ until the nuisance updates (i.e., clever covariate coefficients) are small. The inclusion of the clever covariate ensures the logistic model ``targets" the nuisance fits $P_n^* = (\hat{Q}_n^*, \hat{r}_n^*)$ that solve the estimating equation associated with the EIC. According to the theory of TMLE, under regularity conditions, the plug-in estimator using the targeted nuisances $P_n^*$ achieves the optimal bias-variance tradeoff and will be CAN for the target parameter, $\beta_0$, with asymptotic variance characterized by the variance of the EIC. The following proposition, proven in Section 3.3 of the Supplement, establishes the CAN of the TMLE estimator.

\begin{proposition}[$\hat{\beta}_{TMLE}$ is CAN]
    Suppose Conditions S6-S9 in the Supplement hold. Then
    \begin{align*}
        \sqrt{n}(\hat{\beta}_{\mathrm{TMLE}} - \beta_0) \rightsquigarrow N(0, \text{Cov}(D_P))
    \end{align*}
    Where the convergence in distribution if to a random vector in $\mathbb{R}^d$ and $$\text{Cov}(D_P) := \mathbb{E}[D_P(J,T,V) D_P(J,T,V)^T]$$ is the $d \times d$ covariance matrix of the EIC.
\end{proposition}

\subsection{Estimating strain-specific time-varying VE}\label{subsec:strain-specific}

Since COVE conducted whole genome sequencing of swabs positive for SARS-CoV-2, we can slightly modify our approach to estimate \textit{strain-specific, time-varying VE} \citep{Prentice1978}. We refer the reader to Section S6.3 in the Supplement for full details. We recast our semiparametric \textit{binary} logistic regression model as a semiparametric \textit{multinomial} logistic regression model with probabilities of vaccine-preventable infection ($J=1$) with strain $S=s$ and vaccine-irrelevant infection ($J=0$) given by
\begingroup
  \setlength{\abovedisplayskip}{6pt}   % space above
  \setlength{\belowdisplayskip}{6pt}
\[
\resizebox{0.9\textwidth}{!}{$
\begin{aligned}
    P(J=1, S=s \mid T=t, V=v) &= \frac{p_{0s}(t) \exp\{I(V \leq t) \boldsymbol{\psi}_{s}(t-v) \beta_s + \alpha_0(t)\}}{1 + \sum_{s'=1}^m p_{0s'}(t) \exp\{I(V \leq t) \boldsymbol{\psi}_{s'}(t-v) \beta_{s'} + \alpha_0(t)\}} \\
    P(J=0 \mid T=t, V=v) &= \frac{1}{1 + \sum_{s'=1}^m p_{0s'}(t) \exp\{I(V \leq t) \boldsymbol{\psi}_{s'}(t-v) \beta_{s'} + \alpha_0(t)\}} \\
\end{aligned}
$}
\]
\endgroup
where $\alpha_0(t)$ captures the relative calendar time trends of any vaccine-preventable and irrelevant infection. The model depends on $m$ variant mixture proportions $p_{0s'}(t) := P(S=s' \mid T=t, J=1, Z(t)=0)$ which describe the probability of experiencing a vaccine-preventable infection with strain $S=s'$ conditional on the occurrence of an unspecified vaccine-preventable infection at time $t$ in an unvaccinated person. In our case study, we used external, population-level data to eliminate the mixture proportions and improve model identifiability, a technique referred to as surveillance anchored sieve analysis \citep{Follmann2022}. We estimated the mixture proportions using weekly summaries of SARS-CoV-2 variant proportions in the US from GISAID \citep{khare_gisaids_2021}. In the time period spanning COVE's booster phase, GISAID estimates were based on a weekly average of roughly 62,500 genome submissions and Delta and Omicron were predominant with minimal overlap, hence, we treat the mixture proportions as fixed and known.

With known variant mixture proportions, extending the sieve estimation to the multinomial case is simple (see Section S6.3.1 of the Supplement for details). We approximate $\alpha_0(t)$ using a sieve and estimate a stacked parameter vector $\beta_0 := (\beta_{0s=1}, \ldots, \beta_{0s=m})$ by maximizing the multinomial likelihood. Consistency and asymptotic normality follow from the same arguments used in the binary case. Extending the TMLE approach requires some minor modifications, including deriving a new EIC. Section S6.3.2 of the Supplement contains associated derivations and a description of the adapted TMLE algorithm.

\subsection{Enforcing monotonicity of time-varying VE functions}\label{subsec:monotone}

For many vaccines, barring a short period of time for the vaccine to produce an immune response, it is plausible that VE may remain constant or decrease, but cannot increase in time-since-vaccination. To produce sensible estimators which agree with scientific knowledge and improve finite sample performance, we may wish that our estimates of $f(\cdot)$ be monotonically non-decreasing. As explained in Supplementary Section S6.1, following results by \citet{westling_correcting_2020}, we can obtain a corrected monotonized estimator (and associated confidence interval limits), by running a precision-weighted isotonic regression smoothing of the unconstrained estimator $\hat{f}$ and its confidence limits over a fine grid of values of time-since-vaccination. \citet{westling_correcting_2020} established that the estimation error of the monotonized estimator over the grid is never worse than the unconstrained estimator, that the monotonized confidence limits do not degrade coverage or bandwidth over the grid relative to the unconstrained estimator's limits.

\section{Simulation studies}\label{s:numeric}

The goals of our simulations were (i) to confirm the asymptotic properties of our proposed SLR estimators and (ii) to compare the performance of SLR estimators to standard Cox regression which ignored vaccine-irrelevant infections under several relevant data generating models. We considered four data generating models based on the multiplicative frailty models shown in \ref{eq:frailty}: (i) random vaccination assignment with low risk heterogeneity, (ii) random vaccination assignment with high risk heterogeneity, (iii) random vaccination assignment with post-vaccination behavioral disinhibition, and (iv) non-random vaccination assignment due to prioritizing ``higher risk" groups. 

Our simulation results (see Sections S4 and S5 of the Supplement) confirmed the validity of our asymptotic results for a range of relevant data-generating mechanisms: the SLR estimators showed low bias and maintained nominal confidence interval coverage across all settings and in cases where VE was constant or waned over time. In contrast, large bias and low confidence interval coverage was observed for Cox regression in settings with high risk heterogeneity, behavioral disinhibition, and non-random vaccine assignment. Moreover, we confirmed that bias was magnified by higher infection rates, as this accelerated depletion of susceptibles in the unvaccinated group. In summary, our simulations confirmed that our proposed estimators were more robust to unmeasured confounding and selection bias than standard Cox regression under several realistic data generating mechanisms.

\section{Efficacy of mRNA-1273 booster in a prospective, observational extension of a COVID-19 vaccine trial}\label{sec:casestudy}

The COVE booster analysis set contained 20,404 participants who previously received two doses of mRNA-1273 vaccine (either at randomization or open-label crossover) without virologic/serologic evidence of SARS-CoV-2 infection on or prior to September 23, 2021 \citep{baden_long-term_2024}. Therefore all mRNA-1273 booster VE estimates are relative VE estimates for COVID-19 disease comparing 3 doses (primary series + booster) to 2 doses (primary series only). In response to acute respiratory illness (ARI) symptoms, two nasopharyngeal swabs were collected. One was tested for SARS-CoV-2 using RT-PCR and the other was tested for 20 other respiratory pathogens using a multiplex PCR assay (Biofire$^{\circledR}$ RP2.1). COVE also conducted whole genome sequencing of SARS-CoV-2 positive swabs to identify the infecting strain. In COVE's booster phase, 441 of 1902 (23\%) of swabs positive for SARS-CoV-2 had a missing lineage group. In the US in Winter 2021, Delta was the exclusive circulating lineage for months before being rapidly and fully displaced by Omicron, with only a short period of co-circulation. Hence, lineage could be viewed as a near deterministic function of calendar time, and imputation uncertainty was negligible. Hence, for the strain-specific analysis, missing SARS-CoV-2 lineage groups (Delta or Omicron) were singly imputed by sampling from a binomial distribution using weekly variant mixing proportions obtained from GISAID \citep{khare_gisaids_2021}.

\subsection{Time-varying VE during periods where Delta and Omicron were predominant}

We estimated time-varying VE of the mRNA-1273 booster in COVE for (a) the Delta period (September 23, 2021 -- December 11, 2021) and (b) the Omicron period (December 18, 2021 -- April 05, 2022). We fit the proposed SLR estimators using Sieve and TMLE techniques and a Cox regression estimator for comparison. All estimators used a single parameter to model VE [0,13] days post-boost and allowed VE to vary linearly on the log hazard scale starting 14 days post-boost. For the sieve estimator, we used a B-spline basis with $K=\lfloor n^{(1/3.5)} \rfloor$ interior knots to approximate $\alpha_0(t)$. For the TMLE estimator, $\alpha_0(t)$ and $\mathbf{r}_0(t)$ were estimated using generalized additive models (GAMs). To produce sensible estimators, we applied the precision-weighted isotonic correction to the unconstrained estimates of $f$ as described in Subsection \ref{subsec:monotone}. For comparison, the Cox model ignored data on off-target ARIs and instead stratified by randomization stratum, sex, and underrepresented racial/ethnic minority status and adjusted for a continuous baseline risk score developed using ensemble machine learning \citep{gilbert_immune_2022}. 

\begin{figure}
    \centering
    \includegraphics[width=0.99\linewidth]{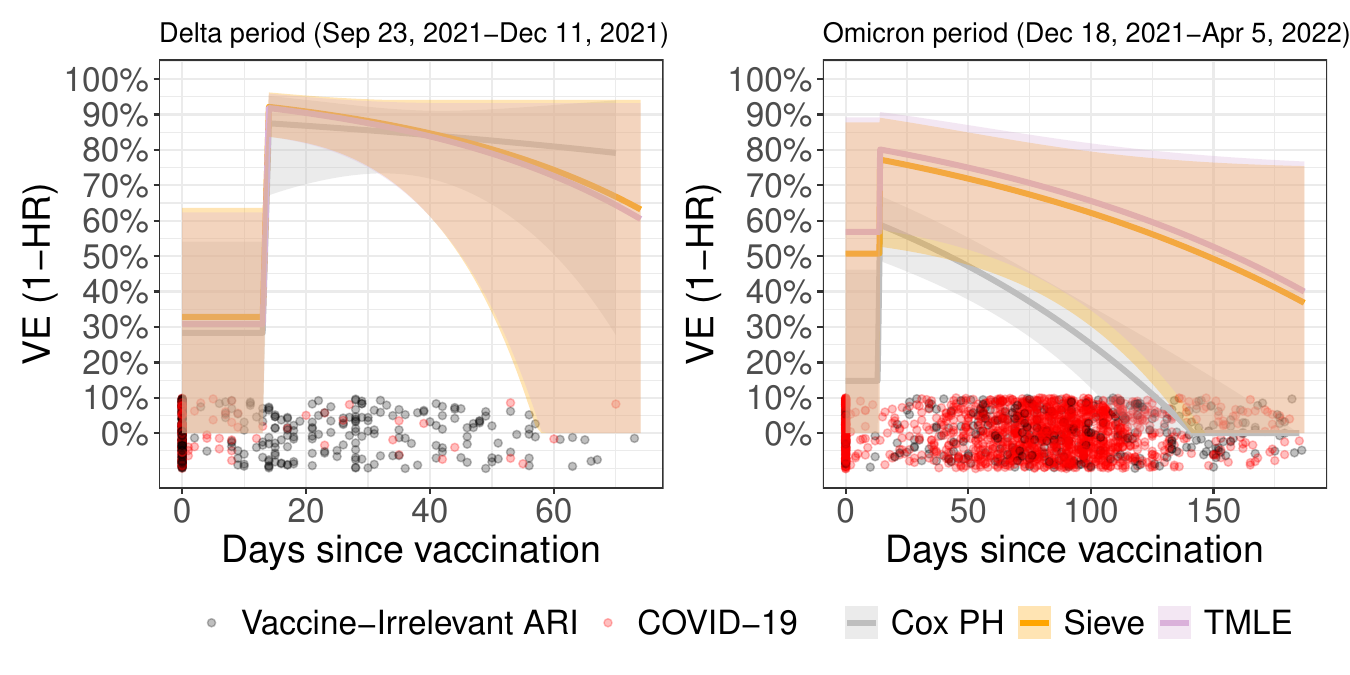}
    \caption{Summary of single mRNA-1273 booster efficacy in Delta (left) and Omicron period (right). Time-varying VE estimated using semiparametric logistic regression model with Sieve and TMLE estimators and Cox regression. Vaccine-irrelevant and COVID-19 illnesses shown as points on bottom of plot. Infections pre-vaccination stacked at $\tau=0$.}
    \label{fig:VE_periods}
\end{figure}

Analysis results are shown in Figure \ref{fig:VE_periods}. During the Delta period, Cox regression reported high initial efficacy (VE at 14 days post-boost = 87.5\%; 95\% CI: 67.2\% to 95.2\%) and durable protection 2 months after the boost (VE at 60 days = 81.0\%; 95\% CI: 50.5\% to 92.7\%). Compared to Cox regression, the TMLE estimator supported slightly higher initial effectiveness (VE at 14 days post-boost = 91.8\%; 95\% CI: 83.7\% to 95.9\%) and more pronounced waning over 2 months (VE at 60 days = 72.6\%; 95\% CI: -11.1\% to 93.2\%). Accounting for vaccine-irrelevant ARIs led to substantially different results in the Omicron period. Cox regression suggested moderate initial effectiveness (VE at 14-days post-boost = 58.9\%; 95\% CI: 48.6\% to 67.1\%) which rapidly decayed over 120 days (VE at 4 months post-boost = 14.0\%; 95\% CI: -2.5\% to 27.9\%). In contrast, the TMLE estimator supported higher initial effectiveness (VE at 14-days post-boost = 80.1\%; 95\% CI: 57.4\% to 90.7\%) and more durable VE (VE at 4 months post-boost = 60.9\%; 95\% CI: 22.8\% to 80.2\%). Results for the Sieve estimator were comparable to the TMLE estimator for both Delta and Omicron period analyses.

\subsection{Strain-specific time-varying VE against Delta and Omicron COVID-19}

Since COVE conducted whole genome sequencing of swabs from COVID-19 cases, we can estimate \textit{strain-specific, time-varying VE} by slightly modifying our approach as described in \ref{subsec:strain-specific}. Strain-specific VE estimates improve upon the period-specific estimates by avoiding strain contamination and extending VE estimation to periods of Delta and Omicron co-circulation. For our proposed estimators, we assumed the variant mixture proportions were known and equaled those provided by GISAID \citep{khare_gisaids_2021}. We fit a semiparametric multinomial logistic regression model using both sieve and TMLE approaches. For the sieve estimator, we approximated $\alpha_0(t)$ using a B-spline basis with 13 interior knots. The TMLE estimator used kernel estimators for the unknown nuisance parameters with an epanechnikov shape and bandwidth chosen using Silverman's rule of thumb. All our proposed estimators assumed a single scalar modeling VE [0,13] days post-boost and assumed a linear specification for $f(\cdot)$ thereafter. For comparison, we fit Cox regression models to Delta and Omicron infections, where the occurrence of the other outcome was treated as censored by a competing risk \citep{Prentice1978}. As before, the Cox regression analyses ignored data on vaccine-irrelevant ARIs but stratified by randomization stratum, sex, and underrepresented racial/ethnic minority status and adjusted for a continuous baseline risk score developed using ensemble machine learning \citep{gilbert_immune_2022}.

\begin{figure}
    \centering
    \includegraphics[width=0.99\linewidth]{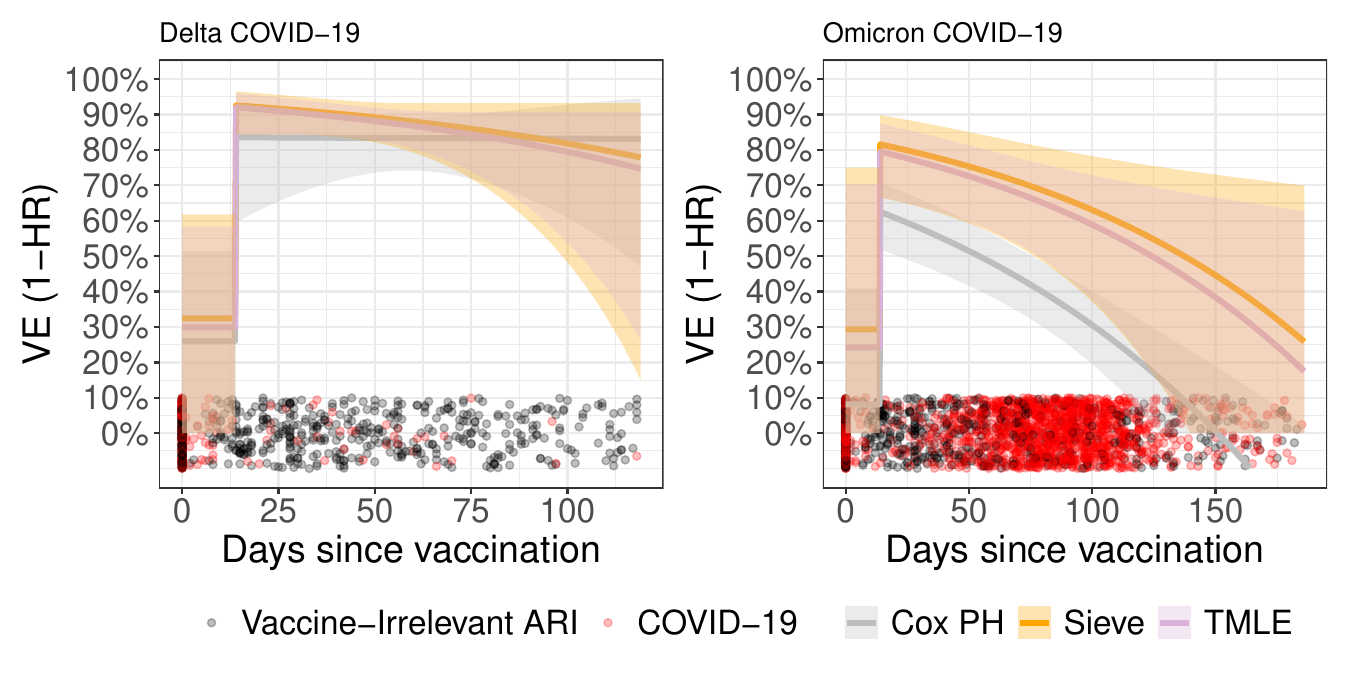}
    \caption{Summary of single mRNA-1273 booster efficacy in Delta (left) and Omicron period (right). Time-varying VE estimated using semiparametric logistic regression model with Sieve and TMLE estimators, and standard Cox PH regression. Vaccine-irrelevant and COVID-19 infections shown as points on bottom of plot. Infections pre-vaccination stacked at $\tau=0$.}
    \label{fig:VE_strains}
\end{figure}

Results can be found in Figure \ref{fig:VE_strains}. Cox regression which ignored vaccine-irrelevant ARIs supported high initial booster VE against Delta COVID-19 (VE at 14 days post-boost = 83.5\%; 95\% CI: 59.4\% to 93.3\%) that remained constant for 100 days (VE at 100 days post-boost = 83.2\%; 95\% CI: 60.9\%, 92.8\%). Compared to Cox regression, the TMLE estimator accounting for vaccine-irrelevant ARIs suggested higher initial VE against Delta (VE at 14 days post-boost = 92.0\%; 95\% CI: 84.3\% to 96.0\%) and more marked VE waning (VE at 100 days post-boost = 79.3\%; 95\% CI: 53.5\%, 90.7\%). Against Omicron COVID-19, Cox regression suggested moderate booster effectiveness (VE at 14 days post-boost = 62.5\%; 95\% CI: 51.8\%, 70.8\%) that waned substantially over 5 months (VE at 150 days post-boost = 0.6\%; 95\% CI: -21.7\%, 18.9\%). In contrast, our proposed TMLE estimator supported that the mRNA-1273 booster was more effective (VE at 14 days post-boost = 79.3\%; 95\% CI: 66.3\%, 87.3\%) and more durable (VE at 150 days post-boost = 38.3\%; 95\% CI: -13.8\%, 66.5\%) against Omicron COVID-19. In both Delta and Omicron analyses, the sieve estimator results were comparable to the TMLE estimator.

\section{Discussion}
\label{s:discuss}

Herein, we present a framework to eliminate unmeasured confounding and selection bias from hazard-based, time-varying vaccine effectiveness (VE) estimates using vaccine-irrelevant infections as negative controls. We leveraged our approach to explore the effectiveness of a single mRNA-1273 booster in an observational extension of a COVID-19 vaccine trial that spanned the Omicron COVID-19 wave and used multiplex PCR testing to detect acute respiratory illnesses (ARIs) caused by SARS-CoV-2 and 20 other ``off-target" respiratory pathogens. Our proposed approach accounting for the vaccine-irrelevant ARIs supported supported waning vaccine effectiveness against Delta COVID-19 and more effective and durable protection against Omicron COVID-19 than suggested by Cox regression analyses which ignored vaccine-irrelevant ARIs.

At first glance, it may seem puzzling why accounting for vaccine-irrelevant ARIs seemingly had a greater impact on time-varying VE estimates for the Omicron period/variant analyses compared to the Delta period/variant analyses. One hypothesis is that the higher force of infection during the Omicron wave accelerated depletion of susceptibles \citep{kanaan_estimation_2002, kahn_identifying_2022}, a phenomenon documented in our simulations. If depletion of susceptibles was higher during in the Omicron period/variant analyses, and vaccine-irrelevant infections were a good proxy for the imbalance in infection risk, we would anticipate a larger bias correction in the Omicron analyses.

We acknowledge some important caveats when interpreting our case study results. First, COVID-19 vaccine effectiveness may depend on viral sequence marks within a single lineage \citep{magaret_quantifying_2024, heng_estimation_2025}, which may lead to minor violations of Assumption 3 if within-lineage viral sequence characteristics change over calendar time. Adapting our approach to accommodate continuous viral marks is left to future work. Second, our developments assumed a participant's infection outcomes do not depend on the vaccination statuses of other study participants or community members. Since study participants were embedded in a much larger population, interactions between study participants were very unlikely and lack of interference between study participants themselves is plausible. However, a study participant's infection outcomes may depend on the vaccination statuses of their community contacts due to herd immunity. Following the logic of \citet{perenyi_variant_2025}, we believe our statistical parameter $f(\cdot)$ could be robust to community interference. Suppose we assume that vaccination of community members influences participants' infection outcomes entirely through contact events and that contact events are necessary to transmit infection. Since $f(\cdot)$ is identified by the log odds ratio of infection causes conditional on infection, $f(\cdot)$ is also identified by a log odds ratio conditional on a contact event. Conditioning on a contact event suggests the parameter could be robust to the interference structure in the population \citep{perenyi_variant_2025}. Exploring our estimator's robustness to interference is an interesting avenue for further research. Lastly, unlike the Cox regression analysis, our proposed estimators suggested that booster VE against Delta COVID-19 waned over time. However, we were unable to evaluate the booster's long-term durability against Delta after Delta ceased circulating in the US.

We anticipate that the approach described here could be efficiently applied to case-only study designs, such as the popular test negative design (TND) -- a variant of a case-control design which conditions on seeking testing in response to acute respiratory illness. In a TND, it is common to define symptomatic illness with a negative test test for the pathogen of interest as the ``control" outcome, which is assumed to be unaffected by vaccination. A recent meta-analysis identified that test-negative ARIs may be compelling choices of NCOs, although care needs to be taken in rigorously defining a test negative event \citep{andrews_evaluating_2025}. Like those measured in COVE, acute respiratory illness with a \textit{positive} test for an unrelated pathogen may be a more objective NCO endpoint by avoiding arbitrary aspects of test-negative event definitions and may exhibit better overlap in unmeasured causes with the primary infection endpoint \citep{Ashby_2025}. Establishing assumptions under which our method provides reliable inference on time-varying VE in TNDs is a promising avenue of future work. Any future efforts to infer time-varying VE from TNDs should endeavor to account for correlated uptake of vaccines against several vaccine-preventable pathogens (e.g., COVID-19, influenza, RSV, etc.), natural immunity caused by prior infection, and imperfect outcome ascertainment. Extending recent double-negative control approaches to estimate time-varying VE in TNDs is another area for future work \citep{li_double_2023}.

Moreover, despite being a popular method among practitioners, the causal interpretation of hazard-based treatment effect estimates remains unclear \citep{Hernan2010, Martinussen2013-cp, Martinussen2020-qm}. We endeavor to describe a set of assumptions under which our statistical parameter can be interpreted as a causal effect. Recent work by \citet{Janvin2024-ly} defined a causal time-varying VE parameter based on period-specific cumulative incidences and proposed assumptions to bounds the unknown parameter. Exploring how to leverage irrelevant infections to tighten the bounds proposed by \citet{Janvin2024-ly}, and developing alternative causal VE waning parameters and identification strategies is a compelling area of future work.

\section*{Ethics and Data Availability Statement}

The study uses de-identified data from the COVE study. The study sponsor, Moderna, Inc., was responsible for conceptualization, trial design, site selection, \& monitoring, The trial was conducted in accordance with the International Council for Harmonisation of Technical Requirements for Registration of Pharmaceuticals for Human Use, Good Clinical Practice guidelines. A central Institutional Review Board (Advarra, Inc., Columbia, MD) approved the protocol and consent forms. All participants provided written informed consent.

Code developed for simulations and case study are publicly available on GitHub. Access to participant-level data and supporting clinical documents by qualified external researchers may be made available upon request and subject to review. A materials transfer and/or data access agreement with the sponsor will be required for accessing shared data.

\section{Acknowledgments and Funding Statement}

We gratefully acknowledge Craig Magaret and all data contributors, the Authors and their Originating laboratories responsible for obtaining the specimens, and their submitting laboratories for generating the genetic sequence and metadata and sharing via the GISAID Initiative, on which this research is based.

The study sponsor, Moderna, Inc., was responsible for conceptualization, trial design, site selection, and monitoring. Statistical analysis was supported by the National Science Foundation Graduate Research Fellowship Program (Grant No. DGE-2140004). Any opinions, findings, conclusions, or recommendations expressed in this material do not necessarily reflect the views of the National Science Foundation. We also would like to acknowledge the NIH R01AI192632 awarded to BZ, HJ, PBG, HES.

%\section{Disclosure Statement}

%The authors report there are no competing interests to declare.

\onehalfspacing
\bibliographystyle{apalike}
\bibliography{biomsample}

\begin{thebibliography}{14}
\providecommand{\natexlab}[1]{#1}
\providecommand{\url}[1]{\texttt{#1}}
\expandafter\ifx\csname urlstyle\endcsname\relax
  \providecommand{\doi}[1]{doi: #1}\else
  \providecommand{\doi}{doi: \begingroup \urlstyle{rm}\Url}\fi

\bibitem[Chen(1995)]{chen_asymptotically_1995}
H.~Chen.
\newblock Asymptotically {Efficient} {Estimation} in {Semiparametric} {Generalized} {Linear} {Models}.
\newblock \emph{The Annals of Statistics}, 23\penalty0 (4):\penalty0 1102--1129, Aug. 1995.
\newblock ISSN 0090-5364, 2168-8966.
\newblock \doi{10.1214/aos/1176324700}.
\newblock URL \url{https://projecteuclid.org/journals/annals-of-statistics/volume-23/issue-4/Asymptotically-Efficient-Estimation-in-Semiparametric-Generalized-Linear-Models/10.1214/aos/1176324700.full}.

\bibitem[Chernozhukov et~al.(2018)Chernozhukov, Chetverikov, Demirer, Duflo, Hansen, Newey, and Robins]{chernozhukov_2018}
V.~Chernozhukov, D.~Chetverikov, M.~Demirer, E.~Duflo, C.~Hansen, W.~Newey, and J.~Robins.
\newblock Double/debiased machine learning for treatment and structural parameters.
\newblock \emph{The Econometrics Journal}, 21, 2018.

\bibitem[Follmann et~al.(2022)Follmann, Fay, and Magaret]{Follmann2022}
D.~Follmann, M.~Fay, and C.~Magaret.
\newblock Estimation of vaccine efficacy for variants that emerge after the placebo group is vaccinated.
\newblock \emph{Statistics in Medicine}, 41\penalty0 (16):\penalty0 3076--3089, July 2022.

\bibitem[Hadfield et~al.(2018)Hadfield, Megill, Bell, Huddleston, Potter, Callender, Sagulenko, Bedford, and Neher]{hadfield_nextstrain_2018}
J.~Hadfield, C.~Megill, S.~M. Bell, J.~Huddleston, B.~Potter, C.~Callender, P.~Sagulenko, T.~Bedford, and R.~A. Neher.
\newblock Nextstrain: real-time tracking of pathogen evolution.
\newblock \emph{Bioinformatics}, 34\penalty0 (23):\penalty0 4121--4123, Dec. 2018.
\newblock ISSN 1367-4803, 1367-4811.
\newblock \doi{10.1093/bioinformatics/bty407}.
\newblock URL \url{https://academic.oup.com/bioinformatics/article/34/23/4121/5001388}.

\bibitem[Johansen(1983)]{johansen_extension_1983}
S.~Johansen.
\newblock An {Extension} of {Cox}'s {Regression} {Model}.
\newblock \emph{International Statistical Review / Revue Internationale de Statistique}, 51\penalty0 (2):\penalty0 165, Aug. 1983.
\newblock ISSN 03067734.
\newblock \doi{10.2307/1402746}.
\newblock URL \url{https://www.jstor.org/stable/1402746?origin=crossref}.

\bibitem[Khare et~al.(2021)Khare, Gurry, Freitas, B~Schultz, Bach, Diallo, Akite, Ho, Tc~Lee, Yeo, {GISAID Core Curation Team}, and Maurer-Stroh]{khare_gisaids_2021}
S.~Khare, C.~Gurry, L.~Freitas, M.~B~Schultz, G.~Bach, A.~Diallo, N.~Akite, J.~Ho, R.~Tc~Lee, W.~Yeo, {GISAID Core Curation Team}, and S.~Maurer-Stroh.
\newblock {GISAID}’s {Role} in {Pandemic} {Response}.
\newblock \emph{China CDC Weekly}, 3\penalty0 (49):\penalty0 1049--1051, 2021.
\newblock ISSN 2096-7071.
\newblock \doi{10.46234/ccdcw2021.255}.
\newblock URL \url{http://weekly.chinacdc.cn/en/article/doi/10.46234/ccdcw2021.255}.

\bibitem[Larson and Dinse(1985)]{Larson1985}
M.~Larson and G.~Dinse.
\newblock A mixture model for the regression analysis of competing risks data.
\newblock \emph{J.R. Stat. Soc. Ser. C. Appl. Stat.}, 34\penalty0 (3):\penalty0 201--211, 1985.

\bibitem[Lawless(2003)]{Lawless_2003}
J.~Lawless.
\newblock \emph{Statistical Models and Methods for Lifetime Data}.
\newblock Jon Wiley and Sons Inc, 2003.

\bibitem[Lin and {Henley}(2016)]{Lin_Henley_2016}
N.~Lin and W.~{Henley}.
\newblock Prior event rate ratio adjustment for hidden confounding in observational studies of treatment effectiveness: a pairwise cox likelihood approach.
\newblock \emph{Statistics in Medicine}, 39:\penalty0 5149–5169, 2016.

\bibitem[Murphy and Van Der~Vaart(2000)]{murphy_profile_2000}
S.~A. Murphy and A.~W. Van Der~Vaart.
\newblock On {Profile} {Likelihood}.
\newblock \emph{Journal of the American Statistical Association}, 95\penalty0 (450):\penalty0 449--465, June 2000.
\newblock ISSN 0162-1459, 1537-274X.
\newblock \doi{10.1080/01621459.2000.10474219}.
\newblock URL \url{http://www.tandfonline.com/doi/abs/10.1080/01621459.2000.10474219}.

\bibitem[Prentice et~al.(1978)Prentice, Kalbfleisch, {Peterson Jr}, Flournoy, Farewell, and Breslow]{Prentice1978}
R.~Prentice, J.~Kalbfleisch, A.~{Peterson Jr}, N.~Flournoy, V.~Farewell, and N.~Breslow.
\newblock The analysis of failure times in the presence of competing risks.
\newblock \emph{Biometrics}, 34\penalty0 (4):\penalty0 541--554, Dec. 1978.

\bibitem[{van der Laan} and {Gilbert}(2025)]{vdl_gilbert_2025}
L.~{van der Laan} and P.~{Gilbert}.
\newblock Semiparametric logistic regression for inference on relative vaccine efficacy in case-only studies with informative missingness.
\newblock arXiv preprint arXiv:2303.11462, 2025.
\newblock URL \url{https://arxiv.org/abs/2303.11462}.

\bibitem[van~der Laan and Rose(2011)]{vanderLaanRose2011}
M.~J. van~der Laan and S.~Rose.
\newblock \emph{Targeted Learning: Causal Inference for Observational and Experimental Data}.
\newblock Springer, New York, 2011.

\bibitem[Westling et~al.(2020)Westling, Laan, and Carone]{westling_correcting_2020}
T.~Westling, M.~J. v.~d. Laan, and M.~Carone.
\newblock Correcting an estimator of a multivariate monotone function with isotonic regression.
\newblock \emph{Electronic Journal of Statistics}, 14\penalty0 (2):\penalty0 3032--3069, Jan. 2020.
\newblock ISSN 1935-7524, 1935-7524.
\newblock \doi{10.1214/20-EJS1740}.
\newblock URL \url{https://projecteuclid.org/journals/electronic-journal-of-statistics/volume-14/issue-2/Correcting-an-estimator-of-a-multivariate-monotone-function-with-isotonic/10.1214/20-EJS1740.full}.
\newblock Publisher: Institute of Mathematical Statistics and Bernoulli Society.

\end{thebibliography}
\end{document}

\maketitle

%\tableofcontents

\section{Key Identification Results}\label{sec:s_deriv}

\subsection{Individually Stratified Partial likelihood derivation under multiplicative frailty model}

Following the multiplicative frailty model, suppose we can write the hazard of vaccine-preventable and vaccine-irrelevant infections as follows. We convert the hazards to the same calendar time scale using the change-of-scale parameter $\alpha_0(t) = \log\{h_{01}(t)/h_{00}(t)\}$.
\begingroup
\setlength{\abovedisplayskip}{6pt}   % space above
\setlength{\belowdisplayskip}{6pt}
\begin{align*}
    h_{0}(t) &= h_{00}(t) U(t)\\
    h_{1}(t) &= h_{00}(t) U(t)\exp\{Z(t) f(t - v) + \alpha_0(t)\}
\end{align*}
\endgroup

Define $h_{0i}(t) := h_{00}(t) U_i(t)$ as an individual-level nuisance parameter. Since the number of nuisance parameters grows linearly with the sample size, it is infeasible to solve for the nuisance parameters directly. However, our main goal is to develop an estimating equation for $f(\cdot)$ that eliminates the need for the participant-level nuisance parameters -- particularly the unobserved frailty $U(t)$ which can produce bias in VE estimates. To achieve this, we will pursue two strategies -- a partial likelihood and a profile likelihood approach -- and demonstrate their equivalence. This result aligns with the appealing theoretical result that the Cox likelihood can be derived as both a partial and profile likelihood.

First, we pursue a partial likelihood approach. The key idea is that elimination of the participant-level frailty, $U(t)$, requires stratifying the partial likelihood at the level of individual patients. Let $T_i$ refer to the first infection time and $J=1$ refer to an indicator of the vaccine-preventable outcome and $J=0$ refer to the indicator of a vaccine-irrelevant outcome respectively. Let $R_i(t)$ refer to the individualized risk set, or the infection outcomes that participant $i$ is at risk of experiencing at calendar time $t$. Hence, $R_i$ can be empty or equal $\{0\}, \{1\}$, or $\{0,1\}$ at any given time. The partial likelihood for the paired survival data is equivalent to the following.
\begingroup
\setlength{\abovedisplayskip}{6pt}   % space above
\setlength{\belowdisplayskip}{6pt}
\[
\resizebox{\textwidth}{!}{$
\begin{aligned}
   &L(f,\alpha) = \prod_{i=1}^n \left(\frac{h_{i1}(T_{i})}{\sum_{j \in R_i(T_i)} h_{ij}(T_{i})}\right)^{J_i} \left(\frac{h_{i0}(T_{i})}{\sum_{j \in R_i(T_i)} h_{ij}(T_{i})}\right)^{1-J_i}\\
   &= \prod_{i=1}^n \left(\frac{h_{0i}(T_{i}) \exp\{Z_i(T_i) f(T_i-V_i) + \alpha(T_i)\}}{h_{0i}(T_i) + h_{0i}(T_{i}) \exp\{Z_i(T_i) f(T_i-V_i) + \alpha(T_i)\}}\right)^{J_i} \left(\frac{h_{0i}(T_{i})}{h_{0i}(T_i) + h_{0i}(T_{i}) \exp\{Z_i(T_i) f(T_i-V_i) + \alpha(T_i)\}}\right)^{1-J_i}\\
   &= \prod_{i=1}^n \left(\frac{\exp\{Z_i(T_i) f(T_i-V_i) + \alpha(T_i)\}}{1 + \exp\{Z_i(T_i) f(T_i-V_i) + \alpha(T_i)\}}\right)^{J_i} 
      \left(\frac{1}{1 + \exp\{Z_i(T_i) f(T_i-V_i) + \alpha(T_i)\}}\right)^{1-J_i}
\end{aligned}
$}
\]
\endgroup
It is worth noting that once a participant experiences their first infection at time $T_i$, the individualized risk set $R_i(t)$ contains only 1 element for all $t > T_i$. Hence, any subsequent infections will contribute a factor of 1 to the individually-stratified partial likelihood, so only the first infection event contributes to the individually-stratified partial likelihood.

A well-known result is that the Cox partial likelihood can be derived as a profile likelihood.\citep{murphy_profile_2000} Let $T_{i1}$ and $T_{i0}$ refer to the time-to-vaccine-preventable and time-to-vaccine-irrelevant infection respectively. The Cox models above admit a likelihood of the full survival data among $n$ individuals in the study \citep{Lawless_2003}. 
\begin{align*}
    L_{full}(f,\alpha) &= \prod_{i=1}^n [h_{0i}(T_{i0})]^{1-J_i} \exp\left(-\int_0^{T_{i0}} h_{0i}(u) du\right) \\
    &\quad [h_{0i}(T_{i1}) \exp\{Z_i(T_{i1}) [f(T_{i1} - V_i)] + \alpha(T_{i1})\}]^{J_i}\\
    &\exp\left(-\int_0^{T_{i1}} h_{0i}(u) \exp\{Z_i(u) [f(u - V_i)] + \alpha(u)\} du\right)
\end{align*}
Since the number of nuisance parameters grows with the sample size, we cannot obtain a consistent estimate of the target parameters. To eliminate the unknown nuisance parameters, we adopt a profile likelihood approach \citep{murphy_profile_2000}. To begin, we profile out the participant-level baseline hazard functions by replacing the infinite-dimensional nuisance with its finite-dimensional values at the observed event times. This essentially treats the hazard as a set of scalar nuisance parameters rather than a full function. We can write the components of the hazard as follows \citep{johansen_extension_1983, Lin_Henley_2016}.
\[
\resizebox{\textwidth}{!}{$
\begin{aligned}
 h_{0i}(T_{i0}) &= (1-J_i) h_{00i} \\
 h_{0i}(T_{i1}) &= J_i h_{01i} \\
 \int_0^{T_{i0}} h_{0i}(u) du &= (1-J_i)h_{0i} + S_i J_i h_{01i} \\
 \int_0^{T_{i1}} h_{0i}(u) \exp\{Z_i(u) [f(u - V_i, \beta)] + \alpha(u)\} du &= J_i h_{01i} \exp\{Z_i(T_{i1}) f(T_{i1} - V_i, \beta) + \alpha(T_{i1})\} \\
 &+ P_i (1-J_i) h_{0i} \exp\{Z_i(T_{i0}) f(T_{i0} - V_i, \beta) + \alpha(T_{i0})\}
\end{aligned}
$}
\]
Where $P_i := I(T_{i0} \leq T_{i1})$ and $S_i := I(T_{i1} \leq T_{i0})$ are the paired event time ranks. Rewriting the log-likelihood in terms of the new scalar nuisance parameters, $h_{00i}, h_{01i}$, we obtain a tractable log-likelihood.
\[
\resizebox{\textwidth}{!}{$
\begin{aligned}
    \ell_{\text{full}} &= \prod_{i=1}^n (1-J_i) \log\{(1-J_i) h_{00i}\} - [(1-J_i) h_{00i} + S_i J_i h_{01i}] + J_i\left(\log\{J_i h_{01i}\} + Z_i(T_{i1}) f(T_{i1} - V_i ; \beta) + \alpha(T_{i1})\right) \\
    & - \left[J_i h_{01i} \exp\{Z_i(T_{i1}) f(T_{i1} - V_i ; \beta) + \alpha(T_{i1})\}\right] -\left[P_i (1-J_i) h_{00i} \exp\{Z_i(T_{i0}) f(T_{i0} - V_i ; \beta) + \alpha(T_{i0})\}\right]
\end{aligned}
$}
\]
We can construct a score equation with respect to the unknown scalar nuisance parameters, $h_{00i}, h_{01i}$, and solve for each nuisance parameter with respect to the other variables.
\begin{align*}
    0 = \frac{\partial \ell_{\text{full}}}{\partial h_{00i}} &= \frac{(1-J_i)}{h_{00i}} - (1-J_i) - P_i (1-J_i) \exp\{Z_i(T_{i0}) f(T_{i0} - V_i ; \beta) + \alpha(T_{i0})\}\\
    &\implies h_{00i} = \frac{1}{1 + P_i \exp\{Z_i(T_{i0}) f(T_{i0} - V_i ; \beta) + \alpha(T_{i0})\}} \\
    0= \frac{\partial \ell_{\text{full}}}{\partial h_{01i}} &= -S_i J_i + \frac{J_i}{h_{01i}} - J_i \exp\{Z_i(T_{i1}) f(T_{i1} - V_i ; \beta) + \alpha(T_{i1})\} \\
    &\implies h_{01i} = \frac{1}{S_i + \exp\{Z_i(T_{i1}) f(T_{i1} - V_i ; \beta) + \alpha(T_{i1})\}}
\end{align*}
Now we compute the MLEs of the nuisance parameters specified above.
\[
\resizebox{\textwidth}{!}{$
\begin{aligned}
h_{0i}(T_{i0}) &= \frac{(1-J_i)}{1 + P_i \exp\{Z_i(T_{i0}) f(T_{i0} - V_i ; \beta) + \alpha(T_{i0})\}} \\
h_{0i}(T_{i1}) &= \frac{J_i}{S_i + \exp\{Z_i(T_{i1}) f(T_{i1} - V_i ; \beta) + \alpha(T_{i1})\}} \\
\int_0^{T_{i0}} h_{0i}(u) du &= \frac{(1-J_i)}{1 + P_i \exp\{Z_i(T_{i0}) f(T_{i0} - V_i ; \beta) + \alpha(T_{i0})\}} + \frac{S_i J_i}{S_i + \exp\{Z_i(T_{i1}) f(T_{i1} - V_i ; \beta) + \alpha(T_{i1})\}} \\
\int_0^{T_{i1}} h_{0i}(u) \exp\{Z_i(u) [f(u - V_i, \beta)] + \alpha(u)\} du &= \frac{J_i  \exp\{Z_i(T_{i1}) [f(T_{i1} - V_i, \beta)] + \alpha(T_{i1})\}}{S_i + \exp\{Z_i(T_{i1}) f(T_{i1} - V_i ; \beta) + \alpha(T_{i1})\}} + \frac{P_i (1-J_i)  \exp\{Z_i(T_{i0}) [f(T_{i0} - V_i, \beta)] + \alpha(T_{i0})\}}{1 + P_i \exp\{Z_i(T_{i0}) f(T_{i0} - V_i ; \beta) + \alpha(T_{i0})\}}
\end{aligned}
$}
\]

Combining the cumulative hazards in the manner specified in the full likelihood leads to cancellation
\[
\resizebox{\textwidth}{!}{$
\begin{aligned}
    &\int_0^{T_{i0}} h_{0i}(u) du  + \int_0^{T_{i1}} h_{0i}(u) \exp\{Z_i(u) [f(u - V_i, \beta)] + \alpha(u)\} du = [1+P_i \exp\{Z_i(T_{i0}) f(T_{i0} - V_i ; \beta) + \alpha(T_{i0})\}] \\
    &\times \left(\frac{(1-J_i)}{1 + P_i \exp\{Z_i(T_{i0}) f(T_{i0} - V_i ; \beta) + \alpha(T_{i0})\}}\right) +\left(\frac{J_i \times [S+\exp\{Z_i(T_{i1}) f(T_{i1} - V_i ; \beta) + \alpha(T_{i1})\}]}{S_i + \exp\{Z_i(T_{i1}) f(T_{i1} - V_i ; \beta) + \alpha(T_{i1})\}}\right) \\
    &= (1-J_i) + J_i
\end{aligned}
$}
\]
Which does not depend on any parameters of interest. Hence, we can ignore the contributions of the cumulative hazards in computing the likelihood.

Now, we plug the MLEs for the nuisance parameters back into the original full data likelihood to profile out the nuisances. We obtain
\[
\resizebox{\textwidth}{!}{$
\begin{aligned}
    L(\beta, \alpha) &= \prod_{i=1}^n \left(\frac{1}{1 + P_i \exp\{Z_i(T_{i0}) f(T_{i0} - V_i ; \beta) + \alpha(T_{i0})\}}\right)^{1-J_i} \left(\frac{\exp\{Z_i(T_{i1}) f(T_{i1} - V_i ; \beta) + \alpha(T_{i1})\}}{S_i + \exp\{Z_i(T_{i1}) f(T_{i1} - V_i ; \beta) + \alpha(T_{i1})\}}\right)^{J_i}
\end{aligned}
$}
\]
Which only depends on the target parameters $(\beta, \alpha)$. The profile likelihood is shown to be exactly equivalent to the individually-stratified Cox partial likelihood. It is also equivalent to the semiparametric logistic regression model shown in the main text.

\subsection{Identification of hazard-based statistical parameter}

As a starting point, we consider the cause-specific hazard ratio of primary infection marginalized over the distribution of $U(t)$. Let $p_{U1}$ refer to the density of $U(t)$ among participants vaccinated $\tau=t-v$ days ago and uninfected at time $t$, and let $p_{U0}$ refer to the density of $U(t)$ among participants unvaccinated at $t$ and uninfected at time $t$.
\[
\resizebox{\textwidth}{!}{$
\begin{aligned}
\frac{h_1(t | Z(t)=1, \tau = t-V)}{h_1(t | Z(t)=0, \tau = 0)} &= \lim_{\epsilon \downarrow 0} \frac{\int P(J=1, T \in [t, t + \epsilon) | Z(t)=1, \tau = t-v, U(t), T \geq t) p_{U1}(u|t) du}{\int P(J=1, T \in [t, t + \epsilon) | Z(t)=0, \tau = 0, U(t), T \geq t) p_{U0}(u|t) du} \\
    &=\lim_{\epsilon \downarrow 0} \frac{\int \frac{P(J=1, T \in [t, t + \epsilon) | Z(t)=1, \tau = t-v, U(t), T \geq t)}{P(J=1, T \in [t, t + \epsilon) | Z(t)=0, \tau = 0, U(t), T \geq t)} P(J=1, T \in [t, t + \epsilon) | Z(t)=0, \tau = 0, U(t), T \geq t) p_{U1}(u|t) du}{\int P(J=1, T \in [t, t + \epsilon) | Z(t)=0, \tau = 0, U(t), T \geq t) p_{U0}(u|t) du} \\
    &= \exp\{f(t-v)\} \left(\lim_{\epsilon \downarrow 0} \frac{E_{U \sim p_{U1}}[P(J=1, T \in [t, t + \epsilon) | Z(t)=0, \tau = 0, U(t)=U, T \geq t)]}{E_{U \sim p_{U0}}[P(J=1, T \in [t, t + \epsilon) | Z(t)=0, \tau = 0, U(t)=U, T \geq t)]}\right)
\end{aligned}
$}
\]
The first equality follows from the definition of cause-specific hazards and iterated expectation, the second follows from a multiply/divide trick, the third follows from simplifying the ratio in the numerator from previous step using Assumption 3 and rewriting the integrals as expectations.

Invoking very similar arguments to the vaccine-preventable hazard ratio, we can simplify the vaccine irrelevant hazard ratio as follows.
\[
\resizebox{\textwidth}{!}{$
\begin{aligned}
    \frac{h_0(t | Z(t)=1, \tau = t-V)}{h_0(t | Z(t)=0, \tau = 0)} &=\lim_{\epsilon \downarrow 0} \frac{\int P(J=0, T \in [t, t + \epsilon) | Z(t)=1, \tau = t-v, U(t), T \geq t) p_{U1}(u) du}{\int P(J=0, T \in [t, t + \epsilon) | Z(t)=0, \tau = 0, U(t), T \geq t) p_{U0}(u) du} \\
    &= \lim_{\epsilon \downarrow 0} \frac{\int \frac{P(J=0, T \in [t, t + \epsilon) | Z(t)=1, \tau = t-v, U(t), T \geq t)}{P(J=0, T \in [t, t + \epsilon) | Z(t)=0, \tau = 0, U(t), T \geq t)} P(J=0, T \in [t, t + \epsilon) | Z(t)=0, \tau = 0, U(t), T \geq t) p_{U1}(u) du}{\int P(J=0, T \in [t, t + \epsilon) | Z(t)=0, \tau = 0, U(t), T \geq t) p_{U0}(u) du} \\
    &= \lim_{\epsilon \downarrow 0} \frac{E_{U \sim p_{U1}}[P(J=0, T \in [t, t + \epsilon) | Z(t)=0, \tau = 0, U(t)=U, T \geq t)]}{E_{U \sim p_{U0}}[P(J=0, T \in [t, t + \epsilon) | Z(t)=0, \tau = 0, U(t)=U, T \geq t)]}
\end{aligned}
$}
\]
Under Assumption 4, the bias terms are equal. Hence, the ratio of observed cause-specific hazard ratios identifies the target parameter.
\begin{align*}
   \frac{h_1(t | Z(t)=1, \tau = t-V)}{h_1(t | Z(t)=0, \tau = 0)} \times \left(\frac{h_0(t | Z(t)=1, \tau = t-V)}{h_0(t | Z(t)=0, \tau = 0)}\right)^{-1} = f(t-v)
\end{align*}
We can further simplify the ratio of hazard ratios.
\[
\resizebox{\textwidth}{!}{$
\begin{aligned}
f(t-v) &= \underset{\epsilon \downarrow 0}{\lim} \frac{P(J=1, T \in [t, t + \epsilon) \mid Z(t)=1, \tau = t-v, T \geq t)}{P(J=1, T \in [t, t + \epsilon) \mid Z(t)=0, \tau = 0, T \geq t)} \times \left(\frac{P(J=0, T \in [t, t + \epsilon) \mid Z(t)=1, \tau = t-v, T \geq t)}{P(J=0, T \in [t, t + \epsilon) \mid Z(t)=0, \tau = 0, T \geq t)}\right)^{-1} \\
&= \underset{\epsilon \downarrow 0}{\lim} \frac{P(J=1 \mid T \in [t, t + \epsilon), Z(t)=1, \tau = t-v, T \geq t)}{P(J=1 \mid T \in [t, t + \epsilon), Z(t)=0, \tau = 0, T \geq t)} \times \left(\frac{P(J=0 \mid T \in [t, t + \epsilon), Z(t)=1, \tau = t-v, T \geq t)}{P(J=0 \mid T \in [t, t + \epsilon), Z(t)=0, \tau = 0, T \geq t)}\right)^{-1} \\
&= \frac{P(J=1 \mid T =t, Z(t)=1, \tau = t-v)}{P(J=1 \mid T=t, Z(t)=0, \tau = 0)} \times \left(\frac{P(J=0 \mid T=t, Z(t)=1, \tau = t-v)}{P(J=0 \mid T=t, Z(t)=0, \tau = 0)}\right)^{-1}
\end{aligned}
$}
\]
Where the first equality is the definition of the observed cause-specific hazards, the second follows from multiplying and dividing by $P(T \in [t, t+\epsilon) | Z(t) =1, \tau=t-v, T \geq t)/P(T \in [t, t+\epsilon) | Z(t) =0, \tau=0, T \geq t)$ and applying conditional probability rules, and the final step involves evaluating the limit.

In conclusion, we have shown that the hazard-based statistical parameter of interest $f()$ is identified by the odds ratio of infection causes among infections at time $T$ between vaccinated and unvaccinated participants.

\subsection{Identification of Causal Effect}

Under Assumptions 5-7, we can write the causal quantity $\theta_C(t-v)$ as follows.
\begingroup
    \setlength{\abovedisplayskip}{6pt}   % space above
    \setlength{\belowdisplayskip}{6pt}
    \begin{align*}
        \theta_C(t-v) &= \underset{\epsilon \downarrow 0}{\lim} \; \frac{P(T_1(v,0) \in [t, t+\epsilon)| T(v,0) \geq t, V=v)}{P(T_1(\infty,t) \in [t, t+\epsilon)| T(v,0) \geq t, V=v)} \\
        &= \underset{\epsilon \downarrow 0}{\lim} \; \frac{P(T_1(v,0) \in [t, t+\epsilon)| T(v,0) \geq t, V=v)}{\alpha_0(t) \times P(T_0(\infty,t) \in [t, t+\epsilon)| T(v,0) \geq t, V=v)} \\
        &= \underset{\epsilon \downarrow 0}{\lim} \; \frac{P(T_1(v,0) \in [t, t+\epsilon)| T(v,0) \geq t, V=v)}{\alpha_0(t) \times P(T_0(v,t) \in [t, t+\epsilon)| T(v,0) \geq t, V=v)} \\
        &= \underset{\epsilon \downarrow 0}{\lim} \; \frac{P(T_1(v,0) \in [t, t+\epsilon)| T(v,0) \geq t, V=v)}{P(T_0(v,0) \in [t, t+\epsilon)| T(v,0) \geq t, V=v)} \times [\alpha_0(t)]^{-1} \\
        &= \underset{\epsilon \downarrow 0}{\lim} \; \frac{P(T_1 \in [t, t+\epsilon)| T \geq t, V=v)}{P(T_0 \in [t, t+\epsilon)| T \geq t, V=v)} \times [\alpha_0(t)]^{-1} \\
        &= \underset{\epsilon \downarrow 0}{\lim} \; \frac{P(J=1, T \in [t, t+\epsilon)| T \geq t, V=v)}{P(J=0, T \in [t, t+\epsilon)| T \geq t, V=v)} \times [\alpha_0(t)]^{-1}
    \end{align*}
    \endgroup
Where the first equality is restating the definition of $\theta_C(t-v)$, the second equality follows from Assumption 5, the third equality from Assumption 6, the fourth from applying Assumption 7 to the denominator, the fifth from consistency of potential outcomes, and the last from rewriting the events $\{T_j \in [t, t+\epsilon)\}$ in an equivalent way as $\{J=j, T \in [t, t+\epsilon)\}$.

To finish the proof, all that is left is to identify $\alpha_0(t)$.
\begin{align*}
    \alpha_0(t) &= \underset{\epsilon \downarrow 0}{\lim} \; \frac{P(T_1(\infty, t) \in [t, t+\epsilon)\mid T(\infty, 0) \geq t, V=\infty)}{P(T_0(\infty, t) \in [t, t+\epsilon)\mid T(\infty, 0) \geq t, V=\infty)} \\
    &= \underset{\epsilon \downarrow 0}{\lim} \; \frac{P(T_1(\infty, 0) \in [t, t+\epsilon)\mid T(\infty, 0) \geq t, V=\infty)}{P(T_0(\infty, 0) \in [t, t+\epsilon)\mid T(\infty, 0) \geq t, V=\infty)} \\
    &= \underset{\epsilon \downarrow 0}{\lim} \frac{P(T_1 \in [t, t+\epsilon)\mid T \geq t, V=\infty)}{P(T_0 \in [t, t+\epsilon)\mid T \geq t, V=\infty)} \\
    &= \underset{\epsilon \downarrow 0}{\lim} \frac{P(J=1, T \in [t, t+\epsilon)\mid T \geq t, V=\infty)}{P(J=0, T \in [t, t+\epsilon)\mid T \geq t, V=\infty)}
\end{align*}
Where the first equality follows from Assumption 5, the second follows from applying Assumption 7 to the numerator and denominator, the third follows from applying consistency of potential outcomes, and the last follows by rewriting the events $\{T_j \in [t, t+\epsilon)\}$ in an equivalent way as $\{J=j, T \in [t, t+\epsilon)\}$.

Hence, after cancellation, $\theta_C(t-v)$ is identified by the observable ratio of cause-specific hazard ratios.
\[
\resizebox{\textwidth}{!}{$
\begin{aligned}
    \theta_C(t-v) &= \underset{\epsilon \downarrow 0}{\lim} \; \frac{P(J=1, T \in [t, t+\epsilon)| T \geq t, V=v)}{P(J=0, T \in [t, t+\epsilon)| T \geq t, V=v)} \times \left(\frac{P(J=1, T \in [t, t+\epsilon)\mid T \geq t, V=\infty)}{P(J=0, T \in [t, t+\epsilon)\mid T \geq t, V=\infty)}\right)^{-1}
\end{aligned}
$}
\]
By applying identical arguments as the previous subsection, we can show that the causal parameter is identified as the odds ratio of infection causes at time $T=t$ between vaccinated and unvaccinated participants.
\begin{align*}
    \theta_C(t-v) &= \frac{P(J=1 \mid T = t, V=v)}{P(J=0 \mid T = t, V=v)} \times \left(\frac{P(J=1 \mid T=t, V=\infty)}{P(J=0 \mid T=t, V=\infty)}\right)^{-1}
\end{align*}

\subsubsection{Sensitivity Analysis}

Assumption 5 posits that in the absence of vaccination, the ratio of vaccine-preventable to vaccine-irrelevant infections will be determined by the relative community infection pressures and will invariant to the particular subgroup or timing of isolation. While we judge this assumption to be reasonable, it may be violated if different subgroups experience fundamentally different community infection pressures. For example, this could occur if participants with different vaccination statuses reside in different geographical areas with different burdens of vaccine-preventable or vaccine-irrelevant infections. Differences in infection burdens could arise from differences in natural immunity caused by prior infection within the communities that vaccinated and unvaccinated participants reside.

To accommodate situations where Assumption 5 may be violated, we propose a sensitivity analysis that relaxes Assumption 5 to assume that the community infection pressures for different subgroups are unequal but proportional. Let $c_0(t)$ refer to a known, time-varying function that links the cross-cause hazard ratio of vaccine-preventable to vaccine-irrelevant infections between subgroups.
\begingroup
\setlength{\abovedisplayskip}{6pt}   % space above
\setlength{\belowdisplayskip}{6pt}
\[
\resizebox{\textwidth}{!}{$
\begin{aligned}
     \frac{P(T_1(\infty, t) \in [t, t+\epsilon) | T(v,0) \geq t, V=v)}{P(T_0(\infty, t) \in [t, t+\epsilon) | T(v,0) \geq t, V=v)} = c_0(t) \times \frac{P(T_1(\infty, t) \in [t, t+\epsilon) | T(\infty,0) \geq t, V=\infty)}{P(T_0(\infty, t) \in [t, t+\epsilon) | T(\infty,0) \geq t, V=\infty)}
\end{aligned}
$}\]
\endgroup
In this case, $\theta_C(t-v)$ would be identified by the odds ratio of infection causes among those infected at time $t$ up to an analyst-defined plausible range of functions $c_0(t)$.

\section{Proof of Proposition 1}

A complete proof of the consistency and asymptotic normality of the semiparametric logistic regression model can be found in \citet{chen_asymptotically_1995}. We review the key elements of their proof.

\subsection{Notation and conditions}

Note that the semiparametric logistic regression model can be written as follows.
\begin{align*}
    E[J|X] = \text{expit}\{\beta_0^T I(V \leq T) \ubar{\psi}(T-V) - \alpha_0(T)\}
\end{align*}
Where $\alpha_0(t)$ is an unspecified function defined on the positive real numbers and $\beta_0 \in \mathbb{R}^d$ is the Euclidean parameter of interest describing the time-varying VE. We use a finite-dimensional basis expansion of time-since-vaccination, $\{\psi_j(\cdot)\}_{j=1}^d$. The simplest possible specification of the basis expansion is a polynomial expansion of degree 1 ($d=2$), meaning that $\psi_1(T-V)=1, \psi_2(T-V)=T-V$, which assumes that VE changes linearly in time-since-vaccination on the log hazard/odds scale.

Suppose without loss of generality that $T \in [0,1]$. The sieve estimator will approximate $\alpha_0$ with a basis expansion that increases with the sample size. The approach will be to divide $[0,1]$ into $K$ equally spaced knots. Let $\ubar{\phi}_{K} \equiv \{\phi_j(v,K)(\cdot)\}_{j=1}^M$ denote the collection of $M = K + v$ B-splines of degree $v$ with $K$ knots.

Suppose we approximate the true unknown function $\alpha_0(t)$ with the basis expansion associated with $\ubar{\phi}_K$.
\begin{align*}
    \alpha_0(t) \approx \sum_{j=1}^M \alpha_j \phi_{j}(t)
\end{align*}
With $\ubar{\alpha} \in \mathbb{R}^M$. Let $\theta_K(x)$ be the linear predictor associated with the logistic regression model with basis expansion for $\alpha_0(t)$ specified above.
\begin{align*}
    \theta_K(t,v) = \beta^T I(v \leq t) \ubar{\psi}(t-v) + \alpha^T \ubar{\phi}_K(t)
\end{align*}
Moreover, define the following matrices
\begin{align*}
    \mathcal{W} \in \mathbb{R}^n \times \mathbb{R}^d \text{ s.t. } \mathcal{W}_{ij} := I(V_i \leq T_i) \psi_j(T_i-V_i) \\
    \mathcal{I} \in \mathbb{R}^n \times \mathbb{R}^M \text{ s.t. } \mathcal{I}_{ij} := \phi_{Kj}(T_i-V_i)
\end{align*}
We define a Hölder class $\mathcal{G}(v_0, \gamma, c_0)$ as the collection of functions $g$ such that
\begin{align*}
    |g^{v_0}(x_1) - g^{(v_0)}(x_0)| \leq c_0 |x_1 - x_0|^{\gamma} \hspace{5mm} \forall x_1, x_0 \in [0,1]
\end{align*}

We introduce the following conditions.

\begin{condition}[Time-varying VE function lies in parametric class]
     Assume that the time-varying VE function $f(\cdot)$ lies in a function class spanned by the $d$-dimensional basis $\ubar{\psi} := \{\psi_j\}_{j=1}^d$ with support on $\mathbb{R}^+$. In other words, for some $\beta_0 \in R^0 \subset \mathbb{R}^d$, where $R$ is a bounded and convex set and $R^0$ is the interior of $R$.
    \begin{align*}
        f(t-v) = \beta_0^T \ubar{\psi}(t-v)
    \end{align*}
\end{condition}

\begin{condition}[Nuisance parameter lies in Hölder Class]
    Suppose that $\alpha_0(t)$ lies in the class of functions $\mathcal{G}(v_0, \gamma, c_0)$ for $\gamma \in [0,1]$ consisting of functions satisfying the Hölder condition.
    \begin{align*}
        |g^{(v_0)}(x_1) - g^{(v_0)}(x_0)| \leq c_0 | x_1-x_0|^{\gamma} \hspace{5mm} \forall \; x_1,x_0 \in [0,1]
    \end{align*}
    Where $c_0$ is a fixed positive constant. Some common choices are Lipschitz functions $(v_0,\gamma)=(0,1)$, functions with bounded derivatives $(v_0, \gamma) = (1,0)$, and functions with bounded higher-order derivatives $(v_0, \gamma)=(r,0)$. Define the smoothness index $p = v_0 + \gamma$. Assume the $\alpha_0(t)$ is also uniformly bounded.
    \begin{align*}
        \sup_{t \in [0,1]} |\alpha_0(t)| < \infty
    \end{align*}
\end{condition}

\begin{condition}[Nuisance basis expansion]
    Let $\ubar{\phi}_K := \{\phi_{j}(\cdot ; K, v)\}_{j=1}^{M}$ refer to collection of $M = K + v$  B-splines of degree $v$ on $[0,1]$ with the following knot placements at $j/K$ for $1 \leq j < K$. Let $S(v,K)$ denote the vector space of dimension $M$ spanned by the basis.
\end{condition}

\begin{condition}[Positive support]
For any $a \in \mathbb{R}^{d}$, $a \neq 0$, and $c \in \mathbb{R}$,
\begin{align*}
P(a^T I(V \leq T) \ubar{\psi}(T-V) \neq c) > 0    
\end{align*}
This condition is required to ensure that the distribution of the linear predictor does not concentrate in a lower-dimensional hyperplane, ensuring $\beta_0$ is identifiable.
\end{condition}

\begin{condition}[Size of sieve space]
    Suppose $K_n = O_p(n^{\lambda})$ for some $0 < \lambda$ and $K_n/n = o_p(1)$.
\end{condition}

\subsection{Proof of Proposition 1}

Recall that $(\hat{\beta}_{K}, \hat{\alpha}_{K})$ is a solution to a series of likelihood equations associated with the semiparametric logistic regression model with linear predictor $\theta_K$. Using the Taylor series expansion of the log likelihood, we can write the score with respect to each parameter as follows
\begin{align*}
    \frac{\partial \ell_n(\beta, \alpha)}{\partial \beta_j}\Bigg|_{\beta = \beta_{K0}\atop
    \alpha=\alpha_{K0}} &= \sum_{i=1}^n \expit{\theta^*_{K}} [\mathcal{W}_i^T (\hat{\beta}_K - \hat{\beta}_{K0}) + \mathcal{I}_{i}^T (\hat{\alpha}_K - \hat{\alpha}_{K0})] I(v_i \leq t_i)  \psi_{j}(t_i-v_i) \\
    \frac{\partial \ell_n(\beta, \alpha)}{\partial \alpha_k}\Bigg|_{\beta = \beta_{K0}\atop
    \alpha=\alpha_{K0}} &= \sum_{i=1}^n \expit{\theta^*_{K}} [\mathcal{W}_i^T (\hat{\beta}_K - \hat{\beta}_{K0}) + \mathcal{I}_{i}^T (\hat{\alpha}_K - \hat{\alpha}_{K0})] \phi_{k}(t_i)
\end{align*}
Where $\theta_K^* = (1-\lambda_n)\hat{\theta}_{K} + \lambda_n \theta_{K0}$ for some $\lambda_n \in [0,1]$. Set $\ubar{Y} = (J_1, \ldots, J_n)^T$, $\ubar{D}_{n} = (-\expit{\theta_{K0}(\ubar{x}_1)},\ldots, -\expit{\theta_{K0}(\ubar{x}_n)})^T$, and let $\mathcal{A}_n$ be a diagonal matrix with $ii$th entry $\exp\{\theta_k^*(x_i)\}/(1+\exp\{\theta_k^*(x_i)\})^2$. Define $\bar{\mathcal{W}}^T = \mathcal{W}^T \mathcal{A}_n^{1/2}$ and $\bar{\mathcal{I}} = \mathcal{I}^T \mathcal{A}_n^{1/2}$. We can rewrite the score equations as follows.
\begin{align*}
    \begin{pmatrix}
        \bar{\mathcal{W}}^T \bar{\mathcal{W}} & \bar{\mathcal{W}}^T \bar{\mathcal{I}} \\
        \bar{\mathcal{I}}^T \bar{\mathcal{W}} & \bar{\mathcal{I}}^T \bar{\mathcal{I}}
    \end{pmatrix} \begin{pmatrix}
        \hat{\beta}_{K} - \beta_{K0} \\
        \hat{\alpha}_{K} - \alpha_{K0}
    \end{pmatrix} &= \begin{pmatrix}
        \bar{\mathcal{W}}^T \\
        \bar{\mathcal{I}}^T
    \end{pmatrix} \mathcal{A}_{n}^{-1/2} [\ubar{Y} + \ubar{D}_n]
\end{align*}
Applying the Frisch–Waugh–Lovell theorem, we can obtain an expression in terms of the parameter of interest, $(\hat{\beta}_{K}-\beta_{K0})$, without dependence on $\alpha$.
\begin{align*}
    (\mathcal{W}^T \mathcal{A}_n^{1/2} (\textbf{I}-\textbf{P}_n) \mathcal{A}_n^{1/2} \mathcal{W})(\hat{\beta}_{K} - \beta_{K0}) = \mathcal{W}^T \mathcal{A}_n^{1/2} (\textbf{I}-\textbf{P}_n) \mathcal{A}_n^{-1/2} (\ubar{Y}-\ubar{D}_n)
\end{align*}
Where $\mathbf{P}_n = \mathcal{A}_n^{1/2} \mathcal{I} (\mathcal{I}^T \mathcal{A}_n \mathcal{I})^{-1} \mathcal{I}^T \mathcal{A}_n^{1/2}$. We recognize that each element of $\ubar{Y}-\ubar{D}_n$ is a residual of the form $Y_i - \expit{\theta_{K0}(X_i)} = Y_i - \expit{\theta_0(X_i)}+ \expit{\theta_0(X_i)} - \expit{\theta_{K0}(X_i)}$. Let the vector $\ubar{\epsilon}_{Kn}$ capture the random error (with elements $Y_i - \expit{\theta_0(X_i)}$) and let the vector $\ubar{b}_n$ reflect the approximation error (with elements $\expit{\theta_0(X_i)} - \expit{\theta_{K0}(X_i)}$). 
\begin{align*}
    (\mathcal{W}^T \mathcal{A}_n^{1/2} (\textbf{I}-\textbf{P}_n) \mathcal{A}_n^{1/2} \mathcal{W})(\hat{\beta}_{K} - \beta_{K0}) = \mathcal{W}^T \mathcal{A}_n^{1/2} (\textbf{I}-\textbf{P}_n) \mathcal{A}_n^{-1/2} (\ubar{\epsilon}_{Kn} + \ubar{b}_n)
\end{align*}
We now note that the asymptotic behavior of $\sqrt{n}(\hat{\beta}_{K} - \beta_{K0})$ depends on three terms.
\begin{enumerate}
    \item $n^{-1} \mathcal{W}^T \mathcal{A}_n^{1/2} (\textbf{I}-\textbf{P}_n) \mathcal{A}_n^{1/2} \mathcal{W}$
    \item $n^{-1/2} \mathcal{W}^T \mathcal{A}_n^{1/2} (\textbf{I}-\textbf{P}_n) \mathcal{A}_n^{-1/2} \ubar{b}_n$
    \item $n^{-1/2} \mathcal{W}^T \mathcal{A}_n^{1/2} (\textbf{I}-\textbf{P}_n) \mathcal{A}_n^{-1/2} \ubar{\epsilon}_{Kn}$
\end{enumerate}

By Lemma 9 in \citet{chen_asymptotically_1995}, the first term converges in probability to a matrix.
\begin{align*}
    n^{-1} \mathcal{W}^T \mathcal{A}_n^{1/2} (\textbf{I}-\textbf{P}_n) \mathcal{A}_n^{1/2} \mathcal{W} \overset{p}{\rightarrow} \mathbf{\Sigma}
\end{align*}
In Section 6.3.1 of \citet{chen_asymptotically_1995}, we can establish a stochastic upper bound on the second term.
\begin{align*}
    |n^{-1/2} \mathcal{W}^T \mathcal{A}_n^{1/2} (\textbf{I}-\textbf{P}_n) \mathcal{A}_n^{-1/2} \ubar{b}_n| = O_P(n^{-1/2} M K^{-p}) + O_p(K^{-p})
\end{align*}
The third term is of particular interest. We note that for any $\ubar{a}^T \in \mathbb{R}^d$, we write 
\begin{align*}
    \ubar{a}^T \mathcal{W}^T \mathcal{A}_0^{1/2} (\textbf{I}-\textbf{P}_0) \mathcal{A}_0^{-1/2} \ubar{\epsilon}_{Kn} = \ubar{a}^T \mathcal{W}^T \mathcal{A}_0^{1/2} (\textbf{I}-\textbf{P}_0) \ubar{e}_n
\end{align*}
Where $e_n$ has elements $e_i = \epsilon_i/\sigma_i$ where $\epsilon_i = J_i - \expit{\theta_0(X_i)}$ and $\sigma_i = [\mathcal{A}_{0}^{1/2}]_{ii}$. We can write $n^{-1/2} \ubar{a}^T \mathcal{W}^T \mathcal{A}_0^{1/2} (\textbf{I}-\textbf{P}_0) \ubar{e}_n$ as $\sum_{j=1}^n c_j e_j$ where $E[e_j]=0$, $\text{Var}(e_j)=1$, and $E[e_j^4]<\infty$. By Lemma 9(a) in \citet{chen_asymptotically_1995}, $\sum_{j=1}^n c_j^2 = O_p(1)$. 

Defining $\ubar{a}^T \mathcal{W}^T \mathcal{A}_0^{1/2} = (g_1, \ldots, g_n)$, it follows from Conditions 2 and 4 that $\max_{1 \leq i \leq n} |g_i| = O_p(1)$ and $\sum_{i=1}^n g_i^2 = O_p(n)$. Let $p_{ij}$ denote the $(ij)$-th element of $\mathbf{P}_0$. It follows from Lemma 8 of \citet{chen_asymptotically_1995} that $\sum_{i=1}^n p_{ij}^2 = p_{ii} = O_p(n^{-1}M)$. Hence,
\begin{align*}
    \sum_{j=1}^n |c_j|^3 &\leq \max_{1 \leq j \leq n} |c_j| \sum_{j=1}^n c_j^2 \\
    &\leq n^{-1/2} \max_{1 \leq j \leq n} \left[|g_j| + \left(\sum_{i=1}^n g_i^2 \sum_{i=1}^n p_{ij}^2\right)^{1/2} \right] \sum_{j=1}^n c_j^2 \\
    &= n^{-1/2}\left[O(1) + (O(n) O_p(n^{-1} M))^{1/2}\right] O_p(1) \\
    &= o_p(1)
\end{align*}
The first inequality is implied by the maximum being an upper bound. The second inequality comes from finding an upper bound on $c_j$ using the triangle inequality. The penultimate equality comes from substituting in the rates for each step, and the final equality comes from manipulating the stochastic order notation.

Hence, by the Cramér-Wald device and the Liapounov central limit theorem, we obtain
\begin{align*}
    n^{-1/2} \mathcal{W}^T \mathcal{A}_0^{1/2} (\textbf{I}-\textbf{P}_0) \ubar{e}_n \rightsquigarrow N(0, \Sigma)
\end{align*}
And equivalently
\begin{align*}
    n^{-1/2} \mathcal{W}^T \mathcal{A}_0^{1/2} (\textbf{I}-\textbf{P}_0) \mathcal{A}_0^{-1/2} \ubar{\epsilon}_{Kn} \rightsquigarrow N(0, \Sigma)
\end{align*}

Rewinding to the start of the proof, recall
{\footnotesize
\begin{align*}
    \underbrace{n^{-1} (\mathcal{W}^T \mathcal{A}_n^{1/2} (\textbf{I}-\textbf{P}_n) \mathcal{A}_n^{1/2} \mathcal{W})}_{\overset{p}{\rightarrow} \mathbf{\Sigma}} \sqrt{n}(\hat{\beta}_{K} - \beta_{K0}) = \underbrace{n^{-1/2}\mathcal{W}^T \mathcal{A}_n^{1/2} (\textbf{I}-\textbf{P}_n) \mathcal{A}_n^{-1/2} \ubar{\epsilon}_{Kn}}_{\star} + \underbrace{n^{-1/2}\mathcal{W}^T \mathcal{A}_n^{1/2} (\textbf{I}-\textbf{P}_n) \mathcal{A}_n^{-1/2} \ubar{b}_n}_{O_p(n^{-1/2} M K^{-p}) + O_p(K^{-p})}
\end{align*}}
By Equation (21) in \citet{chen_asymptotically_1995},
\begin{align*}
    &\mathcal{W}^T \mathcal{A}_n^{1/2} (\textbf{I}-\textbf{P}_n) \mathcal{A}_n^{1/2} \ubar{\epsilon}_{K} \\
    &= \mathcal{W}^T \mathcal{A}_0^{1/2} (\textbf{I}-\textbf{P}_0) \mathcal{A}_0^{-1/2} \ubar{\epsilon}_{K} + \mathcal{W}^T (\mathcal{A}_0 - \mathcal{A}_n) \mathcal{A}_0^{-1/2} \textbf{P}_0 \mathcal{A}_0^{-1/2} \ubar{\epsilon}_{K} \\
    &+ \mathcal{W}^T \mathcal{A}_n \mathcal{I}\left[(\mathcal{I}^T \mathcal{A}_0 \mathcal{I})^{-1} - (\mathcal{I}^T \mathcal{A}_n \mathcal{I})^{-1}\right] \mathcal{I}^T \ubar{\epsilon}_{K}
\end{align*}
The first term on the right hand side looks familiar from the previous CLT result. The second term can be stochastically bounded using Proposition 1 from \citet{chen_asymptotically_1995}.
\begin{align*}
    \mathcal{W}^T(\mathcal{A}_0 - \mathcal{A}_n)\mathcal{A}_0^{-1/2} \mathbf{P}_0 \mathcal{A}_0^{-1/2} \ubar{\epsilon}_{K} = O_P((nK^{-2p}M)^{1/2}) + O_P(M(\log(n)^{3/2})
\end{align*}
The third term can be stochastically bounded using Proposition 2 from \citet{chen_asymptotically_1995}.
\begin{align*}
    \mathcal{W}^T \mathcal{A}_n \mathcal{I}\left[(\mathcal{I}^T \mathcal{A}_0 \mathcal{I})^{-1} - (\mathcal{I}^T \mathcal{A}_n \mathcal{I})^{-1}\right] \mathcal{I}^T \ubar{\epsilon}_{K} &= O_P(n^{-1/2} M^{5/2} \log(n)) + O_P(M^{3/2})
\end{align*}

Hence, in order for $\star \rightsquigarrow N(0, \mathbf{\Sigma})$, we require that each of the stochastically bounded terms go to zero. Recall that $M \approx K \approx n^{\lambda}$.
\begin{enumerate}
    \item $\lim_{n \rightarrow \infty} n^{-1/2} M K^{-p} = \lim_{n \rightarrow \infty}  n^{\lambda (1-p) - 1/2} \implies \lambda(1-p) < 1/2$
    \item $\lim_{n \rightarrow \infty} n^{-1/2} (n K^{-2 p} M)^{1/2} = \lim_{n \rightarrow \infty} n^{\frac{\lambda(1-2p)}{2}} \implies \lambda (1-2p) < 0 \implies p > 1/2$
    \item $\lim_{n \rightarrow \infty} n^{-1/2} M^{3/2} = \lim_{n \rightarrow \infty} n^{3\lambda/2 - 1/2} \implies \lambda < 1/3$
    \item $\lim_{n \rightarrow \infty} n^{-1/2} M \log\{n\}^{3/2} = \lim_{n \rightarrow \infty} n^{\lambda - 1/2} \implies \lambda < 1/2$
    \item $\lim_{n \rightarrow \infty} n^{-1/2} (n^{-1/2} M^{5/2} \log\{n\}) = \lim_{n \rightarrow \infty} n^{-1 + 5\lambda/2} \implies \lambda < 2/5$
    \item $\lim_{n \rightarrow \infty} K^{-p} = \lim_{n \rightarrow \infty} n^{-\lambda p} \implies \lambda, p > 0$
\end{enumerate}
Taken together, we conclude that $0 < \lambda < 1/3$ and $p > 1/2$. When this occurs, the second and third terms on the right hand side of (21) by \citet{chen_asymptotically_1995} go to zero. Hence,
\begin{align*}
    \sqrt{n}(\hat{\beta}_{K} - \beta_{K0}) \rightsquigarrow N(0, \mathbf{\Sigma}^{-1})
\end{align*}
Moreover, from Theorem 1(b) in \citet{chen_asymptotically_1995}, $|\beta_{K0} - \beta_0| = O(K^{-p})$. Hence, 
\begin{align*}
    \sqrt{n}(\hat{\beta}_{K} - \beta_{0}) = \underbrace{\sqrt{n}(\hat{\beta}_{K} - \beta_{K0})}_{\rightsquigarrow N(0, \mathbf{\Sigma}^{-1})} + \underbrace{\sqrt{n}(\beta_{K0}-\beta_0)}_{O(n^{-\lambda p + 1/2})}
\end{align*}
Hence, under the additional undersmoothing condition $\lambda p > 1/2 \equiv p > 1/(2 \lambda)$, the latter term is $o_P(1)$, which resolves the asymptotic distribution of $\sqrt{n}(\hat{\beta}_{K} - \beta_0)$.

\section{Proof of Proposition 2}

Suppose our data of size $n$ are sampled independently and identically from $P_0 \sim M$, where $M$ is the model containing distributions satisfying the following conditional mean equality.
\begin{align*}
    Q_0(V,T) \equiv P(J=1|V,T) = \expit{\beta_0^T I(V \leq T) \cdot \ubar{\psi}(T-V) + \alpha_0(T)} \equiv \expit{\beta_0^T \ubar{Z} + \alpha_0(T)}
\end{align*}
For some $\beta_0 \in \mathbb{R}^d$ and some function $\alpha_0$. In some cases, we may use $\ubar{Z} := I(V \leq T)  \cdot \ubar{\psi}(T-V)$ as shorthand to denote the derived regressors. We will pursue estimation and inference on $\beta_0$ using a \textit{targeted maximum likelihood estimation (TMLE) procedure}. 

The TMLE algorithm first obtains initial estimates of the relevant portions of the distribution of the observed data. Then the initial fits are updated to optimize the bias-variance tradeoff. This is operationalized by updating the initial fits using a logistic working model that solves the efficient score equation -- defined as the efficient influence function -- in the semiparametric model defined above through the use of a ``clever covariate".

\subsection{Derivation of EIC}

A key component of our TMLE algorithm is defining the efficient influence function/efficient influence curve of $\beta_0$ in the semiparametric partially-linear logistic regression model. To identify the efficient influence curve, we note that an influence function that is orthogonal to the nuisance tangent space (i.e., scores obtained by fluctuation $\alpha_0$) is the unique EIF. Hence, a good place to start is identifying the form of the nuisance tangent space. The nuisance tangent space will take the following form.
\begin{align*}
    T_{nuis, \alpha_0}(P_0) = \{h(T) (J - Q_0(V,T)) : h \in L_2\}
\end{align*}
For arbitrary squared integrable function $h$. We wish to construct a path $Q(\epsilon)$ through $Q$ such that the score at $\epsilon=0$ is orthogonal to the nuisance tangent space and is hence the EIC. Consider the candidate paths that fluctuate $\alpha_0(t)$ in the direction indexed by $h_1$ and fluctuate $\beta_0$ in the direction $b$.
\begin{align*}
    Q_{0,b,h_1}(\epsilon) = \expit{(\beta_0^T + \epsilon \cdot b^T) \ubar{Z} + \alpha_0(T) + \epsilon h_1(T)}
\end{align*}
The score of this path at $\epsilon=0$ is given by
\begin{align*}
    \ubar{S}(h_1, \ubar{b}) \equiv \left(\ubar{b}^T \ubar{Z} + h_1(T)\right) (J - Q_0(A,V,T))
\end{align*}
Now we need to identify the forms of $h_1$ and $b$ such that orthogonality holds, meaning for any $h(T)$, we have
\begin{align*}
    0 &= \mathbb{E}[S(h_1,b) \times h(T)(J-Q_0(A,V,T))] \\
    &= \mathbb{E}[\left(\ubar{b}^T \ubar{Z} + h_1(T)\right) (J - Q_0(A,V,T)) \times h(T)(J-Q_0(A,V,T))]  \\
    &= \mathbb{E}[\left(\ubar{b}^T \ubar{Z} + h_1(T)\right) \times h(T) \times (J-Q_0(A,V,T))^2] \\
    &= \mathbb{E}[\left(\ubar{b}^T \ubar{Z} + h_1(T)\right) \times h(T) \times Q_0(A,V,T)(1-Q_0(A,V,T))] ]] \\
    &= \mathbb{E}[\left(\ubar{b}^T I(V \leq T) \ubar{\psi}(T-V) + h_1(T)\right) \times h(T) \times Q_0(A,V,T)(1-Q_0(A,V,T))]
\end{align*}
Where the penultimate equality follows from Tower law with respect to $J$. The unique solution in $h_1$ is given by the following.
\begin{align*}
    h_1^*(Q_0)(t) &= -\ubar{b}^T \frac{\mathbb{E}_V[I(V \leq T) \ubar{\psi}(T-V) Q_0(1-Q_0)(V,T)|T=t]}{\mathbb{E}_V[Q_0(1-Q_0)(V,T)|T=t]}
\end{align*}
Plugging back into the equation for the score, we obtain
\begin{align*}
    \ubar{S}(h^*_1, \ubar{b}) &= \ubar{b}^T \left(I(V \leq T) \ubar{\psi}(T-V) - \frac{\mathbb{E}_V[I(V \leq T) \ubar{\psi}(T-V) Q_0(1-Q_0)(V,T)|T]}{\mathbb{E}_V[Q_0(1-Q_0)(V,T)|T]}\right) (J - Q_0(V,T))
\end{align*}
So for any arbitrary direction $\ubar{b}$, the efficient directional score is given above. This implies that the EIF (or efficient score vector) is proportional to the following.
\begin{align*}
    D(Q_0)(J,V,T) &\propto \left(J - Q_0(V,T)\right) \left(I(V \leq T) \ubar{\psi}(T-V) - r_0(T)\right) \\
    r_0(T) &:= \frac{\mathbb{E}_V[I(V \leq T) \ubar{\psi}(T-V) Q_0(1-Q_0)(V,T)|T]}{\mathbb{E}_V[Q_0(1-Q_0)(V,T)|T]}
\end{align*}
To correctly identify the EIF, we must scale by the information matrix, defined below as follows.
\begin{align*}
    I_{eff} &:= \mathbb{E}[D(Q_0) D(Q_0)^T] \\
    &= \mathbb{E}\left[\left(\ubar{Z} - r_0(T)\right)\left(\ubar{Z} - r_0(T)\right)^T \left(J - Q_0(V,T)\right)^2\right] \\
    &= \mathbb{E}_{T}\left[\mathbb{E}_{V}\left[\left(\ubar{Z} - r_0(T)\right)\left(\ubar{Z} - r_0(T)\right)^T Q_0(V,T)\left(1 - Q_0(V,T)\right) \Big| T \right] \right] \\
    &= \mathbb{E}_{T}\left[\mathbb{E}_{V}\left[\ubar{Z} \ubar{Z}^T Q_0(V,T)\left(1 - Q_0(V,T)\right) \Big| T \right] - r_0(T)r_0(T)^T \mathbb{E}_{V}(Q_0(V,T)\left(1 - Q_0(V,T)\right)|T) \right] \\
     &= \mathbb{E}_{T}\left[\mathbb{E}_{V}(Q_0(V,T)\left(1 - Q_0(V,T)\right)|T)\left(\frac{\mathbb{E}_{V}\left[\ubar{Z} \ubar{Z}^T Q_0(V,T)\left(1 - Q_0(V,T)\right) \Big| T \right]}{\mathbb{E}_{V}(Q_0(V,T)\left(1 - Q_0(V,T)\right)|T)} - r_0(T)r_0(T)^T\right)  \right]
\end{align*}
Where the first equality is definitional, the second substitutes from above, the third follows by iterated expectation, the fourth from reducing the cross terms, and the fifth from using a multiply divide trick.

Hence, the EIF is as follows.
\begin{align*}
    D(Q_0) &= [I_{eff}]^{-1} \left(I(V \leq T) \ubar{\psi}(T-V) - r_0(T)\right) \left(J - Q_0(V,T)\right) 
\end{align*}

\subsection{Description of TMLE estimator}

Now that the EIC is derived, we can compute the TMLE estimator as follows. Compute initial (i.e. flexible) estimators of $Q_0$ (or equivalently, $\beta_0$ and $\alpha(t)$) using flexible models such as generalized additive model or random forest. Let $Q^0$ be defined as your initial estimate and let $p_i^{0} = \expit{I(V_i \leq T_i) \ubar{\psi}(T_i-V_i) \beta^{(0)} + \alpha^{(0)}(T_i)}$. Then for each iteration $k = 0, 1, 2, \ldots$, estimate the following quantities.
\begin{align*}
    w_i^{(k)} &= p_i^{(k)} (1-p_i^{(k)}) \\
    r^{(k)}(t) &= \frac{\mathbb{E}_V[I(V_i \leq T_i) \ubar{\psi}(T_i-V_i) w_i^{(k)}|T=t]}{\mathbb{E}_V[w_i^{(k)}|T=t]}
\end{align*}
The latter quantity can be estimated by kernel smoothing, local regression, or fitting a GAM for each column of $I(V_i \leq T_i) \ubar{\psi}(T_i-V_i)$ with weight $w$. Let $r_i^{(k)} := r^{(k)}(T_i)$ denote the vector at each $T_i$.

Then, compute the clever covariate $H_i^{(k)}$ which is a $n \times d$ matrix.
\begin{align*}
    H_i^{(k)} &:= I(V_i \leq T_i) \ubar{\psi}(T_i-V_i) - r^{(k)}(T_i)
\end{align*}
Next, we regress $J$ on the clever covariate with the previous iteration's probabilities as an offset to solve for the update $\epsilon^{(k)}$.
\begin{align*}
    \text{logit}(p_i^{(new)}) &= \text{logit}(p_i^{(k)}) + \epsilon^T H_i^{(k)}
\end{align*}
After updating the fit, we update $\beta$ and $\alpha(t)$ as follows.
\begin{align*}
    \beta^{(k+1)} &\leftarrow \beta^{(k)} + \epsilon^{(k)} \\
    \alpha^{(k+1)}(t) &\leftarrow \text{smooth}_t\{\text{logit}(p_i^{(new)}) - I(V_i \leq T_i) \ubar{\psi}(T_i-V_i)\beta^{(k+1)}\}
\end{align*}
Where $\text{smooth}_t$ refers to fitting a smoother over $t$ using smoothing spline or a GAM. This represents a profile smoothing update to $\alpha^{(k)}$, as $\beta^{(k+1)}$ is treated as known.

Finally, we recompute the event probabilities.
\begin{align*}
    p_i^{(k+1)} = \expit{I(V_i \leq T_i) \ubar{\psi}(T_i-V_i)^T \beta^{(k+1)} + \alpha^{(k+1)}(T_i)}
\end{align*}

This iterative process is repeated until the magnitude of the nuisance parameter updates, $\epsilon$, is small. Specifically, we repeat until $||\epsilon^{(k)}|| < \text{tol}$ where $\text{tol}$ is a tolerance level (e.g., $10^{-6}$). Our final estimate $\hat{\beta}_{TMLE}$ is obtained when the tolerance criterion is reached, and let $p_i^{final}$ refer to the final targeted fitted probabilities.

The variance of the $\hat{\beta}_{TMLE}$ is estimated by the $n^{-1}$-scaled empirical variance fo the EIF,
\begin{align*}
    \widehat{\text{Var}}(\hat{\beta}_{TMLE}) \approx n^{-1} \left(\frac{1}{n} \sum_{i=1}^n \hat{D}_{P^*_n}(O_i)  \hat{D}_{P^*_n}(O_i)^T \right)
\end{align*}
Where $\hat{D}_{P^*_n}(O_i) = \hat{I}_{eff}^{-1} (J_i - p_i^{final}) H_i^{(final)}$ and $\hat{I}_{eff} = P_n[(J-p^{(final)}) H^{(final)}H^{(final)T}]$.

\subsection{Consistency and Asymptotic normality}

Let $P_n^*$ be the targeted distribution for $\beta(P)$ obtained after updating the initial estimator $P_{n,0} \in \mathcal{M}$. Denote $\tilde{r}_n = \tilde{r}_{P_{n,0}}$ and $Q^*_n := Q_{P_n^*}$. Following the results by \citet{vdl_gilbert_2025}, we propose the following conditions which $\beta(P_n^*)$ is asymptotically linear and efficient for $\beta(P)$. 

\begin{condition}[Uniformly bounded and invertible]
    $\ubar{\psi}(T-V)$ if $P$-uniformly bounded and $\mathbb{E}_P[\ubar{\psi}(T-V) \ubar{\psi}(T-V)^T]$ is an invertible matrix.
\end{condition}

\begin{condition}[Positivity]
    There exists some scalar $\delta > 0$ such that $\mathbb{P}(1-\delta > Q_n^*(V,T) >\delta) = P(1-\delta > Q_0(V,T) > \delta) =1$.
\end{condition}

\begin{condition}[Donsker]
    The estimators $\tilde{r}_n$ and $Q_n^*$ fall with probability one in a uniformly bounded $P$-Donsker function class.
\end{condition}

Note that Condition S8 is not strictly necessary, and can be relaxed to include more flexible machine learning estimators that use sample-splitting and cross-fitting \citep{vanderLaanRose2011, chernozhukov_2018}.

\begin{condition}[Sufficient nuisance convergence rates]
$||\tilde{r}_n - r_0|| = o_P(n^{-1/4})$ and $||Q_n^* - Q_0|| = o_P(n^{-1/4})$ where $||\cdot||$ refers to the $L^2_P$ norm.
\end{condition}

The proof of the CAN of $\beta_{TMLE} := \beta(P_n^*)$ is standard and we refer to \citet{vanderLaanRose2011} for details. It is based on writing the difference between the targeted estimator and targeted parameter in three terms.
\begin{align*}
    \beta(P_n^*) - \beta(P) =\underbrace{-P_n D_{P_n^*}}_{\circled{1}} + \underbrace{(P_n - P) D_{P_n^*}}_{\circled{2}} + \underbrace{\{\beta(P_n^*) - \beta(P) + P D_{P_n^*})}_{\circled{3}}
\end{align*}
The first term is $o_P(n^{-1/2})$ by construction of the TMLE, since we choose our estimate of the relevant components of the data generating distribution to solve the efficient score equation (at least at $o_P(n^{-1/2})$ rate). The third term is $o_P(n^{-1/2})$ based on Lemma 1 of \citet{vdl_gilbert_2025}. Therefore, the expansion takes the form
\begin{align*}
    \beta(P_n^*) - \beta(P) &= (P_n - P) D_{P_n^*} + o_P(n^{-1/2}) \\
    &= (P_n - P) D_{P} + (P_n - P) \{D_{P_n^*}-D_P\} + o_P(n^{-1/2})
\end{align*}
Under Condition S7, $D_P$ is uniformly bounded with finite variance. By the CLT, $\sqrt{n}(P_n - P)D_P$ goes to a normally distributed variable with the desired covariance structure. By Slutsky's theorem, what remains is to show
\begin{align*}
    (P_n - P) \{D_{P_n^*}-D_P\} = o_P(n^{-1/2})
\end{align*}
This empirical process term is negligible when $D_{P_n^*}-D_P$ falls in a Donsker class with probability  1. To establish this, if the nuisance estimators lie in $P$-Donsker classes (Condition S8) and $\tilde{r}_n \rightarrow r_0$ and $Q_n^* \rightarrow Q_0$ converge in $L^2(P)$ at appropriate rates (Condition S9), then the class $\{D(\cdot, Q,r)\}$ is $P$-Donsker. Because the Donsker class is invariant to Lipschitz transformations, $D_{P_n^*}-D_P$ lies in a Donsker class. Stochastic equicontinuity of empirical processes over Donsker classes gives $(P_n - P)\{D_{P_n^*} - D_P\} = o_P(n^{-1/2})$ as desired.

\section{Example Data Application}

Consider a hypothetical study that enrolls $n=10,000$ individuals at a given calendar date and follows them for 1 year. Suppose that $85\%$ of study participants are vaccinated at some point during follow-up and $15\%$ are not. 

Suppose that participant-level hazard of experiencing infection are determined by an unmeasured frailty variable, $U_{i}$. Suppose $U_i \sim \text{logNormal}(0, \sigma_U = 2.5)$ consistent with a very heterogeneous frailty distribution. Suppose that vaccination date is drawn independently from $V \sim \text{Unif}(0, 1.15)$ and any vaccination date greater than 1 is treated as never vaccinated. There was no association between vaccination date and no post-vaccination disinhibition. We assumed 

We assume the true hazards functions for the vaccine-preventable and vaccine-irrelevant infections are as follows
\begin{align*}
    h_{1}(t) &= h_{01}(t) \exp(I(V_i \leq t)[\beta_0 + \beta_1 (t-V_i) + U_{i1}-U_{i0}]+U_{i0}) \\
    h_0(t) &= h_{00}(t) \exp(I(V_i \leq t)[U_{i1}-U_{i0}]+U_{i0})
\end{align*}
Where $h_0^Y(t) = 0.16 \times (1+0.5\sin(t * 2\pi))$ and $h_0^N(t)=0.08$. Hence, the hazard of COVID-19 fluctuates sinusoidally in calendar time while the hazard of off-target infection is constant, implying a sinusoidal change-of-scale parameter $\alpha_0(t)$. The vaccine has no effect on the hazard of vaccine-irrelevant infection but does affect the hazard of vaccine-preventable infections. The time-varying VE curve $f(\tau) = \beta_0 + \beta_1 \tau$ is linear in time-since-vaccination on the log hazard scale. We consider two scenarios. Scenario 1 assumes constant, moderate VE over time ($\beta_0=-1, \beta_1=0$) while scenario 2 assumes a moderate initial VE that wanes linearly on the log-hazard scale to zero after one year ($\beta_0=-1, \beta_1=1$).

In our example data application, we will compare (i) naive Cox regression, (ii) Sieve estimation approach, and (iii) TMLE approach. We assume correctly specified linear models for the VE waning model in each case. We applied a monotone non-decreasing constraint to the parameter estimates obtaind using the sieve and TMLE methods, and obtained associated 95\% confidence intervals by projecting monte carlo samples from the unconstrained asymptotic distribution onto the constrained space.

\begin{figure}
    \centering
    \includegraphics[width=\linewidth]{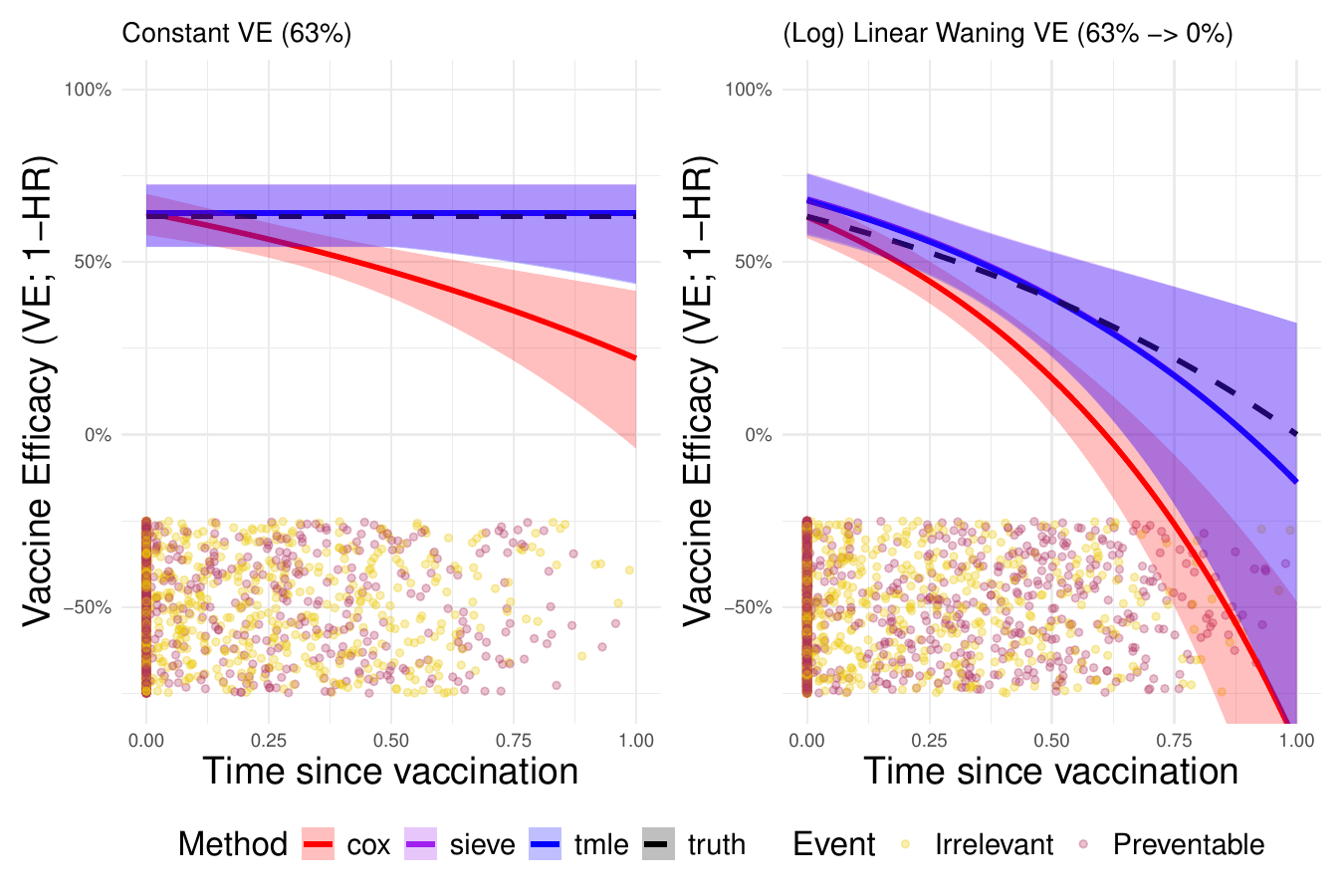}
    \caption{Results of example data application comparing naive Cox regression to our proposed estimators. Sample size $n=10,000$, emulating an observational study where there is hidden association between early vaccination and higher susceptibility to infection and participants exhibit behavioral disinhibition and increased pathogen exposure post-vaccination.}
    \label{fig:vignette}
\end{figure}

We show our results in \ref{fig:vignette}. As anticipated, the standard Cox regression estimator of the time-varying VE curve exhibited exaggerated VE waning due to depletion of susceptibles bias. Qualitatively, the Cox regression model incorrectly inferred that VE was waning even in the setting where VE was constant. In contrast, both the TMLE and sieve estimation approaches which leveraged the vaccine-irrelevant infections both exhibited low bias and correctly identified the true time-varying VE shape under both waning and constant scenarios. Moreover, the confidence intervals for the TMLE and sieve estimators both covered the true VE function well. As anticipated, the confidence intervals for the TMLE and sieve estimators were wider than for cox regression, since they did not make use of censored participants in their estimator. 

\section{Numerical Experiments}

The goals of our simulations were twofold (i) to confirm the asymptotic properties of our proposed estimators and (ii) to compare the performance of our proposed estimators to a standard Cox regression analysis which ignored vaccine irrelevant infections.

We evaluated performance across two settings: a constant VE setting where vaccine efficacy was constant over time at $63\%$ ($f(\tau)=-1$) and a linear VE waning setting where efficacy waned linearly on the log hazard scale from $63\%$ immediately after vaccination to $0\%$ after 1 year ($f(\tau) = -1 + \tau$). In all simulations, we assumed that the baseline hazard of off-target events was constant over time $\lambda_{00}(t) = 0.03$ but the baseline hazard for vaccine-preventable infections followed a sinusoidal pattern, $\lambda_{01}(t) = \lambda_{01} (1-0.5\sin(2\pi t))$, meaning $\alpha_0(t)$ was sinusoidal. All our simulated studies were administrative censored at 1 year and contained 10,000 participants. We assumed that on average, approximately 85\% of participants received the vaccine and 15\% of participants remained unvaccinated by simulating vaccination date $V \sim \text{Unif}(0,1.15)$. If the simulated vaccination date was greater than 1, they were considered unvaccinated.

We varied several different parameters that could challenge our ability to reliably estimate time-varying VE estimates. 
\begin{enumerate}
    \item Unmeasured heterogeneity in infection risk: we simulated frailties $U_i \sim \text{lognormal}(0, \sigma_U)$ and included them as unknown multiplicative factors in the hazard of both vaccine-preventable and vaccine-irrelevant infections.
    \item Background vaccine-preventable infection intensity: prior work has suggested that depletion of susceptibles bias can be magnified during periods of high burden of infection. Hence, we varied the vaccine preventable infection burden by modulating $\lambda_{01} \in \{0.03, 0.06, 0.12\}$ reflecting low, moderate, and high infection burdens.
    \item Behavioral disinhibition: to reflect changes in behavior influencing infection risk after vaccination, $(U_{0}, U_{1})$ from a standard bivariate log-normal distribution with marginal distributions $\text{lognormal}(0, \sigma_U=1)$ and $\text{lognormal}(1, \sigma_U=1)$ and weak correlation $\rho = 0.2$ The associated multiplicative contribution to the hazards is then $I(V \leq t) U_0 + I(V > t) U_1$, where the mean shift from $U_0$ to $U_1$ reflects the case where average participant risk increases after vaccination.
    \item Non-random booster assignment: when booster assignment was random, the frailties $U$ and vaccination date $V$ were simulated separately from normal and uniform distributions. To accommodate a scenario where high-risk, susceptible individuals are prioritized for early vaccination, we simulated $(U, V)$ from a bivariate normal copula model with marginal distributions $\text{lognormal}(0,1)$ and $\text{Unif}(0, 1.15)$ with a negative correlation parameter $\rho = -0.7$. This copula model ensured that early vaccination date was associated with higher values of the frailty.
\end{enumerate}

We considered three different estimation strategies for the time-varying VE function. We considered a Cox regression model estimated using partial likelihood for the vaccine-preventable outcome, which ignored the vaccine-irrelevant infections. The Cox model used a correctly-specified (log) linear model for the time-varying VE function $f(\cdot)$, however, the standard Cox model was misspecified because it did not account for the frailties. The other two estimators were the semiparametric logistic regression (SLR) model using the sieve estimation and TMLE estimation frameworks, with correctly specified linear models for $f(\cdot)$. The sieve estimator approximated the unknown nuisance parameter $\alpha_0(t)$ using a cubic B-spline model with number of knots equal to the cubic root of the number of events. The TMLE estimator estimated $\alpha_0(t)$ and each of the $r_0(T)$ functions using generalized additive model with thin-plate splines with default knots choices and degrees of freedom. After each TMLE update step, estimates of the linear predictors and $\alpha_0(t)$ were updated using the clever covariate using a profile approach which first updated the linear predictor and then updated $\alpha_0(T)$. After each update of $\alpha_0(t)$, we re-smoothed $\alpha_0(t)$ in calendar time using a generalized additive model.

The results of our numerical experiments can be found in Tables \ref{tab:cover} and \ref{tab:MSE}. In Table \ref{tab:cover}, we summarize the empirical coverage of 95\% confidence intervals across 4 different bias scenarios -- low heterogeneity in infection risk, high heterogeneity in infection risk, behavioral disinhibition, and vaccinating vulnerable groups early. Confidence interval coverage for Cox and SLR estimators were both comparable in settings with low risk heterogeneity and were close to the nominal 95\% level. However, in the setting with high risk heterogeneity, the Cox model had coverage estimates that were far below the nominal level while both SLR models had estimated coverages very close to the nominal level. In the remaining two scenarios -- behavioral disinhibition and a hidden association between risk and early vaccination, the standard Cox model had near zero coverage probability while the SLR models maintained 95\% coverage. In Table \ref{tab:MSE}, we observe that all estimators had comparable mean squared errors in the low risk heterogeneity setting, but that the Cox model showed larger bias in the presence of high risk heterogeneity, behavioral disinhibition, and non-random vaccination dates.

\begin{table}[!h]
\centering
\caption{Empirical 95\% confidence interval coverage probabilities across 1000 replicates of different hypothetical COVID-19 vaccine studies. In all trials, $\lambda_{01} = 0.06$ and boost probability is $p = 0.85$ drawn from uniform distribution $[0,1]$. Each simulated study enrolled $n = 10{,}000$.}
\begin{adjustbox}{width=\textwidth}
\centering
\fontsize{10}{12}\selectfont
\begin{tabular}[t]{llrlrrrrrr}
\toprule
\multicolumn{4}{c}{ } & \multicolumn{3}{c}{\textbf{Constant VE}} & \multicolumn{3}{c}{\textbf{Linear VE}} \\
\makecell{Vaccination\\Assignment} & Disinhibition & \makecell{Force of Vaccine-\\Preventable infection ($\lambda_{01}$)} & \makecell{Risk Hetero-\\geneity ($\sigma_U$)} & Cox & \makecell{SLR\\(Sieve)} & \makecell{SLR\\(TMLE)} & Cox & \makecell{SLR\\(Sieve)} & \makecell{SLR\\(TMLE)}\\
\midrule
\cellcolor{gray!10}{Random} & \cellcolor{gray!10}{N} & \cellcolor{gray!10}{0.06} & \cellcolor{gray!10}{Low ($\sigma_U=1$)} & \cellcolor{gray!10}{93.8} & \cellcolor{gray!10}{96.9} & \cellcolor{gray!10}{96.9} & \cellcolor{gray!10}{93.8} & \cellcolor{gray!10}{95.3} & \cellcolor{gray!10}{95.4}\\
Random & N & 0.06 & High ($\sigma_U=2$) & 38.2 & 95.7 & 95.7 & 53.7 & 94.0 & 94.1\\
\cellcolor{gray!10}{Random} & \cellcolor{gray!10}{Y} & \cellcolor{gray!10}{0.06} & \cellcolor{gray!10}{Low ($\sigma_U=1$)} & \cellcolor{gray!10}{0.4} & \cellcolor{gray!10}{96.8} & \cellcolor{gray!10}{96.8} & \cellcolor{gray!10}{0.5} & \cellcolor{gray!10}{96.1} & \cellcolor{gray!10}{96.1}\\
Vulnerable & N & 0.06 & Low ($\sigma_U=1$) & 0.5 & 96.2 & 96.3 & 0.4 & 95.0 & 94.9\\
\bottomrule
\end{tabular}
\end{adjustbox}\label{tab:cover}
\end{table}

\begin{table}[!h]
\centering
\caption{Empirical mean squared error across 1000 replicates of different hypothetical COVID-19 vaccine studies. In all trials, $\lambda_{01} = 0.06$ and boost probability is $p = 0.85$ drawn from uniform distribution $[0,1]$. Each simulated study enrolled $n = 10{,}000$.}
\begin{adjustbox}{width=\textwidth}
\fontsize{10}{12}\selectfont
\begin{tabular}[t]{llrlrrrrrr}
\toprule
\multicolumn{4}{c}{ } & \multicolumn{3}{c}{\textbf{Constant VE}} & \multicolumn{3}{c}{\textbf{Linear VE}} \\
\makecell{Vaccination\\Assignment} & Disinhibition & \makecell{Force of Vaccine-\\Preventable infection ($\lambda_{01}$)} & \makecell{Risk Hetero-\\geneity ($\sigma_U$)} & Cox & \makecell{SLR\\(Sieve)} & \makecell{SLR\\(TMLE)} & Cox & \makecell{SLR\\(Sieve)} & \makecell{SLR\\(TMLE)}\\
\midrule
\cellcolor{gray!10}{Random} & \cellcolor{gray!10}{N} & \cellcolor{gray!10}{0.06} & \cellcolor{gray!10}{Low ($\sigma_U=1$)} & \cellcolor{gray!10}{0.040} & \cellcolor{gray!10}{0.042} & \cellcolor{gray!10}{0.041} & \cellcolor{gray!10}{0.034} & \cellcolor{gray!10}{0.050} & \cellcolor{gray!10}{0.049}\\
Random & N & 0.06 & High ($\sigma_U=2$) & 0.097 & 0.032 & 0.032 & 0.044 & 0.037 & 0.036\\
\cellcolor{gray!10}{Random} & \cellcolor{gray!10}{Y} & \cellcolor{gray!10}{0.06} & \cellcolor{gray!10}{Low ($\sigma_U=1$)} & \cellcolor{gray!10}{1.041} & \cellcolor{gray!10}{0.024} & \cellcolor{gray!10}{0.023} & \cellcolor{gray!10}{0.988} & \cellcolor{gray!10}{0.026} & \cellcolor{gray!10}{0.026}\\
Vulnerable & N & 0.06 & Low ($\sigma_U=1$) & 2.821 & 0.043 & 0.042 & 2.911 & 0.048 & 0.047\\
\bottomrule
\end{tabular}
\end{adjustbox}\label{tab:MSE}
\end{table}

Additionally, we sought to understand the impact of force of vaccine-preventable infection on bias in the time-varying VE curve. The results are shown in Table \ref{tab:foi_effect}. When there was low heterogeneity in infection risk, higher force of vaccine-preventable infections slightly eroded the confidence interval coverage for the standard Cox model, while SLR models exhibited robust coverage. When risk heterogeneity was high, confidence intervals for the standard Cox model were far below the nominal level and rapidly eroded with higher force of infection. In every case, the SLR models exhibited robust coverage near the nominal level.

\begin{table}[!h]
\centering
\caption{Empirical coverage probabilities when $\lambda_{01} = 0.06$ and boost probability is $p = 0.85$ uniformly over $[0,1]$. $n = 10{,}000$.}
\centering
\begin{adjustbox}{width=\textwidth}
\fontsize{10}{12}\selectfont
\begin{tabular}[t]{llrlrrrrrr}
\toprule
\multicolumn{4}{c}{ } & \multicolumn{3}{c}{Constant VE} & \multicolumn{3}{c}{Linear VE} \\
\makecell{Vaccination\\Assignment} & Disinhibition & \makecell{Force of Vaccine-\\Preventable infection ($\lambda_{01}$)} & \makecell{Risk Hetero-\\geneity ($\sigma_U$)} & Cox & \makecell{SLR\\(Sieve)} & \makecell{SLR\\(TMLE)} & Cox & \makecell{SLR\\(Sieve)} & \makecell{SLR\\(TMLE)}\\
\midrule
\cellcolor{gray!10}{Random} & \cellcolor{gray!10}{N} & \cellcolor{gray!10}{0.03} & \cellcolor{gray!10}{Low ($\sigma_U=1$)} & \cellcolor{gray!10}{94.8} & \cellcolor{gray!10}{96.1} & \cellcolor{gray!10}{96.2} & \cellcolor{gray!10}{94.4} & \cellcolor{gray!10}{94.7} & \cellcolor{gray!10}{94.8}\\
Random & N & 0.06 & Low ($\sigma_U=1$) & 93.8 & 96.9 & 96.9 & 93.8 & 95.3 & 95.4\\
\cellcolor{gray!10}{Random} & \cellcolor{gray!10}{N} & \cellcolor{gray!10}{0.12} & \cellcolor{gray!10}{Low ($\sigma_U=1$)} & \cellcolor{gray!10}{89.4} & \cellcolor{gray!10}{96.4} & \cellcolor{gray!10}{96.4} & \cellcolor{gray!10}{90.3} & \cellcolor{gray!10}{94.8} & \cellcolor{gray!10}{94.8}\\
Random & N & 0.03 & High ($\sigma_U=2$) & 66.1 & 96.6 & 96.7 & 75.5 & 95.3 & 95.4\\
\cellcolor{gray!10}{Random} & \cellcolor{gray!10}{N} & \cellcolor{gray!10}{0.06} & \cellcolor{gray!10}{High ($\sigma_U=2$)} & \cellcolor{gray!10}{38.2} & \cellcolor{gray!10}{95.7} & \cellcolor{gray!10}{95.7} & \cellcolor{gray!10}{53.7} & \cellcolor{gray!10}{94.0} & \cellcolor{gray!10}{94.1}\\
\addlinespace
Random & N & 0.12 & High ($\sigma_U=2$) & 22.9 & 96.1 & 96.0 & 28.6 & 95.2 & 95.1\\
\bottomrule
\end{tabular}
\end{adjustbox}\label{tab:foi_effect}
\end{table}

\section{Extensions}\label{s:general}

\subsection{Ensuring monotonicity of time-varying VE}

Domain knowledge supports that after a short period of time for the vaccine to produce an immune response, vaccine efficacy can either remain constant or decrease, but cannot increase in time-since-vaccination. Hence, after a brief period, we expect $f(\cdot)$ to be monotonically non-decreasing. We would like our estimator to respect this constraint to ensure sensible interpretation, mitigate finite-sample bias, and improve precision.

Assume $f(\tau) = \beta_0^T \ubar{\psi}(\tau)$ for some known basis expansion $\{\psi_{j}\}_{j=1}^d$. Define $\hat{f}(\tau) = \hat{\beta}^T \ubar{\psi}(\tau)$ as the unconstrained estimate of time-varying VE, where $\hat{\beta}$ is estimated from sieve/TMLE methods.  Let $\mathcal{T} := \{\tau_1, \ldots, \tau_L\}$ denote a grid of $L$ time-since-vaccination values of interest. Let $\hat{f}_{l}(\tau)$ and $\hat{f}_{u}(\tau)$ denote the upper and lower 95\% confidence intervals of the unconstrained estimator over $\mathcal{T}$ which can be computed using the multivariate delta method.

Following results by \citet{westling_correcting_2020}, we can obtain a corrected monotonized estimator, $\hat{f}_{\text{mono}}(\tau)$, by simply running isotonic regression of the unconstrained estimator $\hat{f}$ over the grid $\mathcal{T}$. This is equivalent to projecting the unconstrained estimator onto the space of estimators satisfying $f(\tau_1) \leq f(\tau_2)$ for any $\tau_1 \leq \tau_2$. We can apply the same isotonic regression procedure to the unconstrained estimator's confidence limits ($\hat{f}_{l}(\tau)$, $\hat{f}_{u}(\tau)$) to obtain valid confidence limits for the monotonized estimator. In practice, we use precision-weighted isotonic regression to ensure that noisy values of $\hat{f}$ do not dominate the smoothing step. \citet{westling_correcting_2020} established that the estimation error of $\hat{f}_{\text{mono}}$ over the grid is never worse than $\hat{f}$ and that the monotonized confidence intervals have no worse coverage than the original confidence intervals, and comes at no cost to the bandwidth over the grid.

\subsection{Incorporating External Disease Surveillance Data}

A careful reader might notice that there are cases where there is low identifiability for the time-varying VE function, $f(\cdot)$ and the change-of-scale parameter $\alpha_0(t)$. This is due to the potential high collinearity between time-since-vaccination, $T-V$ and calendar time $T$. A pathological example is where every participant is vaccinated immediately upon enrollment, leading to perfect correlation between time-since-vaccination and calendar time. In this example, there is no ability to distinguish between waning vaccine effects and changing prevalence of different infections in the community. A more realistic example is a 1 year study where the vast majority of participants are enrolled and vaccinated within a few months. In this example, it may be difficult to distinguish between time-varying vaccine effects and fluctuating incidence of infections in the broader population. In such cases, we may wish to recover identifiability of the model parameters, in particular, the time-varying VE function $f(\cdot)$. 

The approaches we consider leverage external, population-level pathogen surveillance data to profile out the nuisance parameter $\alpha_0(t)$ and concentrate the available information on identifying $f(\cdot)$.

In our first case, we consider an analysis where the analyst has access to an external, population-level disease surveillance data (hereafter denoted as `$\text{ext}$') for both the vaccine-preventable ($J=1$) and vaccine-irrelevant ($J=0$) pathogens. Suppose that adopt the following \textit{trend transport (TT1)} assumption, which claims for all $t \geq 0$, 
\begin{align*}
    &\text{(TT1)} \hspace{1cm} \alpha_0(t) = \log\left\{\frac{h_{01}(t)}{h_{00}(t)}\right\} = \log\left\{\frac{h_{01}^\text{ext}(t)}{h_{00}^\text{ext}(t)}\right\}
\end{align*} 
The TT1 assumption implies that the change-of-scale parameter which describes the time-varying relative differences in baseline hazard between the vaccine-preventable and vaccine-irrelevant infection types also applies to the external population. Importantly, TT1 allows the hazards of infection in the study population to be arbitrarily higher or lower than the external population, as long as the \textit{baseline hazard ratio} for the two outcomes is the same. This is a critical point, since vaccine studies may be conducted in populations with intrinsically higher/lower risk of infection than the broader population.  

Given that population-level case counts are usually summarized over discrete time intervals (per day or week), a natural estimator of the hazard in the external population on date/week $t$ is the nonparametric maximum likelihood estimator of the discrete hazard function, $\hat{h}_{0j}^\text{ext}(t) = \frac{D_j^\text{ext}(t)}{R_j^\text{ext}(t)}$, where $D_j^\text{ext}(t), R_j^\text{ext}(t)$ are the number of incident outcomes of type $j$ detected at date/week $t$ and the number of people in the external population at risk of outcome $j$ on date/week $t$. In most population-level contexts, the size of risk set is very large and minimally impacted by incident cases. Making the very mild assumption that the size of the risk sets for COVID-19 and off-target infection are of similar magnitudes and are much larger than the number of incident cases, we can write the change-of-scale parameter as follows.
\begin{align*}
    \alpha_0(t) \overset{(\text{TT1})}{=} \log\left(\frac{h_{01}^{\text{ext}}(t)}{h_{02}^{\text{ext}}(t)}\right) \approx \log\left(\frac{D_{1}^{\text{ext}}(t)/R_{1}^{\text{ext}}(t)}{D_2^{\text{ext}}(t)/R_2^{\text{ext}}(t)}\right) \approx \log\left(\frac{D_1^{\text{ext}}(t)}{D_0^{\text{ext}}(t)}\right)
\end{align*}
The unknown nuisance parameter can be well-approximated by the (log) ratio of incident vaccine-preventable to vaccine-irrelevant cases at time $t$. Hence, $\log\left(D_1^{\text{ext}}(t)/D_0^{\text{ext}}(t)\right)$ can be substituted in for $\alpha_0(t)$ in a model for the odds of infection, thereby eliminating the nuisance parameter from the model. Hence, even in an entirely vaccinated cohort, the time-varying VE function can still be identified under the TT1 assumption.

In our second case, we consider an analysis where population-level disease surveillance data exists for the vaccine-preventable ($J=1$) but \textbf{not} the vaccine-irrelevant ($J=0$) pathogen. This may be the case where the vaccine-preventable pathogen is of special public-health interest and vaccine-irrelevant pathogens are not tracked. In this case, may be reasonable to assume that the incidence of vaccine-preventable pathogen varies much more in calendar time than the vaccine-irrelevant pathogen, which may exhibit more stable incidence. We formalize this intuition by making the \textit{stable disease assumption (SDA)} on vaccine-irrelevant infection and an alternative \textit{trend transport (TT2)} assumption for the vaccine-preventable infection. Formally,
\begin{align*}
    &\text{(SDA)} \hspace{1cm} h_{00}(t) = a_0 \; \forall t \geq 0
    &\text{(TT2)} \hspace{1cm} \frac{h_{01}(t_1)}{h_{01}(t_2)} = \frac{h_{01}^\text{ext}(t_1)}{h_{01}^\text{ext}(t_2)} \; \forall t_1, t_2 \geq 0
\end{align*}
SDA assumes that the hazard for off-target endpoint is constant over calendar time, although this can be relaxed to allow predictable (seasonal) variations. TT2 allows the hazard for COVID-19 to differ between study and external populations as long as the \textit{calendar-time trend} in COVID-19 incidence is the same between the two populations. Define $\alpha_0 := \log\left(\frac{\left[\prod_{i=1}^M h_{01}^{\text{ext}}(t_i)\right]^{1/M}}{h_{00}(t)}\right)$. We can express the change-of-scale parameter as 
\begin{align*}
    \alpha(t) &\; \overset{\text{(SDA)}}{=} \; \log\left(\frac{h_{01}(t)}{\left[\prod_{i=1}^M h_{01}(t_i)\right]^{1/M}}\right) +\alpha_0 \\
    &\overset{\text{(TT2)}}{=} \; \log\left(\frac{h_{01}^\text{ext}(t)}{\left[\prod_{i=1}^M h_{01}^\text{ext}(t_i)\right]^{1/M}}\right) + \alpha_0 \\
    &\approx \log\left(\frac{D_1^{\text{ext}}(t)}{\left[\prod_{i=1}^M D_1^{\text{ext}}(t_i)\right]^{1/M}}\right) + \alpha_0
\end{align*}
Where $M$ indexes a collection of calendar dates, $\{t_1, \ldots, t_M\}$ where surveillance data were available. Hence, the variation in $\alpha(t)$ over calendar time depends only on the trend in incident cases of the vaccine-preventable pathogen in the external population. The SDA and TT2 assumptions simplify the task of estimating a function-valued nuisance parameter $\alpha_0(t)$ to a single scalar $\alpha_0$, which significantly boosts identifiability of $f(\cdot)$ in the logistic regression model.

\subsection{Strain-specific time-varying VE}

Suppose the vaccine-preventable outcome consists of $m \geq 2$ distinct subtypes, such as infections caused by different strains or variants of the vaccine-preventable infection. We may wish to estimate time-varying VE against each variant separately. One strategy is to adopt a competing risk framework to accommodate different strain-specific hazard functions \citep{Prentice1978}. However, this approach would require estimating a unique change-of-scale parameter for each strain relative to the off-target infection, $\alpha_{s}(t) := \log\{h_{01s}(t)/h_{00}(t)\}$ where $s \in \{1, \ldots, m\}$. Moreover, the change-of-scale parameters are only well-defined over time periods where a strain $s$ and the vaccine-irrelevant pathogen co-circulate, which may be violated during periods where older variants disappear and new variants emerge. Our approach takes inspiration from the work by \citet{Larson1985} by modeling the unconditional (on infection) distribution of vaccine-preventable ($J=1$) versus vaccine-irrelevant ($J=0$) infections. The we model the conditional (on vaccine-preventable infection $J=1$) distribution of failure types caused by different strains.

Let $S$ refer to a categorical random variable denoting the strain of vaccine-preventable infection taking values in $\{1, \ldots, m\}$. For simplicity, we assume the following the multiplicative frailty models hold.
\begin{align*}
    h_{0}(t) &= h_{00}(t) U(t) \\
    h_{1s}(t) &= h_{01s}(t) U(t) \exp\{Z(t)[f_{s}(t-V)]\}
\end{align*}
Suppose we define $h_{01}(t) := \sum_{k=1}^m h_{01k}(t)$ as the sum of each of the baseline hazards over the $m$ vaccine-preventable infection subtypes. By multiplying and dividing by $h_{01}(t)$ and applying the change-of-scale parameter $\alpha_0(t) := \log\{h_{01}(t)/h_{00}(t)\}$ governing the relationship between any vaccine-preventable infection and off-target infection, we can rewrite the hazards above as follows.
\begin{align*}
    h_{0}(t) &= h_{00}(t) U(t) \\
    h_{1s}(t) &= h_{00}(t) p_{0s}(t) U(t) \exp\{Z(t) f_{s}(t-V) + \alpha_0(t)\}
\end{align*}
Where $p_{0s}(t) := \frac{h_{01s}(t)}{\sum_{k=1}^m h_{01k}(t)}$ is the softmax probability of experiencing an vaccine-preventable infection with strain $s$ conditional on experiencing a vaccine-preventable infection at time $t$ and not being vaccinated $Z(t)=0$ and $\log\{U(t)\}=0$. 

Another way to define $p_{0s}(t)$ is by noting that conditional on unvaccinated participant $i$ experiencing a vaccine-preventable infection $J_i=1$, the probability that the infection is of strain $S=k$ is given by
\begin{align*}
    \frac{h_{1k}(t)}{\sum_{s'=1}^m h_{1s'}(t)} &= \frac{h_{00}(t) p_{0k}(t) U(t) \exp\{\alpha_0(t)\}}{\sum_{s'=1}^m h_{00}(t) p_{0s'}(t) U(t) \exp\{\alpha_0(t)\}} \\
    &= \frac{\cancel{h_{00}(t) U(t) \exp\{\alpha_0(t)\}} \cdot p_{0k}(t)}{\cancel{h_{00}(t) U(t) \exp\{\alpha_0(t)\}} \cdot \sum_{s'=1}^m p_{0s'}(t)} \\
    &= \frac{p_{0k}(t)}{\sum_{s'}^m p_{0s'}(t)} = p_{0k}(t)
\end{align*}
Where the first equality follows from plugging in the hazard parametrizations, the second follows from grouping like terms, the third equality follows from canceling the like terms, and the fourth equality follows from recognizing that $\sum_{s'}^m p_{0s'}(t)=1$. Hence, $p_{0s}(t)$ can be interpreted as $P(S=s|T=t, J=1, Z=0)$, the probability of failure from strain $S=s$ among vaccine-preventable infections ($J=1$) occurring at time $T=t$ among unvaccinated ($Z(t)=0$) participants. As before, we can reparametrize the vaccine-preventable hazards according to $\alpha_0(t) = \log\{h_{01}(t)/h_{00}(t)\}$ which refers to the baseline hazard ratio between \textit{any} vaccine-preventable infection and vaccine-irrelevant infection. The variant mixture proportions $p_s(t)$ parametrize the differences between the strain-specific vaccine-preventable hazards.

Given there are $S \in \{1, \ldots, m\}$ possible outcomes, we could imagine running $m$ different logistic regressions comparing vaccine-preventable infections of strain $s$ to vaccine-irrelevant infections.
\begin{align*}
    \log\left\{\frac{P(J=1, S=s|V=v, T=t)}{P(J=0, S=0|V=v, T=t)}\right\} = I(v \leq t) f_s(t-v) + \alpha_0(t) + \log\left\{p_s(t)\right\}
\end{align*}
Where $\log(p_s(t))$ is an offset. Under this model, conditional on infection at time $T=t$, the probability of vaccine-preventable infection of strain $s$ is equal to the softmax probability output.
\begin{equation}\label{eq:multinomial}
    P(J=1, S=s | V=v, T=t) = \frac{p_s(t) \exp\{I(v \leq t)f_s(t-v) + \alpha_0(t)\}}{1+\sum_{s'=1}^m p_{s'}(t) \exp\{I(v \leq t)f_{s'}(t-v) + \alpha_0(t)\}}
\end{equation}

We briefly describe a few strategies on how to estimate the strain-specific mixture probabilities $\{p_s(t)\}_{s=1}^m$. One strategy is to leverage population-level disease surveillance data to obtain estimates of the strain-specific mixture probabilities, an approach referred to as surveillance anchored sieve analysis by \citet{Follmann2022}. For example, NextStrain \citep{hadfield_nextstrain_2018} or GISAID \citep{khare_gisaids_2021} provide high-resolution, population-level data on proportions of breakthrough infections caused by different viral variants. Since these are massive, population-level data sources, we can assume these variant probabilities are estimated with negligible error. As highlighted by \citet{Follmann2022}, external data may be especially useful in identifying model parameters when the majority of participants are vaccinated, since it is impossible to distinguish between waning efficacy against a particular variant and the reduction in the variant's prevalence in the population. If the variant-specific probabilities from the external data sources $p_s^{\text{ext}}(t)$ are believed to be good approximations of the study-specific variant mixture proportions among unvaccinated infections $p_s(t)$, the externally-derived probabilities can be plugged in for the unknown $p_s(t)$ and treated as fixed. This is the strategy we will pursue in the COVE case study.

In principle, it is also possible to treat $\{p_s(t)\}_{s=1}^m$ as random and conduct estimation and inference estimate using a joint likelihood/estimating equation approach. This is left as an area of future research.

In the remaining subsections, we briefly describe how to conduct inference under the sieve estimation and TMLE approaches under multiple strains of vaccine-preventable infections when the strain-specific mixture probabilities $\{p_s(t)\}_{s=1}^m$ are known. 

\subsubsection{Sieve estimation under semiparametric multinomial logistic regression model}

As before, suppose we approximate the change-of-scale parameter using a basis expansion that grows with the sample size.
\begin{align*}
    \alpha_0(t) \approx \sum_{j=1}^{M} \alpha_j \phi_{j, v}(t)
\end{align*}
We assume variant mixture proportions, $\{p_s(t)\}_{s=1}^m$, are borrowed from a population-level surveillance database and are assumed fixed. Moreover, suppose that the time-varying VE function against strain $s$ is given by $f_s(\tau) = \beta_{0s}^T \ubar{\psi}_s(t-v)$ where $\ubar{\psi}$ is a $d$-dimensional basis and $\beta_{0s}$ is the Euclidean parameter of interest. Hence, our model represents a semiparametric multinomial logistic regression model.

We rewrite the nonparametric multinomial likelihood in \ref{eq:multinomial} according to our semiparametric model as follows
\begin{align*}
    P(J=0 \mid  V=v, T=t) = \frac{1}{1+\sum_{s'=1}^m p_{s'}(t) \exp\{I(v \leq t) \beta_{0s'}^T \ubar{\psi}_{s'}(t-v) + \sum_{j=1}^{M} \alpha_j \phi_{j, v}(t)\}} \\
    P(J=1, S=s \mid V=v, T=t) = \frac{p_s(t) \exp\{I(v \leq t) \beta_{0s}^T \ubar{\psi}_s(t-v) + \sum_{j=1}^{M} \alpha_j \phi_{j, v}(t)\}}{1+\sum_{s'=1}^m p_{s'}(t) \exp\{I(v \leq t) \beta_{0s'}^T \ubar{\psi}_{s'}(t-v) + \sum_{j=1}^{M} \alpha_j \phi_{j, v}(t)\}}
\end{align*}

Then estimation and inference on $(\beta_{0s}^T\}_{s=1}^m$ and $\alpha_0^T$ can be accomplished by maximizing the multinomial likelihood associated with the models above. 

\subsubsection{TMLE under semiparametric multinomial logistic regression model}

We provide a brief overview of the approach. Recall that we can define our semiparametric model for our data as follows.
\begin{align*}
    P(J=1, S=s|V=v, T=t) &= \frac{p_s(t) \exp\{I(v \leq t) f_s(t-v) + \alpha_0(t)\}}{1 + \sum_{s'=1}^m p_{s'}(t) \exp\{I(v \leq t) f_{s'}(t-v) + \alpha_0(t)\}} \\
    P(J=0 | V=v, T=t) &= \frac{1}{1 + \sum_{s'=1}^m p_{s'}(t) \exp\{I(v \leq t) f_{s'}(t-v) + \alpha_0(t)\}}
\end{align*}
Let $Y_s = I(J=1, S=s)$ and let $Y_0 = I(J=0)$. Define the vector $\ubar{Y} = (Y_1, \ldots, Y_m)^T$ and define the corresponding unconditional event probabilities based on the formula above.
\begin{align*}
    Q_{0s}(t,v) &:= P(J=1, S=s | V=v, T=t)  \\
    Q_{0}(t,v) &:= \sum_{s=1}^m Q_{0s}(t,v)
\end{align*}

The multinomial logistic regression model is equivalent to a series of binary regressions, meaning for $s=1, \ldots, m$.
\begin{align*}
    \log\left\{\frac{Q_{0s}(t,v)}{1-Q_0(t,v)}\right\} &= \log\{p_s(t)\} + I(v \leq t) f_s(t-v ; \beta_s) + \alpha_0(t)
\end{align*}
Where $\alpha_0(t)$ is an unknown nuisance, $p_s(t)$ are known offsets, and $f_s(\cdot \; \beta_s)$ is known the time-varying VE function against strain $s$ which is known up to a scalar parameter $\beta_s$.

Suppose that $f_s(t-v ; \beta_s) = \ubar{\psi}_{s}(t-v) \beta_s$. The log-likelihood of a single contribution is given by.
\begin{align*}
    \ell &= \sum_{s=1}^m Y_s \{\log\{p_s(T)\} + I(V \leq T) \ubar{\psi}_s(T-V) \beta_s + \alpha_0(t)\} \\
    &- \log\left\{1+ \sum_{r=1}^m p_r(T)\exp\{I(V \leq T) \ubar{\psi}_r(T-V) \beta_r + \alpha_0(t)\} \right\}
\end{align*}
The \textit{parametric} score with respect to $\beta_s$ is given by
\begin{align*}
    S_{\beta_s}(O) &= I(V \leq T) \ubar{\psi}_s(T-V) \left(Y_s - Q_{0s}(T,V)\right)
\end{align*}
which follows from multinomial calculus.

To identify the efficient influence curve, we note that an influence function that is orthogonal to the nuisance tangent space (i.e., scores obtained by fluctuation $\alpha_0$) is the unique EIF. Next up is to derive the nuisance tangent space of $\alpha_0(t)$. The nuisance tangent space will take the following form.
\begin{align*}
    T_{nuis, \alpha_0}(P_0) = \{h(T) \sum_{r=1}^m (Y_r - Q_{0r}(V,T)) : h \in L_2\}
\end{align*}

Note that we wish to obtain the \textit{efficient} score by taking the parametric score and subtracting off its projection onto the nuisance tangent space. Note that the projection of $S_{\beta_s, 0}$ onto $T_{nuis, \alpha_0}$ will take the form
\begin{align*}
    a_s(T) \sum_{r=1}^m (Y_r - Q_{0r}(V,T))
\end{align*}
Where the particular choice of $a(t)$ is such that the residual is orthogonal to every $h(t)$. That is,
\begin{align*}
    a_s(t) &= \frac{\mathbb{E}[S_{\beta_{0s}} \sum_{r=1}^m (Y_r - Q_{0r}) | T=t]}{\mathbb{E}[\left(\sum_{r=1}^m (Y_r - Q_{0r})\right)^2 | T=t]} \\
    &= \frac{\mathbb{E}[I(V \leq T) \ubar{\psi}_s(T-V) (Y_s - Q_{0s}) \sum_{r=1}^m (Y_r - Q_{0r}) | T=t]}{\mathbb{E}[\left(\sum_{r=1}^m (Y_r - Q_{0r})\right)^2 | T=t]} \\
    &= \frac{\mathbb{E}[I(V \leq T) \ubar{\psi}_s(T-V) (Y_s - Q_{0s}) (J - Q_{0}) | T=t]}{\mathbb{E}[\left(J - Q_{0}\right)^2 | T=t]} \\
    &= \frac{\mathbb{E}[I(V \leq T) \ubar{\psi}_s(T-V) \mathbb{E}[(Y_s - Q_{0s}) (J - Q_{0}) \mid | T=t, V] \mid T=t]}{\mathbb{E}[\mathbb{E}[\left(J - Q_{0}\right)^2 \mid T=t, V] \mid T=t]} \\
    &= \frac{\mathbb{E}[I(V \leq T) \ubar{\psi}_s(T-V) \mathbb{E}[Y_s - Y_s Q_{0} - J Q_{0s} + Q_{0s} Q_0 \mid | T=t, V] \mid T=t]}{\mathbb{E}[ \mathbb{E}[J - 2 J Q_0 + Q_0^2 \mid T=t, V] \mid T=t]} \\
    &= \frac{\mathbb{E}_V[I(V \leq t) \ubar{\psi}_s(t-V) Q_{0s}(V,t) (1- Q_{0}(V,t)) \mid T=t]}{\mathbb{E}_V[ Q_0(V,t) (1-Q_0(V,t))\mid T=t]} 
\end{align*}
Where the first equality is to achieve orthogonality, the second follows from plugging in the formula for the score, the third follows from noting that $J = \sum_{r=1}^m Y_r$ and $Q_{0} = \sum_{r=1}^m Q_{0r}$, the fourth from iterated expectation, the fifth follows from expanding the products and simplifying binary terms, and the sixth follows from evaluating the inner expectations.

Hence, the efficient score function for $\beta_{0s}$ is given by 
\begin{align*}
    S_{\text{eff}, \beta_{0s}}(O) &= I(V \leq T) \ubar{\psi}_{s}(T-V)\{Y_s - Q_{0s}(T,V)\} - a_s(T) \sum_{r=1}^m (Y_r - Q_{0r}(V,T)) \\
    &= I(V \leq T) \ubar{\psi}_{s}(T-V)\{Y_s - Q_{0s}(T,V)\} - a_s(T) (J - Q_{0}(V,T))
\end{align*}
The EIC is proportional to the efficient score up to a scaling matrix. The EIC can be derived by scaling the efficient score by the inverse of its variance.

Before we derive the TMLE algorithm, it is important to note that a multinomial logistic regression model can be expressed as a series of independent binary regressions which compare each vaccine-preventable infection with strain $S$ to the ``pivot" outcome of vaccine-irrelevant infections. As above, we can write the model as
\begin{align*}
    \log\left(\frac{Q_{0s}(t,v)}{1-Q_0(t,v)}\right) &= \log\{p_s(t)\} + I(v \leq t) f_s(t-v; \beta_s) + \alpha_0(t)
\end{align*}
The TMLE algorithm associated with our model is writeable as follows.
\begin{enumerate}
    \item Produce initial estimators of $\{\hat{Q}^{(0)}_{0s}\}_{s=1}^m$ by estimating $\alpha_0(t)$ and $\{\beta_{0s}\}_{s=1}^m,$ using initial estimators $\hat{\alpha}_0(t)$ and $\hat{\beta}^{(0)}_{s}$, and treating the variant mixture proportions as known.
    \item Estimate the unknown nuisance parameters $\{a_s(T)\}_{s=1}^m$. We propose estimating this quantity using a Nadaraya–Watson (NW) local kernel smoothing estimator, which essentially assigns a local (kernel) weight to values of $T$ close to $t$. For instance, the N-W estimator of $a_S(T)$ is given by
    \begin{align*}
        \hat{a}^{NW}_{s}(t) &= \frac{\frac{1}{n} \sum_{i=1}^n K_h(T_i - t) I(V_i \leq t) \ubar{\psi}_s(t-V_i) \hat{Q}^{(0)}_{0s}(V_i, t) (1-\hat{Q}^{(0)}_0(V_i, t))}{\frac{1}{n} \sum_{i=1}^n K_h(T_i - t) \hat{Q}^{(0)}_0(V_i, t) (1-\hat{Q}^{(0)}_0(V_i, t))}
    \end{align*}
    Where $K_h(x) = \frac{1}{h} K\left(\frac{x}{h}\right)$ and $K$ is the kernel shape. For instance, a Gaussian kernel $K(x) = \frac{1}{\sqrt{2 \pi}} \exp\{-x^2/2\}$.
    \item For each $s$, construct the clever covariate vector.
    \begin{align*}
        H_{s}(T,V) &:= I(V \leq T) \ubar{\psi}_{s}(T-V) - \hat{a}_s^{NW}(T)
    \end{align*}
    \item Using the available study data, fluctuate the initial model predictions using a multinomial logit model which includes the original model predictions as offsets.
    \begin{align*}
        \text{log}\left(\frac{Q^{\epsilon}_{0s}(t,v)}{1-Q^{\epsilon}_0(t,v)}\right) = \text{log}\left(\frac{\hat{Q}^{(0)}_{0s}(t,v)}{1-\hat{Q}^{(0)}_0(t,v)}\right) + \epsilon_s H_s
    \end{align*}
    \item For each $s \in \{1, \ldots, m\}$, update the parameter estimates using the coefficient on the associated clever covariates: $\hat{\beta}^{(1)}_s = \hat{\beta}^{(0)}_s + \hat{\epsilon}_s H_s$. Define the updated parameter estimates $\hat{\beta}^{(1)} := (\hat{\beta}^{(1)}_{s=1}, \ldots, \hat{\beta}^{(1)}_{s=m})$.
    \item Using and $\hat{\alpha}_0(t)$ and $\hat{\beta}^{(1)}$, compute the updated event probabilities $\hat{Q}_{0s}^{(1)}$ using the multinomial model above.
    \item Iterate until the updates $(\epsilon_1, \ldots, \epsilon_m)$ are sufficiently small (below some tolerance level), which implies that the mean of the EIC is close to zero.
\end{enumerate}

\onehalfspacing
\bibliographystyle{apalike}
\bibliography{supplement_final.bbl}